\documentclass[hyper]{JHEP3}
\usepackage{amssymb}
\usepackage{amsmath}
\usepackage{dsfont}
\newcommand{\Scal}[1]{\Bigl ({#1} \Bigr )}
\newcommand{\scal}[1]{\bigl ({#1} \bigr )}
\def\bea{\begin{eqnarray}}
\def\eea{\end{eqnarray}}
\newcommand{\be}{\begin{equation}}
\newcommand{\ee}{\end{equation}}
\def\ie{{\it i.e.}\ }
\def\eg{{\it e.g.}\ }
\def\N{\mathcal{N}}
\newcommand{\CR}{\nonumber \\*}
\DeclareMathAlphabet{\mathpzc}{OT1}{pzc}{m}{it}

\newcommand{\ord}[1]{{\scriptscriptstyle (#1)}}
\newcommand{\Tgp}{{\mathrm T}^\scp}
\newcommand{\dTgp}{d{\mathrm T}^\scp}
\newcommand{\Tgm}{{\mathrm T}^\scm}
\newcommand{\dTgm}{d{\mathrm T}^\scm}
\newcommand{\Tgpm}{{\mathrm T}^\pm}
\newcommand{\dTgpm}{d{\mathrm T}^\pm}

\def\gl{\mathfrak{gl}}
\def\g{\mathfrak{g}}
\def\k{\mathfrak{k}}
\def\sl{\mathfrak{sl}}

\def\un{{\mathpzc{1}}}
\def\deux{{\mathpzc{2}}}
\def\trois{{\mathpzc{3}}}

\def\cF{{\mathcal F}}
\def\cH{{\mathcal H}}
\def\cR{{\mathcal R}}
\def\cV{{\mathcal V}}
\def\cK{{\mathcal{K}}}
\def\csK{{\scriptscriptstyle \cK}}
\def\scp{{\scriptscriptstyle +}}
\def\scm{{\scriptscriptstyle -}}
\newcommand{\Iprod}[2]{\langle {#1}, {#2} \rangle}

\def\Gammamun{{\Gamma^\ord{{\mbox{\tiny -}}1}}}
\def\Gammamtrois{{\Gamma^\ord{{\mbox{\tiny -}}3}}}
\def\kmun{{k^\ord{{\mbox{\tiny -}}1}}}

\title{Duality covariant multi-centre black hole systems}

\author{Guillaume Bossard$^{\sf{a}}$  and Stefanos Katmadas$^{\sf{a}, \sf{b}}$
\\ {$\sf{a}$ Centre de Physique Th\'eorique, \'Ecole Polytechnique, CNRS, 91128
Palaiseau, France}
\\ {$\sf{b}$ Institut de Physique Th\'eorique, CEA Saclay, CNRS-URA 2306, \\ \hspace{.14cm}
  91191 Gif sur Yvette, France }
\\ \email{guillaume.bossard [at] cpht.polytechnique.fr},\\
   \email{stefanos.katmadas  [at] cpht.polytechnique.fr}}

\abstract{We present a manifestly duality covariant formulation of the
composite non-BPS and almost-BPS systems of multi-centre black hole solutions in four dimensions.
The method of nilpotent orbits is used to define the two systems in terms
of first order flow equations that transform covariantly under the duality
group. Subsequently, we rewrite both systems of equations in terms of
real, manifestly duality covariant, linear systems of Poisson equations.
Somewhat unexpectedly, we find that the two systems are naturally
described by the same equations involving space dependent abelian isometries
that are conjugate to T-dualities by similarity transformations.}
\preprint{{CPHT-RR-026.0413} \\{IPhT-T13/057} }


\begin{document}

\newpage
\section{Introduction and overview}
\label{sec:intro}

The structure of black hole solutions in the supergravity effective description
of string/M-theory compactifications has long been a useful tool in understanding
their microscopic realisation in string theory. In the supersymmetric (BPS) case,
the supergravity black hole solutions \cite{Ferrara:1995ih,Ferrara:1996dd, Strominger:1996kf} have been understood microscopically to be described by D-branes wrapping supersymmetric cycles \cite{Strominger:1996sh, Maldacena:1997de}. Nonetheless, all microscopic BPS configurations cannot correspond to single centre solutions in the effective $\N=2$ supergravity description, and it has been understood \cite{Denef:2000nb} that the latter then describe composite bound states of BPS black holes  \cite{Behrndt:1997ny, Denef:2000nb,Bates:2003vx}. The detailed description of these solutions and their domain of stability in moduli space was very important in order to correctly reproduce the corresponding microscopic results \cite{Dabholkar:2005dt,Denef:2007vg}.

In the aim of generalising this understanding to more realistic non-supersymmetric black holes, the first non-trivial step is to consider non-supersymmetric extremal black holes.  One can classify extremal black hole solutions  into two main classes: the over-rotating class, for which the angular momentum saturates the extremality bound, and the under-rotating class, for which the electro-magnetic charges saturate the extremality bound, and which includes the supersymmetric solutions. Given that a stationary space-time defines a time fibration over a space-like three-dimensional base, one may distinguish the various classes by their corresponding base space. The characteristic of all known under-rotating extremal solutions is that the three-dimensional base space is flat, \ie $\mathbb{R}^3$ with the Euclidean metric, whereas the over-rotating extremal solutions all admit the singular three-dimensional base of the extremal Kerr solution. In this paper we will only discuss under-rotating solutions admitting a 
flat 
three-dimensional base space.\footnote{In the five dimensional uplift these solutions 
feature more elaborate base spaces, which however still include a flat
$\mathbb{R}^3$.}

Within the class of under-rotating solutions with an $\mathbb{R}^3$ base,
there are two known interacting systems of non-BPS black holes. The
composite non-BPS system \cite{Bossard:2011kz, Bossard:2012ge, Bossard:2012xsa}
describes the interactions of black holes that are non-BPS in isolation,
while the almost-BPS system \cite{Goldstein:2008fq, Bena:2009ev, Bena:2009en},
allows for configurations of centres that are be both BPS and
non-BPS in isolation. Both these systems have been studied in some detail
and, in several cases, the known explicit solutions are general enough
to allow for the most general solution to be obtained by dualities (\ie symmetries of the three-dimensional theory).

An important feature of the existing formulations of both these non-BPS
systems is the existence of distinguished directions in charge space, such that some are associated to harmonic functions as in the BPS system, whereas others are associated to functions solving Poisson equations. The different directions being then intrinsically inequivalent, one cannot straightforwardly rotate one into another by duality, as for BPS
solutions. It is therefore customary to solve these equations and
construct solutions parametrised by integration constants that can
only be related to charges and asymptotic moduli a posteriori. Of
course, one may straightforwardly apply a duality rotation on the solutions
to change the identification of pre-determined charges, thus covering
all possible charge configurations. This process is however not only
cumbersome, but more importantly it obscures the overall structure
of these systems, which should be manifest in a fully covariant
formulation.

A further motivation for generalising the existing formulations is that
non-su\-per\-sym\-me\-tric under-rotating extremal black holes seem to admit a
microscopic description similar to their BPS cousins \cite{Emparan:2006it,Dabholkar:2006tb}.
Understanding the domains of existence and stability of these solutions will
eventually be important in order to understand in more detail their microscopic
description. However, despite the existence of explicit solutions describing bound
states of such non-BPS black holes, their domain of existence in moduli space has
not been studied in detail. One of the main obstacles in carrying out this program
originates from the property that these solutions are described in terms of
parameters that are not the physical charges themselves, as explained above. In
order to solve this problem, it is important to obtain the general
solutions associated to fixed charge configurations. Being able to do this in
a straightforward manner requires the definition of covariant equations that
are not constrained to a fixed duality frame. 

In this paper, we provide a manifestly duality covariant formulation of both
the composite non-BPS and the almost-BPS systems that unifies their description, by
rewriting these systems of linear differential equations in terms of duality covariant
quantities, similar to the BPS system \cite{Denef:2000nb}. Although these systems of
equations turn out to be more complicated than the BPS one, the formalism permits to compute the
most general solutions associated to fixed charge configurations. The construction
of explicit solutions within these systems will be discussed in a forthcoming publication.

Somewhat surprisingly, we find that both systems are naturally formulated by considering
space dependent translations along abelian isometries of the
scalar manifold. As all equations are written in terms of objects transforming
linearly under electromagnetic dualities, it is natural to introduce
local generators for the abelian subgroups corresponding to these isometries.
The relevant equations in both the composite non-BPS and the almost-BPS
can then be expressed in terms of a covariant
derivative that contains a nontrivial connection in the Lie algebra of the
relevant abelian isometries.

In a specific frame, these abelian isometries coincide with the isometries that generalise the action of T-dualities combined with large gauge transformations of the $p$-form gauge fields present in string theory, for continuous parameters. In general, they will define abelian subgroups that are homomorphic to the latter by similarity transformations. For simplicity, we will refer to them as T-dualities in this paper, despite the fact that they are not in general associated to any duality in string theory. 
The set of all abelian isometries in the scalar target space of symmetric models
has been considered before in the context of (non-)BPS black holes in different guises,
see \eg \cite{Bena:2008wt, Dall'Agata:2010dy}. 

The paper is organised as follows. The remainder of this introductory
section is devoted to an informal presentation of the method we use to obtain the
duality covariant formulation of the composite non-BPS and almost-BPS systems.
In doing so, we also explain in more detail way how the T-dualities arise
in both systems and in fact allow us to define their action on the relevant quantities
without introducing explicit matrix representations. In section \ref{sec:T-duality}
we present the general action of arbitrary T-dualities in terms of symplectic
vectors and discuss the corresponding decomposition of the charge vector space.
In section \ref{sec:syst-def} we consider the three dimensional Euclidean coset
non-linear sigma model describing stationary black hole solutions. We then explicitly
solve the nilpotency conditions on the scalar momentum as a Lie algebra element
for both the composite non-BPS and the almost-BPS system, to obtain the first order
flow equations describing each class. We then go on in sections \ref{sec:composite}
and \ref{sec:almost} to rewrite these flow equations as a linear system of Poisson
equations for a set of symplectic vectors parametrising the solutions in an
arbitrary duality frame. The reader interested in applications can find summaries
of the two systems in sections \ref{sec:summ-5} and \ref{sec:summ-7}, which are self
contained and only require the definition of T-dualities in section
\ref{sec:t-dual-real}. We conclude in section \ref{sec:concl}, where we discuss some
of the implications of the local T-dualities acting on both systems and comment on
further generalisations.

\subsection{Overview of results}

In order to find systems describing stationary black hole solutions, we consider
the reduction along time to a three-dimensional Euclidean theory. In this setting,
one can straightforwardly dualise all vector fields to scalars, to obtain a
non-linear sigma model coupled to Euclidean gravity. In this paper, we only consider
$\N=2$ supergravity theories with a symmetric scalar space, so that the scalar fields
of the resulting non-linear sigma model over a pseudo-Riemannian symmetric
space. In the case of solutions with a flat base space, there is a powerful method
for obtaining first order flow equations that solve the full equations of motion,
starting from the observation that the Einstein equation
\begin{equation}
  \mbox{Tr}\scal{ P_\mu P_\nu} = R_{\mu\nu}  = 0 \,,
\end{equation}
can be solved by assuming the scalar momentum, $P$, to be a nilpotent element of
the Lie algebra. It appears then that the scalar fields equations of motion reduce to a solvable system of differential equations. Based on standard group theoretical considerations, one shows that the nilpotency of $P$ implies that there exists a second element of the algebra, ${\bf h }$,
such that an eigenvalue equation of the type
\begin{equation}\label{eigen-gen}
\prod_{i=1}^n ( {\bf h } - i ) \, P = 0 \ ,
\end{equation}
holds, where ${\bf h } P \equiv [  {\bf h } , P ]$ and $n$ is a positive integer.
It then follows that $P$ must be a linear combination of eigenvectors of ${\bf h }$
with positive eigenvalues. In section \ref{sec:syst-def} we consider the relevant
eigenvalue equations for both the composite non-BPS and almost-BPS systems, giving
the precise solutions for both the momentum $P$ and the auxiliary fields
${\bf h }$. For the composite non-BPS system, this analysis has been given in
\cite{Bossard:2012xsa}, but is included here for completeness. The
corresponding flow equations for the almost-BPS system have been derived
for the STU model in \cite{Ferrara:2012qm} by reduction of the second order
equations of motion, while we present the general analysis for any symmetric
model.

The presence of the aforementioned auxiliary fields is a central feature of both
systems, as will become clear from our treatment. In fact, we find it most
convenient to keep them throughout, as variables characterising the solutions,
despite the fact that one may in principle solve for them in terms of physical
fields. In the systems we shall consider, ${\bf h}$ can be parametrised in terms of a distinguished direction in the charge vector space (plus a  phase for the almost-BPS system). This distinguished direction is associated to a constant so-called 
very small vector, or in other words a one-charge vector, in the sense that it
can always be brought by dualities to a canonical form where it has only one
charge, as a pure $D6$ charge for example. 
However, one must keep this very small vector arbitrary in order to be able to consider generic charge configurations. 

Treating the system in terms of electromagnetic charges and auxiliary vectors
implies that one has to work with real quantities, rather than the complex
quantities that appear naturally in the Euclidean 3-dimensional non-linear sigma model. It turns out
that this is essential in solving the system in terms of local functions,
since a linear structure only appears after writing all flow equations in the
real basis. As mentioned above, this change of basis, that permits to solve
non-linear first order equations in terms a solvable system of linear differential
equations, is the main technical result of this paper.

In order to appreciate the importance of the real formulation, the simple
example of the analogous situation for the BPS system is instructive. In
this case the auxiliary element ${\bf h }$ in \eqref{eigen-gen} is much
simpler, as it is parametrised by a single phase, $e^{i\alpha}$. Taking
the static case for simplicity, \eqref{eigen-gen} leads to the standard
BPS equations
\begin{equation}\label{BPS-static}
 \partial_rU  = r^{-2}e^U \mbox{Re}(e^{-i\alpha}Z(\Gamma))\,, \qquad 
 \partial_rt^i  = r^{-2}e^U e^{i\alpha} \bar{Z}^i(\Gamma)\,.
\end{equation}
Here, $r$ is the distance from the horizon, $e^U$ is the single function
parametrising the metric, $t^i$ are the vector multiplet scalars and
$Z(\Gamma)$, $Z_i(\Gamma)$ are the central charge of the charge
vector $\Gamma$, and its K\"{a}hler derivative. In this form, it seems a rather non-trivial task to
solve \eqref{BPS-static} explicitly. It is however possible to combine
the scalar degrees of freedom in a single vector, $e^{-U}e^{-i\alpha}\cV$,
where $\cV$ is the so called symplectic section, which parametrises the
scalars $t^i$. The two equations \eqref{BPS-static} can then be written
as in \cite{Denef:2000nb},
\begin{equation}\label{BPS-eqn}
 2\,\partial_r\mbox{Im}\left(e^{-U}e^{-i\alpha}\cV\right) = \frac{\Gamma}{r^{2}}\,,
\end{equation}
using standard special geometry identities. It is now trivial to solve the last
equation as 
\begin{equation}\label{BPS-soln}
 2\,\mbox{Im}\left(e^{-U}e^{-i\alpha}\cV\right) = - \cH\,,
\end{equation}
where $\cH$ is a vector of harmonic functions, whose poles are identified with
the charges $\Gamma$. One still has to solve a set of algebraic equations to
obtain the scalars $t^i$ and $e^U$ in terms of harmonic functions
\cite{Bates:2003vx}, but it is important to stress that \eqref{BPS-soln}
is equivalent to a solution for the physical scalars.

The analogous computation for the non-BPS case includes the auxiliary
vector pa\-ra\-me\-tri\-sing the element ${\bf h }$ and is therefore
considerably more involved. For the restricted case of single centre
under-rotating black holes, this was done in detail in \cite{Bossard:2012xsa}
and the result was given in terms of the objects appearing in
\eqref{BPS-soln}, together with the vectors in \eqref{eigen-gen} above.
This class is naturally part of both the multi-centre systems considered
in this paper and its description is crucial for understanding the structure
of the more general classes. Explicitly, for the single centre class we have
\cite{Galli:2010mg, Bossard:2012xsa}
\begin{equation}\label{scals-single}
 2\,\mbox{Im}\left(e^{-U}e^{-i\alpha}\cV\right) = 
- \cH + 2\,\tfrac{\Iprod{\cH}{R^*}}{\Iprod{R}{ R^*}}\, R
 +  \tfrac{M}{\Iprod{\cH}{R^*}}\, R^* \,,
\end{equation}
where $\cH$ is again a vector of harmonic functions describing the charges
of the black hole, while three new objects appear, namely the function $M$
and the two constant very small symplectic (pseudo-charge) vectors $R$ and
$R^*$, which are not mutually local, \ie $\Iprod{R}{R^*}\neq 0$.

The presence of two different very small vectors in
\eqref{scals-single} can be somehow surprising, but it is important to note that
they do not define independent parameters. Indeed, we find that these vectors
are determined by the electromagnetic charges $\Gamma$, up to duality
transformations leaving $\Gamma$ invariant \cite{Bossard:2012xsa}, so that they
are determined in terms of $\Gamma$ and the asymptotic scalar fields $t^i_{\infty}$.
Conversely, one can view the charges, $\Gamma$, and the harmonic vector $\cH$ as
being constrained to lie in a Lagrangian subspace determined by $R$ and $R^*$.
The terms proportional to the two very small vectors are worth
discussing in more detail. First, note that the term proportional to $R$ is simply
a projection on that component of $\cH$ with an additional factor of 2,
implying that this particular component appears with a flipped sign. This
is a very general feature of non-BPS solutions that has been observed in
many examples in the literature \cite{Cvetic:1995kv, Ortin:1996bz, Kallosh:2006ib}.
On the other hand, the component along $R^*$ in \eqref{scals-single}
contains the function, $M$, which represents the only genuinely new term
in the expression for the scalars, and is constrained to be a dipole harmonic
function characterising the angular momentum of the under-rotating non-BPS black hole 
\cite{Bena:2009ev}.

In sections \ref{sec:composite} and \ref{sec:almost}, we explicitly solve the
composite non-BPS and almost-BPS non-linear first order systems derived in section
\ref{sec:syst-def} in terms of real vectors of local functions. As we shall see, the
generalisation to multi-centre systems can be performed by allowing one of $R$ or $R^*$
to vary in space, while keeping their symplectic product fixed  $\Iprod{R}{R^*}=4$.
The space dependence of the non-constant vector can then be reabsorbed into a space
dependent duality transformation that leaves the constant very small vector invariant.
Because this duality transformation lies in an abelian subgroup that is conjugate to
the group of T-dualities by similarity transformations, we shall simply refer to them
as T-dualities.  In particular, the composite non-BPS system is described by the
T-dualities, $\Tgp$, defined as leaving $R$ invariant, while the almost-BPS system is
described by the T-dualities, $\Tgm$, leaving $R^*$ invariant. One then shows that the 
symplectic section $\cV$ still takes the form (\ref{scals-single}), with $\cH$
constrained to lie in the same Lagrangian subspace determined by $R$ and $R^*$. Since
one of the very small vectors is not constant, $\cH$ is not harmonic anymore, but
satisfies the Poisson equation 
\begin{equation}\label{Poiss-gen}
 d\star d \scal{ \exp[-\Tgpm] \cH}  =\dTgpm \wedge\star \dTgpm  \exp[-\Tgpm]  \cH\,,
\qquad
 d\star \dTgpm =0\,,
\end{equation}
where the second equation imposes that the $n_v$ parameters of the T-dualities are
given by arbitrary harmonic functions. The function $M$ in \eqref{scals-single} is
also specified by a T-duality covariant equation, given by
\begin{equation}
 \star d \omega - d M = 
\Iprod{ \cH }{d \cH - 2\,\dTgpm\,\cH }\,,
\end{equation}
which also fixes the angular momentum one-form, $\omega$. Although it is not manifest
from these general equations, the graded structure of the
vector space is such that the system is solvable. Because the two T-dualities in the right hand side of \eqref{Poiss-gen} have a nontrivial kernel, 
it follows that the source in \eqref{Poiss-gen} does not contain some of the components of
$\exp[-\Tgpm] \cH$, which therefore includes both harmonic and non-harmonic components.
It turns out that $\Tgpm$ act as raising or lowering operators, so that only the harmonic
components turn out to source the non-harmonic ones, \ie the non-harmonic components
of $\exp[-\Tgpm] \cH$ do not source themselves. These properties
will be discussed in detail in sections \ref{sec:composite} and \ref{sec:almost}.

By definition, T-dualities are represented in terms of symplectic matrices
in the $2(n_v+1)$-dimensional vector space of charges. However, as will be
shown in detail in the following sections, the $n_v$ parameters of a T-duality 
can be arranged into a symplectic vector obeying a number of constraints. One
can therefore write the action of the corresponding T-duality in terms of this
vector, together with $R$ and $R^*$, using the symplectic product and the quartic
symmetric invariant $I_4$ (which defines the entropy of extremal static black holes).
In this way, we obtain explicit expressions for the sources in \eqref{Poiss-gen},
in terms of $\exp[-\Tgpm]\cH$ and the parameters of the T-dualities,
that can be evaluated explicitly once a model and its associated quartic invariant
are specified. These equations reduce to the systems introduced in
\cite{Bossard:2011kz, Goldstein:2008fq}, for a particular choice for the two
very small vectors above.

This concludes our short presentation of the main results in this paper.
For the convenience of the reader, we provide an account of the results
appearing in the following sections, which can be read independently of
each other with the exception of section \ref{sec:T-duality}, that is
basic to most applications. In section \ref{sec:T-duality} we give a
detailed discussion of the general T-dualities. We show that given two very
small vectors $R$ and $R^*$ that do not mutually commute, one can explicitly
define a graded decomposition of the symplectic vector space, on which the
T-dualities $\Tgpm$ can be respectively defined as  raising and lowering operators. 
The structure of this decomposition is essential for all applications in this paper.

We then go on in section \ref{sec:syst-def} to define the two systems
of non-BPS multi-centre black hole solutions in terms of the non-linear sigma model
in three dimensions, obtained after dimensional reduction over the
time direction. After an overview of the main properties of this
three-dimensional Euclidean theory, we discuss in detail the eigenvalue equations
\eqref{eigen-gen} for the two systems at hand. This results to two
sets of first order flow equations that completely describe the
composite non-BPS and almost-BPS systems in four dimensions.

Sections \ref{sec:composite} and \ref{sec:almost} are mirror copies
of each other, wherein we present in detail the change of variables that
transforms the flow equations of section \ref{sec:syst-def} to two
linear systems. The reader can find a concise summary
of the two multi-centre systems in the real formulation in sections
\ref{sec:summ-5} and \ref{sec:summ-7} respectively, where we also discuss
some general properties of the solutions, deferring a more detailed
presentation and explicit examples for a forthcoming publication
\cite{Bossard:tocome}.

\section{T-dualities}
\label{sec:T-duality}

In this section, we provide a detailed discussion of the abelian isometries on the scalar
target space of $\N=2$ supergravity,\footnote{We refer to the appendix for a concise review
of our conventions and notations.} that leave a given charge vector invariant.
As will be shown in later sections of this paper, the precise action of these isometries, in
their most general form, is of central importance in the construction of multi-centre non-BPS
black holes. We start with an informal discussion of the simplest example of such abelian
isometries. Subsequently, we review some properties of symmetric special
K\"{a}hler spaces with a cubic prepotential in section \ref{sec:4d-coset}. Section
\ref{sec:t-dual-real} is devoted to the definition of the abelian isometries
we shall refer to as T-dualities and their explicit action in terms of the associated real
vectors. In section \ref{sec:t-dual-def} we discuss the realisation of the same isometries
in the complex basis defined by the central charge and its K\"{a}hler derivatives. 

Throughout this paper, we study extremal multi-centre black hole solutions in $\N=2$
supergravity coupled to $n_v$ vector multiplets labeled by an index $i=1,\dots, n_v$,
whose scalar fields, $t^i$, parametrise a symmetric special K\"ahler target space, $\mathcal{M}_4$.
These spaces were classified some time ago \cite{deWit:1995tf} and include minimally
coupled vector multiplets, which are not of interest in this work, and theories with a
cubic prepotential, specified by a completely symmetric tensor $c_{ijk}$  \cite{Gunaydin:1983bi} (cf. 
\eqref{prep-def}). The target space geometry is governed by a K\"ahler potential
(cf. \eqref{Kah-pot}), which manifestly depends only on the imaginary part of the scalars, $t^i$.
It follows that the real parts of the scalars are coordinates along $n_v$ isometries of
the scalar manifold, acting as
\begin{equation}\label{abel-iso}
 t^i \rightarrow t^i + k^i\,,
\end{equation}
where $k^i$ is a vector of $n_v$ constant real parameters. These isometries are generic in
all cubic models and clearly form an abelian algebra. For theories originating from
Calabi--Yau string compactifications, the operation \eqref{abel-iso} can be viewed as
large gauge transformations on the higher dimensional tensor gauge fields along internal
cycles, combined with T-dualities.

While this description is useful in characterising the symmetries themselves, for reasons
that will become clear below, in this work we are interested in the embedding of these
isometries in the symplectic group, which acts on the electric and magnetic gauge fields
in four dimensions. The convenient variable to use in order to make the action of the
isometries in \eqref{abel-iso} transparent is the so called scalar symplectic
section, $\cV$, which is a somewhat redundant way of repackaging
the scalars, as
\begin{equation} \label{eq:sym-sec-0}
\cV=\begin{pmatrix} X^I\\ F_I\end{pmatrix}\,, \qquad
t^i = \frac{X^i}{X^0}\,,
\end{equation}
where $F_I$ are the derivatives of the prepotential with respect to the $X^I$. The index
$I=\{0,i\}$, runs over one more entry than $n_v$ and enumerates all the gauge fields in the
theory, \ie the vector multiplet gauge fields and the graviphoton. Note that $\cV$
changes under K\"ahler transformations by a phase and is subject to the constraint
\eqref{eq:D-gauge}, so that it encompasses only $2\,n_v$ degrees of freedom,
identified with the physical scalars $t^i$s. The advantage
of this variable is that, unlike the physical scalars, it transforms linearly under
electric/magnetic duality transformations, in exactly the same way as the electromagnetic
charges.

For instance, the isometries in \eqref{abel-iso} are described by a linear transformation
acting on the charges as
\begin{equation} \label{eq:spec-flow}
 \exp[{\mathrm T}_k]
\begin{pmatrix}  p^0 \\ p^i \\ q_i \\ q_0  \end{pmatrix}
\to
\begin{pmatrix}
 p^0 \\   p^i + k^ip^0 \\
 q_i + c_{ijk} k^j p^k +\tfrac12\,c_{ijk} k^j k^k p^0 \\
 q_0 - k^i q_i - \tfrac12\,c_{ijk} k^j k^k p^i
   - \tfrac{1}{6} c_{ijk} k^i k^j k^k p^0
\end{pmatrix}
\,,
\end{equation}
where we defined the abelian generators ${\mathrm T}_k$ for later convenience.
One can now easily verify that the same operation \eqref{eq:spec-flow} acting on
the section in \eqref{eq:sym-sec-0} leads to \eqref{abel-iso} for the physical scalars.
In this formulation, the connection of the isometries \eqref{abel-iso} to higher
dimensional gauge transformations is more transparent, since it has a natural action
on the electromagnetic gauge fields. Moreover, this particular set of abelian
transformations also arises in the form of spectral flows in conformal field theories describing black holes microscopically.

The crucial feature of the symplectic embedding of the abelian isometries is, however,
that one may generate an infinite number of inequivalent sets of abelian isometries by
conjugating the matrix $\exp[{\mathrm T}_k]$ in \eqref{eq:spec-flow} by a general
$U$-duality transformation, as in \cite{Dall'Agata:2010dy}.
\footnote{Following a common misuse of language, we call $U$-duality transformations
all the continuous isometries of the scalar field symmetric space, independently of
whether they define or not actual string theory dualities for appropriate integral
coefficients.} These sets of isometries are more complicated than the one in
\eqref{abel-iso} and do not commute with it. From a higher dimensional
point of view, some of these more general isometries can also be viewed either as
large gauge transformations conjugated with generic $T$- and/or $U$-dualities, or
as (generalised) spectral flows in a dual conformal field theory. Here, we refer to them simply as
T-dualities for brevity and we focus on the case of symmetric scalar manifolds,
which allows for the most general transformations to be described explicitly.

The representation of a generic T-duality in terms of matrices is of course a rather tedious 
task, which can be circumvented in a natural way, intrinsically tied to the
systems of non-BPS black holes we consider. The crucial observation is that there is
always a graded decomposition of the vector space in four components, generalising
the clear distinction between the various components in \eqref{eq:spec-flow}, based on
their transformation rule under T-dualities. Indeed, general T-dualities act
consistently on each component of the charge space with a fixed homogeneity in the
parameters $k^i$, which can never exceed three. In particular, there is a distinguished
direction that is invariant under the action of any given T-duality, as for example the
electric charge $q_0$ is left invariant in \eqref{eq:spec-flow}. The $q_0$ charge is
rather special, since it is an example of a so-called very small vector. We will recall
the precise definition of such a vector in what follows, but loosely speaking a very small
vector can be defined as a `one charge vector', in the sense that it is $U$-dual to a
pure $q_0$ charge. Clearly, the distinguished direction that is left invariant under a
generic T-duality must then always be a very small vector, given that all such
transformations are $U$-dual to the above example.

The relevance of very small vectors for extremal non-BPS solutions arises already
in the single centre class \cite{Galli:2010mg, Bossard:2012xsa, Hristov:2012nu}, which is naturally
described in terms of an auxiliary pair of mutually nonlocal constant very small vectors,
constrained by the physical charge vector. As we will show explicitly in sections
\ref{sec:composite} and \ref{sec:almost} below, the corresponding multi-centre
systems are naturally described by promoting one of these very small vectors to be
not constant. In order to see the connection to T-dualities, consider the very small
vector $\hat{R}$ defined such that its only non-vanishing component is $q_0 =4$. A
general very small vector can be parametrised as
\begin{equation} S = c \left( 1 ,\, s^i   \,; \, 
\tfrac{1}{2} c_{ijk} s^j s^k, \,  - \tfrac{1}{6} c_{ijk} s^i s^j s^k \right)^T   ,
\end{equation}
where $c$ and $s^i$ are allowed to take singular values as long as the components of $S$ are well defined in the limit.\footnote{For example one recovers $\hat{R}$ in the limit $c\rightarrow 0$, $s^i \rightarrow \infty $ with $ - c \tfrac{1}{6} c_{ijk} s^i s^j s^k =4 $.} It follows that 
a general very small vector $\hat{R}^*$ satisfying $\Iprod{\hat{R}}{\hat{R}^*}=4$ can be parametrised as
\begin{equation}\label{gen-small}
 \hat{R}^*= \exp[\Tgp_s]\, R^*_0 =  \left( 1 ,\, s^i   \,; \, 
\tfrac{1}{2} c_{ijk} s^j s^k, \,  - \tfrac{1}{6} c_{ijk} s^i s^j s^k \right)^T   ,
\end{equation}
where the only non-vanishing component of $R^*_0$ is $p^0=1$. In the case of a constant vector $\hat{R}^*$, \eqref{gen-small} is simply a
convenient parametrisation, but in the more general case when $\hat{R}^*$ is not
constant, one can assume the parameters $s^i$ to be functions of space,
to obtain
\begin{equation}\label{der-small}
 d \hat{R}^* = d\Tgp_s\, \hat{R}^* = \Tgp_{ds}\, \hat{R}^* \,,
\end{equation}
where we used the abelian property of \eqref{eq:spec-flow}. As we will show explicitly
in later sections, the composite non-BPS system is naturally
characterised by two very small vectors, one constant $\hat{R}$ and one non-constant $\hat{R}^*$, which have a non-vanishing constant symplectic product $\Iprod{\hat{R}}{\hat{R}^*}=4$. The almost-BPS system is similarly described by two very small vectors, only the role of $\hat{R}$ and $\hat{R}^*$ are interchanged. The constant very small vector is left invariant by the relevant T-dualities (as for example \eqref{eq:spec-flow}), which are therefore different for each system.
The non-constant very small vector can be expressed as in \eqref{gen-small} for a
constant vector $R^*_0$ in the composite non-BPS system (or respectively $R_0$ for
the almost-BPS system), and it follows that it satisfies \eqref{der-small}. In the
rest of the paper $\hat{R}$ and $\hat{R}^*$ will be generic very small vectors, and
the associated T-dualities will define abelian subgroups conjugate to the one
described in \eqref{eq:spec-flow}.

\subsection{Symmetric special K\"{a}hler spaces}
\label{sec:4d-coset}

In this paper we consider $\N=2$ supergravity theories defined in  \cite{Gunaydin:1983bi} for which the special K\"ahler
target space, $\mathcal{M}_4$, is a symmetric space and that can be obtained as Kaluza--Klein
reductions of corresponding five dimensional theories.\footnote{This excludes theories
with minimally coupled vector multiplets, which do not include systems of the type we
consider here.} In this case, $\mathcal{M}_4$ is a coset space of the four-dimensional duality
group, $G_4$, by its maximal compact subgroup $U(1)\times K_4$
\begin{equation}
\mathcal{M}_4\cong ( U(1)\times K_4 ) \backslash G_4  \, .
\label{4d-coset}
\end{equation}
For the class of theories we consider, the scalar target space is a symmetric space even after
dimensional reduction/oxidation to three/five dimensions, so that \eqref{4d-coset}
is part of the sequence of embeddings
\begin{equation}
K_5 \backslash G_5 
\hookrightarrow ( U(1)\times K_4 ) \backslash G_4\hookrightarrow
 ( SL(2)\times G_4 ) \backslash G_3\,,
\end{equation}
where by $G_d$, $K_d$ we denote the duality group and (part of) the isotropy group in
$d$ dimensions respectively. Note that the divisor group in three dimensions is
non-compact because we consider the time-like reduction to three
dimensions,\footnote{For a spacelike reduction, one obtains $SU(2)\times K_3$, where
$K_3$ is the compact real form of $G_4$.} as that is relevant for the applications we
consider later on. Note that $K_4$ is the compact real form of $G_5$, by property of $\N=2$ supersymmetry.

One can always define a set of vielbeine associated to the K\"ahler metric
$g_{i\bar \jmath} $ on $\mathcal{M}_4$
\begin{equation}  g^{i\bar \jmath}   = e_a^i e^{a\, \bar \jmath} \label{eq:kah-viel}, \end{equation}
such that the constant symmetric tensor
\begin{equation} c_{abc} =  i  \,  e^{\cK} e_a^i e_b^j e_c^k\,  c_{ijk}, \label{Ctangent} \end{equation}
where $c_{ijk}$ is the $G_5$ invariant tensor defining the prepotential (cf. \eqref{prep-def}), is left
invariant by $K_4$. Then, the contravariant symmetric tensor $c^{abc}$ in the conjugate
representation satisfies the Jordan identity \cite{Gunaydin:1983bi}
\begin{equation} c_{f(ab} c_{cd)g} c^{efg}  = \frac{4}{3} \delta^e_{(a} c_{bcd)}\,.
 \label{symmetric}\end{equation}

In a complex basis, the Lie algebra of $G_4$, denoted $\mathfrak{g}_4$ (and respectively $\mathfrak{k}_4$ for $K_4$), naturally decomposes as
\begin{equation}\label{g4-dec}
 \mathfrak{g}_4 \cong \mathfrak{u}(1) \oplus \mathfrak{k}_4 \oplus  \mathds{C}^{n_v}\ .
\end{equation}
It follows that the relevant parameters
are given by those corresponding to the elements of $\k_4$, denoted by $G^a{}_b$, a real
scalar $\gamma$ and a complex vector $\Lambda_a$. The corresponding algebra is realised
in terms of anticommuting parameters with the nilpotent differential\footnote{The nilpotent
differential $\delta$ in these equations acts as
$\delta X_A T^A  = \tfrac12\,X_A X_B [T^A,T^B]$ on a Lie algebra element expressed in
a basis of generators ${T^A}$ using a set of anticommuting parameters ${X_A}$.}
\begin{align}
\delta \Lambda_ a = - G^b{}_a \Lambda_b + 2 i \gamma \Lambda_a
\qquad & \delta \gamma = \frac{i}{3} \bar \Lambda^a \Lambda_a \ , \CR
\delta G^a{}_b = G^a{}_c
G^c{}_b + c^{ace} c_{bde} \Lambda_c \bar \Lambda^d +& \bar \Lambda^a \Lambda_b +
\frac{1}{3} \bar \Lambda^c \Lambda_c \delta^a_b \ .
\label{g4-algebra}
\end{align}
Note that the statement of invariance of the tensor $c_{abc}$ under $K_4$ implies
that $\N[\bar Z] \equiv \frac{1}{6} c_{abc} \bar Z^a \bar Z^b \bar Z^c$ is $K_4$
invariant for any vector $\bar Z^a$ transforming in the relevant $n_v$-dimensional complex representation of
$K_4$. One can check that the variation of $G^a{}_b$ in \eqref{g4-algebra}
indeed leaves invariant the cubic norm $\N[Z] \equiv \overline{\N[\bar Z]}$ for an anticommuting $\Lambda_a$.

The invariance of the cubic norm $\N[Z]$ can be used to define duality invariants 
and restricted charge vectors, a concept that is of central importance for the
applications we consider later in this paper. First, we introduce the quartic
invariant for a charge vector $\Gamma$, in terms of its central charges,
$Z \equiv Z(\Gamma)$, $Z_a \equiv  Z_a(\Gamma)$, as
\begin{align} 
I_4(\Gamma) =&\, \left( Z \, \bar Z - Z_a \, \bar Z^a \right)^2 
- c_{eab} \bar Z^a \bar Z^b \, c^{ecd} Z_c  Z_d 
+  4\, \bar Z \, \N[Z] + 4\, Z \,\N [\bar Z]\,.
\label{I4-def}
\end{align}
This expression can be verified to be invariant under the $\mathfrak{g}_4 $ generators
of \eqref{g4-algebra}, so that it is moduli independent. This is manifest by the corresponding
real form of this invariant, which is given solely in terms of charges by
\begin{eqnarray}
I_4(\Gamma)&=& \frac{1}{4!} t^{MNPQ}\Gamma_M\Gamma_N\Gamma_P\Gamma_Q
\nonumber\\
         &=& - (p^0 q_0 + p^i q_i)^2 + \frac{2}{3} \,q_0\,c_{ijk} p^i p^j p^k- \frac{2}{3} \,p^0\,c^{ijk} q_i q_j q_k 
             + c_{ijk}p^jp^k\,c^{ilm}q_lq_m\,, \label{I4-ch}
\end{eqnarray}
where we also defined the completely symmetric tensor $t^{MNPQ}$ for later reference.
Again, one can easily check that \eqref{I4-ch} is invariant under the example
T-duality in \eqref{eq:spec-flow}, and it is more generally invariant under an arbitrary $G_4$ transformation. 

We are now in a position to introduce the concept of charge vectors of restricted rank.
A generic vector leads to a nonvanishing invariant \eqref{I4-def}-\eqref{I4-ch} and
is also referred to as a rank-four vector, due to the quartic nature of the invariant. 
Similarly, a rank-three vector, $\Gamma_\trois$, is a vector for which the quartic invariant vanishes,
but not its derivative. An obvious example is a vector with only $p^i\neq 0$ and all
other charges vanishing, so that the derivative $I_4^\prime(\Gamma_\trois)$ is nonzero and
proportional to the cubic term $\N[p]$.

There are two more classes of restricted vectors, defined analogously as rank-two (small)
and rank-one (very small) vectors. A rank-two vector, $\Gamma_\deux$, is defined such that both
$I_4(\Gamma_\deux)= I^\prime_4(\Gamma_\deux)=0$, and a simple example is provided by
a vector with all entries vanishing except the $p^i$, with the additional constraint
that $\N[p]=0$. Finally, a very small vector, $\Gamma_\un$, is defined such that
\begin{gather}
 I_4(\Gamma_\un)= I^\prime_4(\Gamma_\un)=0\,, 
\CR
\frac{1}{4} I_4(\Gamma_\un , \Gamma_\un , \Gamma , \Gamma ) 
\equiv \frac{1}{4} t^{MNPQ} \Gamma_{\un M} \Gamma_{\un N} \Gamma_P \Gamma_Q
 = - \Iprod{\Gamma_\un}{\Gamma}^2\,, \label{VerySmallI4}
\end{gather}
for any vector $\Gamma$. 
In this case, we can also give a general definition in terms of the complex basis, which
is in fact independent of the values of the scalar fields. In this paper we will often
make use of a rank one vector, $R$, that we choose without loss of generality such that
$|Z(R)|=1$. One shows that such a very small vector satisfies 
\begin{equation}\label{N-Om-def}
 Z(R)=\N[\Omega]\,, \qquad Z_a(R) = \Omega_a\,,
\end{equation}
where $\N[\Omega]$ is a phase by construction. The remaining central charges $\Omega_a$ are such that
\begin{equation}\label{small-def}
 \frac12\,c^{abc}\Omega_b\Omega_c = \N[\Omega]\,\bar\Omega_a\,, \qquad
\bar \Omega^a \Omega_a = 3  \ .
\end{equation}
A general very small vector can be obtained by rescaling both $\N[\Omega]$ and $\Omega_a$
by a real function. Examples of very small vectors were already given above, as vectors
where only the $q_0$ or $p^0$ component is nonzero, while the parametrisation
given in \eqref{gen-small} is generic up to a possibly singular rescaling.

\subsection{Freudenthal ternary algebra realisation of $G_4$}
\label{sec:t-dual-real}

We now proceed in describing the duality group $G_4$, as defined above, in terms of real
vector parameters. This is essential for discussing the T-dualities, which are contained
in $G_4$ as subgroups and can therefore also be described in terms of real vector parameters
in the general case, similar to the example \eqref{eq:spec-flow} above. 

The central object for the definition of $G_4$ in the real basis is the quartic invariant in
\eqref{I4-ch} and its derivatives. It is convenient to define a symplectic vector out the
first derivative, $I_4^\prime(\Gamma)$, of the quartic invariant so that
\begin{equation}
 \Iprod{\Gamma}{I^\prime_4(\Gamma)} = 4  I_4(\Gamma)  \ , \qquad I^\prime_4(\Gamma,\Gamma,\Gamma) = 6 I_4^\prime(\Gamma) \ .
\end{equation}
Using the definition \eqref{I4-ch} of the quartic invariant and the properties of the rank-three symmetric tensor $c_{ijk}$, one shows the following quintic identity
\begin{equation}
 I_4^{\prime} (\Gamma,\Gamma,I_4^\prime(\Gamma)) = - 8 I_4(\Gamma) \Gamma \  , \label{Freudenthal} 
\end{equation}
for a generic charge vector $\Gamma$. This identity is equivalent to the property that the Freudenthal ternary product 
\begin{equation}
( X,Y,Z) \equiv \frac{1}{4} I_4^\prime(X,Y,Z) + \frac{1}{2} \Iprod{X}{Y} Z - \frac{1}{2} \Iprod{Z}{X} Y + \frac{1}{2} \Iprod{Y}{Z} X \ , \label{Ternary}
\end{equation}
satisfies the four axioms defined in \cite{Faulkner:1971}, and inversely, any Freudenthal ternary product is necessarily of the form \eqref{Ternary}, for a completely symmetric rank four tensor satisfying \eqref{Freudenthal}. One can therefore define the $\mathfrak{g}_4$ Lie algebra as in \cite{Faulkner:1971}. We shall not use the Freudenthal ternary product, but rather the quintic identity \eqref{Freudenthal}. We refer to \cite{Faulkner:1971} for the more formal definition of the $\mathfrak{g}_4$ Lie algebra from the ternary product itself.  It is straightforward to combine \eqref{Freudenthal} with the symmetry properties of the sextic invariant $\Iprod{I^\prime(\Gamma_1)}{I^\prime(\Gamma_2)}$ to
show that for any two vectors $J_1$ and $J_2$, the linear transformation 
\begin{equation}
{\bf g}(J_1,J_2) \Gamma  \equiv  \frac{1}{2} I_4^{\prime}( J_1,J_2 ,\Gamma)   -J_1 \Iprod{J_2}{\Gamma}  - J_2  \Iprod{J_1}{\Gamma} \  ,\label{g4Freud}
\end{equation}
preserves both the symplectic product and the quartic invariant. One concludes that \eqref{g4Freud}
defines a generator of $\mathfrak{g}_4$, and all $\mathfrak{g}_4$ generators can in fact be defined
in this way. It follows that the Lie algebra takes the form
\begin{equation}
[ {\bf g}(J_1,J_2) , {\bf g}(J_3,J_4) ] =  {\bf g}({\bf g}(J_1,J_2)  J_3,J_4) +  {\bf g}(J_3, {\bf g}(J_1,J_2) J_4) \ ,\label{TripleAlgebra}
\end{equation}
as shown in \cite{Faulkner:1971}.

A special case arises for two rank 1 vectors, denoted $R$ and $R^*$, which are assumed to be
mutually non-commuting. In this case, one can define the corresponding $\mathfrak{g}_4$
generator as in \eqref{g4Freud}
\begin{equation}
{\bf h}_T\, \Gamma \equiv \Iprod{R}{R^*}^{-1} \Scal{  \frac{1}{2} I_4^{\prime}( R,R^* ,\Gamma)  + \Iprod{\Gamma}{R^*} R  -R^* \Iprod{R}{\Gamma}} \  , \label{hTReal} 
\end{equation}
which is central in the description of T-dualities. It is clear from \eqref{VerySmallI4} that for any
rank 1 vector $R$ (or respectively $R^*$) one has
\begin{equation}
 I_4^\prime( R,R,\Gamma) = 4 \Iprod{R}{\Gamma} R \  .
\end{equation}
This generator admits therefore $R$ and $R^*$ as eigenvectors, with eigenvalues $+3$
and $-3$ respectively, and the remaining eigenvectors
of ${\bf h}_T$ can be characterised as follows. Using \eqref{Freudenthal} and
\eqref{VerySmallI4}, one can show that
\begin{equation}
\frac{1}{4} I_4^{\prime}( R,R^* ,I_4^{\prime}(R,R^*,\Gamma)) = \Iprod{R}{R^*}^2 \, \Gamma + 3 \Iprod{R}{R^*} \scal{ \Iprod{\Gamma}{R^*} R + R^* \Iprod{R}{\Gamma} } \  ,\label{FreudCons} 
\end{equation}\eqref{FreudCons}
from which follows the action of the square of ${\bf h}_T$, as
\begin{equation}\label{hT-projet}
 {\bf h}_T^{\; 2 } \Gamma  =  \Gamma + 8  \Iprod{R}{R^*}^{-1} \scal{  \Iprod{\Gamma}{R^*} R  +R^* \Iprod{R}{\Gamma}}  \ .
\end{equation}
This equation implies that $R$ and $R^*$ are the unique eigenvectors with eigenvalues $+3$
and $-3$ respectively, and also leads to the characteristic equation 
\begin{equation}\label{hT-eigen}
 \scal{ {\bf h}_T^{\; 4 } - 10 {\bf h}_T^{\; 2} + 9} \Gamma  = 
( {\bf h}_T-3 ) ( {\bf h}_T-1) ( {\bf h}_T+1) ({\bf h}_T+3)\Gamma
= 0 \ .  
\end{equation}
In view of the fact that ${\bf h}_T$ is symplectic, it follows from \eqref{hT-projet}, \eqref{hT-eigen}
that the $2n_v+2$ electromagnetic charge vector space decomposes into
\begin{equation} \label{eq:vec-decomp}
\mathds{R}^{2n_v+2} \cong \mathds{R}^\ord{-3} \oplus  ({\mathds{R}}^{n_v})^\ord{-1}  \oplus  ({\mathds{R}}^{n_v})^\ord{1} \oplus \mathds{R}^\ord{3} \ ,
\end{equation}
where the two distinguished vectors $R$ and $R^*$ are by definition the components of
grade $3$ and $-3$, respectively. This decomposition is clearly relevant to the T-dualities
as described in \eqref{eq:vec-decomp}, as it allows to identify four eigenspaces, based on
two very small vectors. The $n_v$ eigenvectors of eigenvalue $+1$ and the $n_v$
eigenvectors of eigenvalue $-1$ can be obtained by defining the corresponding projectors
to the four eigenspaces of ${\bf h}_T$, as
${\bf h}_T \Gamma^\ord{n} = n \Gamma^\ord{n}$ for $n=-3,-1,1,3$, \ie
\begin{eqnarray}
 \Gamma^\ord{3}  &=& \Iprod{R}{R^*}^{-1} \Iprod{\Gamma}{R^*} R\ ,  \CR
 \Gamma^\ord{1} &=& \frac{1}{2} \Gamma+ \frac{1}{2}\Iprod{R}{R^*}^{-1}  \Scal{  \frac{1}{2} I_4^{\prime}( R,R^* ,\Gamma)  - 3 \Iprod{\Gamma}{R^*} R  +R^* \Iprod{R}{\Gamma}} \ , \CR
  \Gammamun &=& \frac{1}{2} \Gamma- \frac{1}{2}\Iprod{R}{R^*}^{-1}  \Scal{  \frac{1}{2} I_4^{\prime}( R,R^* ,\Gamma)  -  \Iprod{\Gamma}{R^*} R  +3R^* \Iprod{R}{\Gamma}} \ , \CR
 \Gammamtrois  &=& \Iprod{R}{R^*}^{-1} \Iprod{R}{\Gamma} R^* \ . \label{RProjector} 
\end{eqnarray}
Note that the $\Gamma^\ord{\pm1}$ can simply be identified as the solutions to
\be\label{pmunConstraint} \frac{1}{2}  I^\prime_4(R,R^*,\Gamma^\ord{\pm1}) = \pm \Iprod{R}{R^*} \Gamma^\ord{\pm 1} \ . \ee
These expressions will be very useful in evaluating the action of T-dualities on
general symplectic vectors in the following sections. A further practical advantage of this
decomposition is the fact that all inner products must respect the grading, leading
to strong constraints on the possible nontrivial combinations. For instance, the grading
implies that
\begin{equation}
I^\prime(\Gammamun,\Gammamun,R^*)
=  0 \ , \qquad I^\prime(\Gamma^\ord{1},\Gamma^\ord{1},R)=  0  \ ,
\end{equation}
since there is no vector of weight $\pm 5$ that
these cubic terms could be equal to. Similar considerations apply to scalar products,
which necessarily vanish unless the sum of grades of the vectors involved vanishes.

In addition to the decomposition \eqref{eq:vec-decomp} of the vector space, the generator ${\bf h}_T$
implies a corresponding decomposition of the duality group generators. Indeed, ${\bf h}_T$ commutes
with $\mathfrak{g}_5 \subset \mathfrak{g}_4$ and defines the following graded decomposition of
$\mathfrak{g}_4$ 
\begin{equation} \label{gl4-decomp}
\mathfrak{g}_4 \cong ({\mathds{R}}^{n_v})^\ord{-2} \oplus \scal{ \mathfrak{gl}_1 \oplus \mathfrak{g}_5}^\ord{0} \oplus   ({\mathds{R}}^{n_v})^\ord{2} \ , 
\end{equation}
where the $\mathfrak{gl}_1$ corresponds to ${\bf h}_T$ itself. Clearly, the $2 n_v$ generators
of eigenvalue $\pm 2$ with respect to ${\bf h}_T$ can be used as raising and lowering operators
on the eigenspaces in \eqref{eq:vec-decomp}. As the reader might already understand, these grade $2$ generators are related to the transformations  \eqref{eq:spec-flow} by similarity transformations in $G_4$.

In terms of the explicit expression \eqref{g4Freud} for the action of $\mathfrak{g}_4$, one may
consider any grade $-1$ vector of parameters $\kmun$ to define the grade 2 generators as
\begin{equation}
 \Tgp_k \Gamma \equiv - \frac{1}{2} \Iprod{R}{R^*}^{-1}  \Scal{  \frac{1}{2} I_4^{\prime}( R,\kmun ,\Gamma)  + \Iprod{\Gamma}{\kmun} R  -\kmun \Iprod{R}{\Gamma}} \  . \label{Tkplus}  
\end{equation}
This generator is manifestly of grade $2$ because of the grading of $R$ and $\kmun$ themselves and the algebra \eqref{TripleAlgebra}. It is convenient to write it in a way that makes the grading explicit 
\begin{equation}
 \Tgp_k \Gamma = \Iprod{R}{R^*}^{-1}  \Scal{  \kmun \Iprod{R}{\Gamma^\ord{-3}} -   \frac{1}{4} I_4^{\prime}( R,\kmun ,\Gammamun)    - \Iprod{\Gamma^\ord{1}}{\kmun} R } \  , \label{Tp1} 
\end{equation}
where we used the projections in \eqref{RProjector} and the fact that $\kmun$ is of grade $(-1)$. 
All these generators clearly commute between themselves for different $\kmun$'s. 
Similarly, one defines the grade $-2$ generator in terms of a grade $1$ vector $k^\ord{1}$ 
\bea \Tgm_k \Gamma &\equiv&  \frac{1}{2} \Iprod{R}{R^*}^{-1}  \Scal{  \frac{1}{2} I_4^{\prime}( R^*,k^\ord{1} ,\Gamma)  - \Iprod{k^\ord{1}}{\Gamma} R^*  +k^\ord{1} \Iprod{\Gamma}{R^*}}\CR
 &=& \Iprod{R}{R^*}^{-1}  \Scal{k^\ord{1} \Iprod{\Gamma^\ord{3}}{R^*}  + \frac{1}{4} I_4^{\prime}( R^*,k^\ord{1} ,\Gamma^\ord{1})  - \Iprod{k^\ord{1}}{\Gammamun} R^*  } \  . \label{Tm1} \eea 
The normalisations we have chosen are such that 
\begin{equation}\label{t-dual-params}
 \Tgp_k R^* = \kmun \ , \qquad \Tgm_k R = k^\ord{1} \ ,   
\end{equation}
while one easily computes that 
\begin{equation}
\Tgp_k R = 0 \ , \qquad \Tgm_k R^* = 0 \ .
\end{equation}
In this form, one easily computes that these generators are nilpotent of order 4, as
\begin{equation}
 (\Tgpm_k)^{4} \Gamma = 0 \ ,
\end{equation}
consistent with the grading \eqref{eq:vec-decomp}, which only allows for four eigenspaces.
Explicitly, we find the following expressions for the two sets of generators
\begin{eqnarray}
 (\Tgp_k)^{ 2} \Gamma&=& - \frac{1}{4}  \Iprod{R}{R^*}^{-2} \Scal{ I_4^\prime(R,\kmun,\kmun) \Iprod{R}{\Gamma} + I_4(R,\kmun,\kmun,\Gamma) R } \  , \label{Tp2} \\
(\Tgp_k)^{3} \Gamma &=&- \frac{1}{4}  \Iprod{R}{R^*}^{-3}  I_4(R,\kmun,\kmun,\kmun) \Iprod{R}{\Gamma} R \ , \label{Tp3}\\
(\Tgm_k)^{ 2} \Gamma&=&  \frac{1}{4}  \Iprod{R}{R^*}^{-2} \Scal{ I_4^\prime(R^*,k^\ord{1},k^\ord{1}) \Iprod{\Gamma}{R^*} - I_4(R^*,k^\ord{1},k^\ord{1},\Gamma) R^* } \  , \label{Tm2}\\
(\Tgm_k)^{3} \Gamma &=&-\frac{1}{4}  \Iprod{R}{R^*}^{-3}  I_4(R^*,k^\ord{1},k^\ord{1},k^\ord{1}) \Iprod{\Gamma}{R^*} R^* \ , \label{Tm3} 
\end{eqnarray}
and moreover
\begin{equation} 
[ {\bf h}_T , \Tgpm_k ] =  \pm 2 \Tgpm_k \ , \label{TGrading} 
\end{equation} 
as in \eqref{gl4-decomp}.

Finally, one also computes using \eqref{TripleAlgebra} and the grading that for any
grade $-1$ vector $e$ and grade $1$ vector $f$
\begin{equation} 
[ \Tgp_e , \Tgm_f ] = \frac{1}{2} \frac{1}{\Iprod{R}{R^*}} \Scal{ {\bf g}(e,f)  + \Iprod{e}{f} {\bf h}_T } \ . \end{equation}
One can straightforwardly check that 
\begin{equation} 
\Scal{  {\bf g}(e,f)  + \tfrac{1}{3}  \Iprod{e}{f} {\bf h}_T } R = 0 \ , \qquad  \Scal{  {\bf g}(e,f)  + \tfrac{1}{3}  \Iprod{e}{f} {\bf h}_T } R^* = 0 \ , 
\end{equation}
so that the latter transformation lies in the $\mathfrak{g}_5 \subset \mathfrak{g}_4$
subalgebra, consistently with the graded decomposition \eqref{gl4-decomp}. One can
indeed check that these transformations preserve the cubic norm
$I_4(R,\Gammamun,\Gammamun,\Gammamun)$ for an arbitrary grade $-1$ vector $\Gammamun$.
It turns out that the identity \eqref{Freudenthal} implies the associated Jordan identity
\begin{equation}
I_4^\prime(R^*,I_4^\prime(R,\Gammamun,\Gammamun), I_4^\prime(R,\Gammamun,\Gammamun)) =  \frac{64}{3} \Iprod{R}{R^*} I_4(R,\Gammamun,\Gammamun,\Gammamun) \Gammamun \ ,
\end{equation}
which generalises \eqref{symmetric}. These equations are clearly valid upon replacing
$R$ with $R^*$ and $\Gammamun$ by a grade $(+1)$ vector $\Gamma^\ord{1}$ throughout.

One may now use the above formulae to identify $\Tgpm$ with T-dualities explicitly.
Indeed, one can easily check that upon identifying $R$ with the very small vector whose only
nonvanishing component $q_0$ and $R^*$ with its magnetic dual along $p^0$, the exponentiated
transformations 
\begin{equation}\label{expTdual}
\exp[\Tgp_k] = 1 + \Tgp_k + \tfrac12\, (\Tgp_k)^2 + \tfrac16\, (\Tgp_k)^3\,, 
\end{equation}
are identical to the spectral flow shown in \eqref{eq:spec-flow}.
The corresponding set of generators $\Tgm$ then generate the T-dualities one obtains
by conjugating \eqref{eq:spec-flow} by an electric/magnetic duality and leave $R^*$ invariant.
The grade $(-1)$ and $(+1)$ components are then easily seen to be given by the magnetic, $p^i$,
and electric components, $q_i$, respectively.

In the general case, we can identify all possible sets of T-dualities as given by a
choice of $R$ or $R^*$, as above, as the generators $\Tgpm_k$ are entirely determined
by the rank 1 vector they leave invariant. Indeed, the characteristic feature of these abelian
subgroups is that there is always a unique (up to rescaling) very small vector (\eg $R$) that
they leave invariant, whereas they act transitively (up to a rescaling) on the set of very small
vectors (\eg $R^*$) that are not mutually commuting with the former. In the specific example of
\eqref{eq:spec-flow}, any very small vector that is not mutually local with $R$ (along $q_0$)
can be obtained by acting with a finite transformation $\exp(\Tgp_k)$ on $R^*$ (along $p^0$).

In the following sections, we will consider the action of general T-dualities, as we find
it convenient to describe multi-centre black hole solutions in terms of two auxiliary very small vectors
$R$ and $R^*$ that arise naturally from the equations of motion, as mentioned below
\eqref{der-small}. Therefore, we will always consider a T-duality as given explicitly by
an exponential as in \eqref{expTdual}, where the explicit action of each order is given
by \eqref{Tp1}-\eqref{Tm3} above, rather than the equivalent matrix similar to the one in
\eqref{eq:spec-flow} that has to be defined explicitly.

This concludes our discussion of T-dualities in the real basis. In the next section,
we consider the same transformations in the complex basis, for later use. The reader interested
in constructing solutions can however safely skip this technical discussion.

\subsection{T-dualities in the complex basis}
\label{sec:t-dual-def}
In this section we discuss the realisation of T-dualities in the complex basis
defined by the central charge $Z = Z(\Gamma)$ and its K\"{a}hler derivative
$Z_a = Z_a(\Gamma)$. The discussion here is parallel to the one of the previous
section, in the real basis, and is complementary to it. However, the construction of
the T-duality generators in the complex basis will be necessary to solve the first
order equations describing black hole composites in the following. 

For this purpose, we define the very small vector $R$ as in \eqref{N-Om-def}-\eqref{small-def},
while $R^*$ is defined from $R$ using an arbitrary phase, $e^{i\alpha} \ne \N[\Omega] $, as
\begin{equation}\label{R-star-bas}
 Z(R^*) = e^{3i\alpha/2}\N[\bar\Omega]^{1/2}\,, \qquad
 Z_a(R^*) = e^{i\alpha/2}\N[\bar\Omega]^{1/2}\Omega_a\,,
\end{equation}
where the choice of the phases is done for later convenience. It will also be useful to define
the complex function $Y$
\begin{equation} \label{Y-def-gen}
Y  \equiv \frac{2}{1-e^{-i\alpha} \N[\Omega]}
   = -i\, \frac{e^{i\alpha/2} \N[\bar\Omega]^{1/2}}{\mbox{Im}\left(e^{i\alpha/2} \N[\bar\Omega]^{1/2}\right)}
   = 1 + i\,e^{2U} M\ ,
\end{equation}
that has unit real part and the specific parametrisation of the imaginary part will
become meaningful in the following. The vector \eqref{R-star-bas} is by construction
mutually nonlocal with $R$ because
\begin{equation} 
\Iprod{R^*}{R} =\Scal{ 2 \mbox{Im}\left(e^{i\alpha/2} \N[\bar\Omega]^{1/2}\right)}^3 \ ,
\end{equation}
and defines a natural magnetic dual to $R$.

We will first determine the T-dualities $\Tgp$ that leave $R$ invariant. We note that $\Omega_a$ is by construction \eqref{small-def} invariant with respect to a subgroup $K_5\subset K_4$. In order to describe the action of $\mathfrak{g}_4$ in \eqref{g4-dec}
on this vector, we will parametrize the remaining $n_v-1$ generators of $\mathfrak{k}_4$, which describe
the coset component $\mathfrak{k}_4\ominus\mathfrak{k}_5$, in terms of a vector $Q^a$. Requiring the matrix
\be -G^b{}_a(Q) = c_{ace} c^{bde} \Omega_d Q^c -  \Omega_a Q^bZ_b \ , \ee
to be anti-Hermitian and to preserve $\N[Z]$ fixes the relative coefficients and implies the constraints 
\begin{equation}\label{q-constr}
  \Omega_a Q^a = 0\,, \qquad Q^a = \N[\bar \Omega] c^{abc} \Omega_b \bar Q_c\,.
\end{equation}
Similarly, we parametrize $\mathfrak{u}(1)$ in \eqref{g4-dec} by $\gamma$ and $\mathds{C}^{n_v}$
by a complex vector $P^a$, such that the final result we find that the action of $\mathfrak{g}_4$ on
a general vector reads
\begin{align}\label{g4-act-P}
 \delta Z =&\, P^a Z_a + 3i\, \gamma \, Z \ ,  \CR
 \delta Z_a = &\, \bar P_a Z + c_{abc} P^b \bar Z^c + i\,\gamma\,Z_a
      +  c_{ace} c^{bde} \Omega_d Q^c\, Z_b
      -   \Omega_a Q^bZ_b\ .
\end{align}
In order to describe T-dualities, we must impose that these transformations
leave $R$ invariant, which can be shown to hold if \footnote{Note that $R$ is also
invariant if
\[
 \gamma=Q^a=0\,, \quad P^a \Omega_a=0,, \quad \bar P_a = - \N[\bar \Omega]\, c_{abc} \bar \Omega^b P^c\,.
\]
These generators, along with the $K_5$ subgroup of $K_4$ and the generators described
by \eqref{eq:P-g-Q}, account for the full $G_5\!\ltimes\! \mathbb{R}^{n_v}$ subgroup of
$G_4$ leaving invariant a given very small vector \cite{Ferrara:1997uz}.}
\begin{equation}\label{eq:P-g-Q}
 P^a=-i\,\gamma\,\N[\Omega]\,\bar \Omega^a - \N[\Omega]\,Q^a\ .
\end{equation}
One can now verify that the resulting transformations
\begin{align}\label{T-duality}
 \delta Z \equiv \Tgp_{\gamma, Q}  Z
     =&\,3\, i\, \gamma \, Z   -i\,\gamma\,\N[\Omega]\,\bar \Omega^aZ_a  - \,\N[\Omega]\,Q^a Z_a\CR
 \delta Z_a \equiv \Tgp_{\gamma, Q} Z_a
     = &\, i\,\gamma\,Z_a+  i\,\gamma\,\N[\bar \Omega]\,\Omega_a Z
      - i\,\gamma\,\N[\Omega]\,c_{abc} \bar \Omega^b \bar Z^c
      -\,\N[\bar\Omega]\,\bar Q_a Z
\CR & -\,\N[\Omega]\,c_{abc}Q^b \bar Z^c + \, c_{ace} c^{bde} \Omega_d Q^c\, Z_b
      - \, \Omega_a Q^bZ_b\,,
\end{align}
leave $R$ invariant and commute with each other. Furthermore, one shows that
$\Tgp_{\gamma, Q}$ is nilpotent of order four, as is clear from the example
\eqref{eq:spec-flow} where terms at most cubic in the parameters $k^i$ appear.
Indeed, $\Tgp_{\gamma, Q}$ can be identified with the corresponding generators
in \eqref{Tp1}, which act as raising operators on the decomposition
\eqref{eq:vec-decomp} and the vector $R$ is the highest weight vector, to
which we assign weight $3$.

At this point it is important to appreciate the fact that, while we used the complex
scalar dependent basis to define T-dualities, the following relations hold
\be  \Tgp_{\gamma, Q} Z(\Gamma) = Z(  \Tgp_{k} \Gamma) \ , \quad   \Tgp_{\gamma, Q} Z_a(\Gamma) = Z_a(  \Tgp_{k} \Gamma)\ , \ee
where $\Tgp_{k}$ denote the representation of these generators in the real basis, parametrised
in terms of a grade $-1$ vector $k$, as in \eqref{Tkplus}-\eqref{Tp1}. It follows that the
parameters $\gamma$ and $Q^a$ depend on the $n_v$ constant parameters $k$ and the scalar fields.

As alluded to above, the second very small vector \eqref{R-star-bas} plays a role dual 
to that of $R$, as one can check that $R^*$ is never a zero mode of the T-duality operator
defined in \eqref{T-duality}, and in particular
\begin{eqnarray}
Z(\Tgp_k R^*) &=& Z(k)=
-3\,Y^2 \N[ \Omega]\,\gamma\,,
\CR
Z_a(\Tgp_k R^*) &=& Z_a(k) = \left( 3\,|Y|^2-2\,Y\right) \Omega_a \,\gamma
 -4\,i\,Y\, \bar Q_a
\,,
\label{rstar-flow}
\end{eqnarray}
where we also used \eqref{t-dual-params} to given the explicit relation of the vector
$k$ to the parameters $\gamma$ and $Q_a$.
In addition, one can verify that $(\Tgp_{\gamma, Q})^3 R^* \propto R$, as in \eqref{Tp3},
so that the vector $R^*$ can be identified with the lowest weight vector of the operators
$\Tgp_{\gamma, Q}$, with assigned weight $-3$.

As expected from the example in \eqref{eq:spec-flow}, the parameters $k^i$ must be
a rank-three vector for a general T-duality. Indeed, the vector defined by
\eqref{rstar-flow} is of rank three, as can be shown by computing its quartic invariant.
Furthermore, one can verify that this vector satisfies the reality constraint
\bea
 \bar Z^a  - \N[\Omega] \bar \Omega^a \bar Z =
 e^{-i \alpha} \scal{ c^{abc}
\Omega_b Z_c  + \bar \Omega^a ( Z - \N[\Omega]  \bar \Omega^b Z_b )}
\label{ZAconstraint} \,,
\eea
introduced in \cite{Bossard:2012xsa} in the study of single centre solutions. This
defines a Lagrangian subspace that includes the small vector $R$, whereas one also verifies that
\bea
\bar Z(\Tgp R^*)  &=& e^{-2i\alpha} Z(\Tgp R^*) \ , \CR
\bar \Omega^a Z_a(\Tgp R^*) &=& ( 2 e^{-i\alpha} + \N[\bar \Omega] ) Z(\Tgp R^*)\ ,  \eea
which implies that $\langle \Tgp R^* , R^*\rangle = 0$.

Making use of the above structure based on the original very small vector $R$, one can proceed to
define similar structures for the dual very small vector $R^*$, in exactly the same way. This
seems redundant at first sight, since one can always identify the small vector invariant under
the T-dualities with $R$, as above. However, this is more natural in view of the discussion in
the real basis in the previous section, as well as for the applications we are interested in,
where both vectors appear simultaneously. It is therefore useful to have a dual description in terms of
$R^*$ throughout.

The T-dualities leaving $R^*$ invariant can then be shown to be defined as in \eqref{g4-act-P}
with parameters given by
\be P^a = - e^{i\alpha} ( i \gamma\bar \Omega^a + Q^a ) \ , \ee
where the $Q^a$ are again constrained by \eqref{q-constr}, so that the resulting transformations read
\begin{align}\label{T-duality-Rstar}
 \delta Z \equiv \Tgm _{\gamma, Q} Z
     =&\,  3i\, \gamma \, Z -i\,\gamma\,e^{i\alpha}\,\bar \Omega^aZ_a
          - \,e^{i\alpha} \,Q^a Z_a  \CR
 \delta Z_a \equiv \Tgm _{\gamma, Q} Z_a
     = &\, i\,\gamma\,e^{-i\alpha}\,\Omega_a Z
      - i\,\gamma\,e^{i\alpha}\,c_{abc} \bar \Omega^b \bar Z^c + i\,\gamma\,Z_a
      -\,e^{-i\alpha}\,\bar Q_a Z
\CR & -\,e^{i\alpha}\,c_{abc}Q^b \bar Z^c
      + \, c_{ace} c^{bde} \Omega_d Q^c\, Z_b
      - \, \Omega_a Q^bZ_b\,.
\end{align}
As expected, $R$ is never invariant under \eqref{T-duality-Rstar}, with transformation rule
\bea\label{r-flow}
Z(\Tgm_k  R) &=& Z(k)= -6\,i\,\frac{e^{i\alpha}}{Y}  \,\gamma
\CR
Z_a(\Tgm_k  R) &=& Z_a(k)=
-2 i\,|Y|^{-2} \, ( \bar Y - 2 Y ) \Omega_a \,\gamma
 +4\,|Y|^{-2} \, \bar Q_a
\,,
\eea
where we show the relation of the vector $k$ in \eqref{t-dual-params} to the parameters
in the complex basis. Again, one can compute that $(\Tgm_{\gamma, Q})^3 R \propto R^*$,
consistent with \eqref{Tm3}. It follows that $\Tgm_{\gamma, Q}$
are lowering operators with $R^*$ and $R$ as their lowest and highest weight vectors respectively.
Furthermore, one can define a Lagrangian subspace that
includes $R^*$ and $\Tgm  R$, through the constraint dual to \eqref{ZAconstraint}, as
\begin{equation}
\bar Z^a - e^{i\alpha}\, \bar \Omega^a \bar Z =
\N[\bar\Omega]  c^{abc} \Omega_b Z_c + \bar \Omega^a ( e^{-i\alpha} Z - \bar \Omega^b Z_b )
\label{ZAconstraintStar} \,.
\end{equation}
In addition, the vector \eqref{r-flow} satisfies the relations
\bea
\bar Z(\Tgm  R)  &=& e^{-i\alpha}  \N[\bar\Omega] Z(\Tgm R) \ , \CR
\bar \Omega^a Z_a(\Tgm R) &=& ( 2 \N[\bar \Omega] + e^{-i\alpha}  ) Z(\Tgm R)\ ,  \label{ZAConstraintK}\eea
which imply that $\langle \Tgm R , R\rangle = 0$.

One can obtain the action of the relevant generator ${\bf h}_T$ defined in \eqref{hTReal} in the complex basis,
as the commutator of a T-duality leaving $R$ invariant and a T-duality leaving $R^*$
invariant. In general, such a commutator will also give rise to an element of the grade
zero component $\mathfrak{g}_5$, as in \eqref{gl4-decomp}. Choosing the parameters
$\gamma$ and $Q^a$ for the two transformations to be identical, one obtains
\begin{equation}\label{T-comm}
 [\Tgp , \Tgm ] =  {\bf G}(\gamma , Q^a) \,,
\end{equation}
where $ {\bf G}(\gamma , Q^a)$ is a generator of $\gl_1 \oplus \mathfrak{g}_5$ bilinear
in $\gamma$ and $Q^a$. The $\gl_1$ component, which is to be identified with the operator
${\bf h}_T$, corresponds to the transformation of parameter
$\frac{1}{2} \gamma^2 + \frac{1}{3} Q^a \bar Q_a$, whereas the $\mathfrak{g}_5$
transformation is parametrised by $\gamma Q^a$ and
$Q^a \bar Q_b - \frac{1}{n_v-1} ( \delta^a_b - \frac{1}{3} \bar \Omega^a \Omega_b ) Q^c \bar Q_c $.
To project to the $\gl_1$ component, one can simply identify the terms quadratic in $Q^a$ as
\begin{equation}
 Q^a \bar Q_b \sim \frac{1}{n_v-1} \scal{  \delta^a_b - \tfrac{1}{3} \bar \Omega^a \Omega_b } Q^c \bar Q_c  \ ,
\end{equation}
so as to cancel all the terms in $\mathfrak{g}_5$. Equivalently, this identification can
be understood as an average obtained by acting on the parameter $Q^a$ with the
$K_5\subset K_4$ subgroup leaving $\Omega_a$ invariant, and integrating out the result
over $K_5$. By definition, none of the generators of $\mathfrak{g}_5$ are $K_5$ singlets,
and the resulting expression is necessarily proportional to the $\gl_1$ generator ${\bf h}_T$. Because
of the reality constraint \eqref{q-constr} on the $Q^a$, this average furthermore implies
\begin{equation}
Q^a Q^b \sim \frac{1}{n_v-1} \bar Q_d Q^d \Scal{ \N[\bar \Omega] c^{abc} \Omega_c - \tfrac{2}{3}\bar  \Omega^a \bar \Omega^b } \ ,
\end{equation}
and therefore
\begin{equation}\label{constr-hT}
c_{abc} Q^b Q^c = \frac{1}{3}  \bar Q_b Q^b \ \N[\bar \Omega]  \Omega_a \ ,
\end{equation}
where we used \eqref{symmetric} to show that $c^{acd} c_{bcd} = \frac{n_v+3}{3} \delta^a_b$. 
In practice, \eqref{constr-hT} is the only constraint one needs to use when computing the
commutator in \eqref{T-comm}.

After imposing the relations above, one finds
\begin{align}
{\bf h}_T  Z =&\,  \frac{1}{4} |Y|^2  \Scal{ 3\,(e^{i\alpha}\N[\bar\Omega] - e^{-i\alpha}\N[\Omega])\,Z
             + 2\,(\N[\Omega] - e^{i\alpha})\,\bar\Omega^a Z_a} \,,
\CR
{\bf h}_T Z_a = &\,\frac{1}{4} |Y|^2  \Bigl(   (e^{i\alpha}\N[\bar\Omega] - e^{-i\alpha}\N[\Omega])\,Z_a
             + 2\,(\N[\bar\Omega] - e^{-i\alpha})\,Z\,\Omega_a \Bigr .
\CR      &\, \hspace{65mm}
 \Bigl . + 2\,(\N[\Omega] - e^{i\alpha})\,c_{abc}\bar\Omega^b \bar Z^c \Bigr ) \,.
\label{grade-operator}
\end{align}
One can now check that \eqref{TGrading} is indeed satisfied in the complex basis. Similarly, the vector space of charges splits into 4 subspaces of eigenvalue $\{-3, -1,1,3\}$ with
respect to this linear operator.

By construction, $R$ and $R^*$ are the unique vectors of eingenvalue
$3$ and $-3$ respectively, up to an overall rescaling. The remaining $2 n_v$ directions can then 
be simply identified with the parameters of the two T-dualities $\Tgpm$, as given above. 
It follows that the eigenspace of eigenvalue $-1$ is spanned by the $n_v$ vectors that satisfy the
constraint \eqref{ZAconstraint} and are mutually local\footnote{This is has to be imposed since $R$
itself is the trivial solution of \eqref{ZAconstraint}} with $R^*$, as $\Iprod{R^*}{\Gammamun}=0$,
which is equivalent to
\bea
\bar Z(\Gammamun)  &=& e^{-2i\alpha} Z(\Gammamun) \ , \CR
\bar \Omega^a Z_a(\Gammamun) &=& ( 2 e^{-i\alpha} + \N[\bar \Omega] ) Z(\Gammamun)\ , \CR
c^{abc} \Omega_b Z_c(\Gammamun) &=& e^{i\alpha} \bar Z^a + e^{-i\alpha} \N[\Omega] \bar \Omega^a Z(\Gammamun)\ . \eea
Similarly, the eigenspace of eigenvalue $1$ is spanned by the $n_v$ vectors that satisfy
the dual constraint \eqref{ZAconstraintStar} and $\Iprod{R}{\Gamma^\ord{1}}=0$, which lead to
\bea
\bar Z(\Gamma^\ord{1})  &=& e^{-i\alpha}  \N[\bar\Omega] Z(\Gamma^\ord{1}) \ , \CR
\bar \Omega^a Z_a(\Gamma^\ord{1}) &=& ( 2 \N[\bar \Omega] + e^{-i\alpha}  ) Z(\Gamma^\ord{1})\ , \CR
c^{abc} \Omega_b Z_c(\Gamma^\ord{1}) &=& \N[\Omega]  \bar Z^a +  \bar \Omega^a Z(\Gamma^\ord{1})\ .  \eea
These equations can be identified from \eqref{RProjector} in the real basis as (\ref{pmunConstraint}).

\section{The $c^*$-map, nilpotent orbits and first order systems}
\label{sec:syst-def}

In this section we consider the first order systems describing multi-centre
non-BPS black holes in $\N=2$ supergravity coupled to $n_v$ vector
multiplets labelled by an index $i=1,\dots, n_v$. We refer to the appendix
for a short overview of our conventions on $\N=2$ supergravity, which
coincide with the ones in \cite{Bossard:2012xsa}, to which we refer for further
details.

The systems of black holes we are interested in can be constructed systematically
in the special case when the special K\"ahler manifold, $\mathcal{M}_4$,
parametrised by the vector multiplet scalars, $t^i$, is symmetric. Moreover,
we exclusively consider stationary solutions, \ie we always assume a timelike
isometry. In this case, one can consider a timelike dimensional reduction to three
dimensions and dualise all vector fields to scalars \cite{Breitenlohner:1987dg},
to obtain an effective euclidean sigma model describing stationary black hole
backgrounds. The resulting equations of motion are still rather complicated,
so that it is common to consider special linear systems
that solve the full equations of motion, but do not provide a full list of
possible solutions. There are two such systems known, namely the composite
non-BPS system \cite{Bossard:2011kz} and the almost-BPS system
\cite{Goldstein:2008fq}, which together account for a representative majority of
the explicitly known multi-centre solutions featuring a flat three-dimensional
base space. The purpose of this section is to define these systems in terms of
four-dimensional, manifestly duality covariant quantities, aiming for a clear
description of their general structure.

To this end, we make use of the duality symmetries of the three-dimensional theory
resulting from the dimensional reduction, through the formalism developed in
\cite{Bossard:2010mv,Bossard:2012xsa}. We therefore first describe the basics
of this effective theory in section \ref{sec:3d-coset}, followed by a discussion
of the method of nilpotent orbits in section \ref{sec:nil-orb}. We then present
the derivation of the first order flow equations for the composite non-BPS system
and the almost-BPS system in sections \ref{sec:comp-flow} and \ref{sec:alm-flow}
respectively. Note that, while the derivation of the almost-BPS system has not
appeared before, our section \ref{sec:comp-flow} is essentially a review of
the derivation of the same system in \cite{Bossard:2012xsa}, which we include
for completeness.

\subsection{The three-dimensional non-linear sigma model}
\label{sec:3d-coset}
In order to describe stationary asymptotically flat extremal black holes, we
introduce the standard Ansatz for the metric
\begin{equation}
 \label{metricMltc}
 ds^2=- e^{2U}( d t+\omega )^2 + e^{-2U}   d {\bf x} \cdot  d{\bf x} \,,
\end{equation}
in terms of a scale function $U(x)$ and the Kaluza--Klein one-form
$\omega(x)$ (with spatial components only), which are both required to asymptote to zero at
spatial infinity. Here and henceforth, all quantities are independent of
time, so that all scalars and forms are defined on the flat three-dimensional
base. The $n_v+1$ gauge fields of the theory $F^{I} = d A^I$ for $I=0,\dots n_v$
include the graviphoton and the vector multiplet gauge fields. Together with their
magnetic duals, these can be arranged in a symplectic vector, $\cF$, as in
\eqref{eq:dual-gauge}, which transforms linearly under electric/magnetic duality. 
For a background as in \eqref{metricMltc}, the appropriate decomposition 
of the gauge fields takes the form
\begin{equation} 2  \mathcal{A} = \zeta ( d t+\omega) + w \end{equation}
and accordingly for the field strengths
\begin{equation}
 \label{gauge-decop}
2  \cF= d \zeta \, ( d t+\omega) + F   \,, \qquad
   F= \zeta\, d  \omega + d w \,,
\end{equation}
where we defined the gauge field scalars $\zeta$, arising as the time component
of the gauge fields, and the one-forms $w$ describing the charges. Here, $F$ is
defined as the spatial component of the field strength, which is not closed but
satisfies
\begin{equation} d F = e^{2U} \star d \omega \wedge \mathrm{J} F \; , \label{F-closed}  \end{equation}
according to \eqref{cmplx-sdual}, which can be written as
\begin{equation}\label{cmplx-self}
 d \zeta = e^{2U}  \mathrm{J}\star F \,.
\end{equation}
Note that this first order equation determines the $\zeta$ in terms of the vector fields $w$ and the
scalars.

Upon dimensional reduction over the time direction, the set of moduli $t^i$ parametrising the
coset space \eqref{4d-coset} are extended to include the scaling factor $U$ and the scalar dual to
the angular momentum $\omega$ in \eqref{metricMltc}, as well as the fields $\zeta$ in \eqref{gauge-decop}, which altogether parametrize the para-quaternionic
symmetric space \footnote{Here para-quaternionic refers to the property that the holonomy group of
$\mathcal{M}_3$ $SL(2) \times G_4 \subset SL(2) \times Sp(2n_v+2,\mathds{R})$.}
\begin{equation} \label{M3}
\mathcal{M}_3 \cong G_3 / \scal{ SL(2) \times G_4 }  \ .
\end{equation}
This defines the so-called $c^*$-map, which can be related to the standard $c$-map
\cite{Ferrara:1989ik} by analytic continuation. The three-dimensional symmetry group
Lie algebra $\g_3$ decomposes as
\begin{equation} \label{3d-alg-decomp}
\mathfrak{g}_3 \cong \mathbf{1}^\ord{-2}\oplus \mathfrak{l}_4^\ord{-1}
  \oplus (\gl_1 \oplus \mathfrak{g}_4)^\ord{0}
  \oplus \mathfrak{l}_4^\ord{1} \oplus \mathbf{1}^\ord{2}\,,
\end{equation}
where the weights refer to the
eigenvalues under the adjoint action of the $\gl_1$ generator. The grade one
generators in $\mathfrak{l}_4^\ord{1}$ are associated to the gauge invariance
with respect to a constant shift of the scalars $\zeta$, and accordingly the
grade two generator corresponds to the shift of integration constant defining the scalar dual to $\omega$.

As discussed in more detail in \cite{Bossard:2012xsa}, the equations of motion
for the scalar fields pa\-ra\-me-tri\-sing the symmetric space $\mathcal{M}_3$ are
expressed in terms of the corresponding Maurer--Cartan form
\begin{equation} v^{-1} d v = P + B\,. \end{equation}
Here, $v \in G_3$ is a coset representative describing the scalar fields,
while $P$ is the coset component of the Maurer--Cartan form, defining the scalar
momenta. Similarly, $B$ is the $\sl_2 \oplus \g_4$ component defining the pulled
back spin connection. In components, the scalar momenta are defined as
\begin{equation} \label{P-component}
{\rm w} \equiv - d U - \frac{i}{2} e^{2U} \star d \omega \,, \quad \Sigma^a =- e^a_i d t^i\,, \quad
Z \equiv e^U Z(\star F)\,, \quad Z_a \equiv e^U Z_a(\star F)\ ,
\end{equation}
where we introduce some shorthand notations, based on the central charges in
\eqref{central-ch-def}, that will be used for the remainder of the section.

At this stage it is important to introduce some properties of the $\mathfrak{g}_3$ algebra.
The components of an element of the Lie algebra $\g_3$ in the coset component
$\g_3 \ominus ( \sl_2 \oplus \g_4) $, as given in \eqref{P-component} above, correspond
to $U(1) \times U(1) \times K_4 $ irreducible representations as the two complex parameters ${\rm w}$
and $Z$ and the two complex vectors $\bar Z^a , \Sigma^a$ transform in the $\mathds{C}^{n_v}$
representation of $K_4$, according to the decomposition \eqref{g4-dec}, the same as the scalar
field momenta $e^a_i d t^i$ in four dimensions. As one can check explicitly, the quadratic
trace invariant which defines the $SL(2) \times G_4$ invariant norm
\begin{equation}
| {\rm w} |^2 - |Z|^2  - Z_a \bar Z^a + \Sigma^a \bar \Sigma_a \label{Tracesquare}\ ,
\end{equation}
is equivalent to the effective Lagrangian in the background \eqref{metricMltc}.

The $\sl_2$ algebra is realised on these components as
\begin{equation}
\delta  {\rm w} =i \rho  {\rm w}  + \bar \lambda Z\,, \quad
\delta Z =-  i \rho Z  +  \lambda {\rm w}\,,  \quad
\delta \bar Z^a = i \rho \bar Z^a + \bar \lambda \Sigma^a\,, \quad
\delta \Sigma^a = -i \rho \Sigma^a + \lambda \bar Z^a\,,
\label{sl2-alg}\end{equation}
where $\rho$ and $\lambda$ are a real and a complex parameter respectively,
parametrising the $\sl_2$ group.
Similarly, the action of $\g_4$ can be written as using the parameters defined below
\eqref{g4-dec}, as
\begin{equation}\begin{split} 
\delta {\rm w} &= \Lambda_a \bar Z^a + 3 i \gamma {\rm w}\,,\\
\delta Z &= \Lambda_a \Sigma^a + 3 i \gamma Z \,, \end{split}\qquad
\begin{split} \delta \Sigma^a &= \bar \Lambda^a Z + c^{abc} \Lambda_b Z_c +
G^a{}_b \Sigma^b + i \gamma \Sigma^a \ , \\
\delta \bar Z^a &=  \bar \Lambda^a {\rm w}  + c^{abc} \Lambda_b \bar \Sigma_c +
G^a{}_b\bar Z^b + i \gamma \bar Z^a \ .  \end{split}\end{equation}
Note that the $\g_4$ action defined by these equations corresponds
to the divisor group in \eqref{M3}, rather than the original $\g_4$ as given in four
dimensions in \eqref{g4-algebra}, the two being related by a conjugation
in $\g_3$.

Finally, we give the components of the $\sl_2 \oplus \g_4$ component
of the Maurer-Cartan form, $B$, along $\sl_2$
\begin{equation}
\rho(B) = -\tfrac{1}{4} \mathrm{e}^{2U} \star d \omega - \tfrac{1}{2} Q \,, \qquad
\lambda(B) =  e^U Z(\star F)  \,, \label{Bsl2} \end{equation}
and along $\g_4$
\begin{gather}
\gamma(B) = -\tfrac{1}{4} \mathrm{e}^{2U} \star d \omega + \tfrac{1}{6} Q \,, \qquad
\Lambda_a(B) = e^U Z_a(\star F) \,,  \label{Bg4} 
\\
G^a{}_b(B) = e^a_i \partial_{\bar \jmath} e^i_b \, d \bar t^{\, \bar \jmath}
- e_{\bar \jmath b} \partial_i e^{\bar\jmath a} d t^i - \frac{2i}{3} \delta^a_b Q \ ,
\label{Bk4}
\end{gather}
where $G^a{}_b(B)$ defines the $\k_4$ valued traceless\footnote{To prove that
$G^a{}_b(B)$ is indeed traceless, one can use \eqref{Ctangent} and \eqref{symmetric}
to show that $c^{acd} c_{bcd} = \frac{n_v+3}{3} \delta^a_b$.} component of the pulled
back spin connection on $\mathcal{M}_4$ and $Q$ is the pulled back K\"{a}hler
connection \eqref{Kah-conn}.

\subsection{Nilpotent orbits}
\label{sec:nil-orb}

The basic observation for constructing black hole solutions in four dimensions using the
three-dimensional Euclidean theory describing stationary solutions is that regular stationary solutions of $\N=2$ supergravity
with a flat three-dimensional base metric can be described by a three-dimensional momentum $P$ that is nilpotent as a Lie algebra element. This implies in particular that $P$ can be
written in terms of the basis element ${\bf e}_\alpha$ of a nilpotent subalgebra of $\g_3$.
Such a subalgebra is always associated to a semi-simple element \footnote{Semi-simple means
that it is in the conjugation class of an element of the Cartan subalgebra.} ${\bf h}$ of
$\sl_2 \oplus \g_4$ such that
\begin{equation}  \label{nil-mom}
{\bf h }\,  {\bf e}_\alpha  :=  [  {\bf h } , {\bf e}_\alpha ]  = p_\alpha {\bf e}_\alpha \ , \qquad 1\le p_\alpha \le n \ , \end{equation}
where $n$ defines the maximal possible eigenvalue of $\mbox{ad}_{\bf h}$ in $\g_3$.  This implies for instance the equation
\begin{equation} \prod_{i=1}^n ( {\bf h } - i ) \, P = 0 \label{FirstOrderP} \ , \end{equation}
which defines a first order constraint on the components of $P$. In order for \eqref{FirstOrderP} to be consistent with the equations of motion and the Bianchi identity, the covariant derivative of the generator ${\bf h}$ must satisfy
\begin{equation}
\sum_{i=1}^n \prod_{j=i+1}^n ({\bf h} -j) d_B {\bf h}  \prod_{k=1}^{i-1} ( {\bf h} -k) \wedge P  = 0 \ , \quad  \sum_{i=1}^n \prod_{j=i+1}^n ({\bf h} -j) d_B {\bf h}  \prod_{k=1}^{i-1} ( {\bf h} -k) \star  P  = 0  \ . \end{equation}
These equations are satisfied if $d_B {\bf h}$ also lies in the nilpotent algebra defined by ${\bf h}$,  or equivalently if one imposes that
\begin{equation}
\prod_{i=1}^n ( \mbox{ad}_{\bf h} - i ) d_B {\bf h} = 0 \ . \label{AuxilH}
\end{equation}
Note that this is not necessarily the most general solution for the generator ${\bf h}$, as it might be possible
to construct special solutions, for which \eg there are preferred spacetime directions described
by a nonzero derivative. Considering this type of solutions can be understood as the natural generalisation of
the BPS black hole solutions in \cite{Behrndt:1997ny, Denef:2000nb} to include supersymmetric string solutions.
These are not accounted for by the standard black hole ansatz, but are are allowed by the requirement of
preserved supersymmetry. Similarly, we will only consider the generic situation in \eqref{AuxilH}, which should
be satisfied for all composite black hole solutions with a flat three-dimensional base.  

Without loss of generality, one can always choose the generators ${\bf h}\!\in\!\sl_2 \oplus \g_4$
such that only its components $\lambda$ and $\Lambda_a$ do not vanish, as these are the only
generators with a positive Cartan norm\footnote{This is similar to the canonical choice of
$\partial/\partial t$ as a timelike vector in Minkowski spacetime, \ie there are more general
timelike vectors but they can be obtained from the canonical one by symmetries.}. Equation
\eqref{AuxilH} can then be viewed as first order equations for these auxiliary components, which
can be solved to determine their evolution in space in terms of the physical fields. With this
information at hand, equation \eqref{FirstOrderP} then defines first order equations for the
physical fields, which contain the auxiliary components $\lambda$ and $\Lambda_a$ and determine
${\rm w}$ and $\Sigma^a$ of \eqref{P-component} in terms of $e^{U} Z(\star F)$ and
$e^{U} Z_a(\star F)$, plus some possible constraints on the latter if the dimension of the
coset component of the nilpotent algebra defined by ${\bf h}$ is strictly less than $2n_v + 2$.

Given the definitions in the previous section, it is possible to make \eqref{FirstOrderP}-\eqref{AuxilH} explicit, in order to determine the auxiliary components and the first order flow directly. A simple example is given by the BPS system, which is characterised by an element of ${\bf{h}}_*\!\in\!\sl_2$, \ie $\Lambda_a({\bf{h}}_*) = 0 $. It follows that its only nonvanishing component is $\lambda({\bf{h}}_*) = e^{i \alpha}$, where this phase is identified with the phase $\alpha$ that defines the covariantly constant spinors as in \cite{Denef:2000nb}.
The action of this generator follows from \eqref{sl2-alg} as
\begin{equation} \label{h-star-gen}
{\bf h}_* {\rm w} = e^{-i \alpha} Z\,, \quad
{\bf h}_* Z =  e^{i \alpha}  {\rm w}\,, \qquad
{\bf h}_* \bar Z^a = e^{-i \alpha}  \Sigma^a\,, \quad
{\bf h}_* \Sigma^a = e^{i \alpha} \bar Z^a\,,
\end{equation}
while the relevant eigenvalue equation \eqref{FirstOrderP} is simply
\begin{equation}\label{bps-eigen}
 {\bf h}_* P = P\,.
\end{equation}
The last relation clearly imposes a linear relation between the derivatives of the
four-dimensional scalars and the gauge fields upon using \eqref{h-star-gen}. The phase
$ e^{i \alpha}$ is determined by \eqref{AuxilH}, which can be shown to reduce to
\begin{equation} \label{bps-phase}
d \alpha  + Q + \frac{1}{2} e^{2U} \star d \omega + 2\,\mbox{Im}(e^{-i \alpha} Z) = 
d \alpha  + Q - \frac{1}{2} e^{2U} \star d \omega = 0 \, ,
\end{equation}
where in the second equality we used the first of \eqref{h-star-gen} and \eqref{bps-eigen}.
These equations are easily seen to be equivalent to the BPS system of
\cite{Behrndt:1997ny, Denef:2000nb} and we refer to \cite{Bossard:2010mv} for a more
detailed analysis of the eigenvalue equation \eqref{nil-mom} for this generator, leading
to the system of equations describing multi-centre BPS black holes. Similarly, non-BPS
solutions with vanishing central charge at the horizons are described in a similar fashion,
with $\lambda = 0 $ and a normalised rank one
$\Lambda_a$ (\ie $c^{abc} \Lambda_b \Lambda_c = 0 $ and $\Lambda_a \bar \Lambda^a = 1$)
\cite{Bossard:2010mv}.

Other examples of such constructions using nilpotent orbits have been used to obtain the
systems describing respectively single centre non-BPS black holes with a non-vanishing central
charge at the horizon and the composite non-BPS multi-centre system \cite{Bossard:2012xsa}.
In the next two sections we discuss in some detail the construction of the two multi-centre
non-BPS systems, the composite non-BPS and the almost-BPS, which form the basis of this paper.

\subsection{Composite non-BPS flows}
\label{sec:comp-flow}
The composite non-BPS system \cite{Bossard:2011kz, Bossard:2012ge, Bossard:2012xsa},
describes configurations of interacting non-BPS centres and corresponds to an
eigenvalue equation as in \eqref{nil-mom}, where the relevant element ${\bf h}_{\text{\tiny C}}$
belongs to $\g_4$. As the scalar momentum $P$ must be of positive grade with
respect to ${\bf h}_{\text{\tiny C}}$ (given that $p_\alpha>0$), we consider the relevant
decomposition of the coset component, parametrising
$ {\bf 2} \otimes \mathds{R}^{2n_v+2}$ through \eqref{P-component},
in terms of grades with respect to an element ${\bf h}_{\text{\tiny C}} \in \g_4$.
The relevant graded decomposition is in this case 
\begin{equation} \label{decomp-comp-g4}
\g_4 \cong (\mathds{R}^{n_v} )^\ord{-2} \oplus \scal{ \gl_1 \oplus \g_5 }^\ord{0} \oplus (\mathds{R}^{n_v} )^\ord{2}\,, \end{equation}
for $\g_4$ itself and
\begin{equation}  \label{decomp-comp}
{\bf 2} \otimes \mathds{R}^{2n_v+2} \cong {\bf 2}^\ord{-3} \oplus ({\bf 2}\otimes\mathds{R}^{n_v} )^\ord{-1} \oplus ({\bf 2}\otimes\mathds{R}^{n_v} )^\ord{1}\oplus  {\bf
2}^\ord{3}\,, \end{equation}
for the coset component, {\it \ie} $P \in({\bf 2}\otimes\mathds{R}^{n_v} )^\ord{1}\oplus  {\bf
2}^\ord{3}$. We can choose ${\bf h}_{\text{\tiny C}}$ to be Hermitian (\ie to lie in $\g_4
\ominus ( \mathfrak{u}(1) \oplus \k_4 )$), so that it is realised for
\begin{equation}
\gamma({\bf h}_{\text{\tiny C}}) =0\,, \quad \Lambda_a({\bf h}_{\text{\tiny C}})=\Omega_a\,, 
\quad G^a{}_b({\bf h}_{\text{\tiny C}})=0\,,
\end{equation}
where $\Omega_a$ describes a very small vector, as in \eqref{small-def}.
Equivalently, $\Omega_a$ is in the $U(1) \times K_4 $ orbit of the Jordan algebra identity.

More explicitly, one finds the following action on the coset component
\begin{equation} \label{h-com-gen}
{\bf h}_{\text{\tiny C}} {\rm w} = \Omega_a \bar Z^a \quad {\bf h}_{\text{\tiny C}} Z = \Omega_a \Sigma^a
\quad {\bf h}_{\text{\tiny C}} \bar Z^a = \bar \Omega^a  {\rm w} + c^{abc} \Omega_b \bar \Sigma_c
\quad {\bf h}_{\text{\tiny C}} \Sigma^a = \bar \Omega^a  Z  + c^{abc} \Omega_b Z_c \,,\end{equation}
which is to be identified with the elements of the coset component, through
\eqref{nil-mom} for $p_\alpha=1$ and $p_\alpha=3$, according to \eqref{decomp-comp}. 
Considering a general linear combination of the grade one and three solutions as in
\cite{Bossard:2012xsa}, one can express ${\rm w}$ and $\Sigma^a$ in terms of $Z$ and $Z_a$ as
\begin{equation} {\rm w} = \frac{1}{2} \scal{ \Omega_a \bar Z^a - \N[\Omega] \bar Z }\,, \qquad
\Sigma^a = c^{abc} \Omega_b Z_c + \frac{1}{2} \bar \Omega^a \scal{ Z -
\N[\Omega] \bar \Omega^b Z_b }\,, \label{NonBPSfirstOrder} \end{equation}
which are the explicit first order relations for the scalar momenta, upon using
the definitions in \eqref{P-component}.

These flow equations contain the auxiliary components $\N[\Omega]$ and $\Omega_a$ of ${\bf h}_{\text{\tiny C}}$,
which satisfy \eqref{small-def} and define a very small vector $R$ of unit mass through
\begin{equation} Z(R) = \N[\Omega] \qquad Z_a(R) = \Omega_a\,. \label{R-def} \end{equation}
The flow equations for these auxiliary fields are given by \eqref{AuxilH}, which in this
system reduces to
\begin{equation} 
[ {\bf h}_{\text{\tiny C}} , d_B {\bf h}_{\text{\tiny C}} ] = 2\, d_B {\bf h}_{\text{\tiny C}} \ , \label{GradeDeux}  
\end{equation}
in view the fact that $d_B {\bf h}_{\text{\tiny C}}$ lies in $\g_4$ and is inert under $\g_5$ by definition,
and the decomposition in \eqref{decomp-comp-g4}.
Using the explicit form of $B$ in \eqref{Bsl2}-\eqref{Bk4} and the first order flow \eqref{NonBPSfirstOrder}, one computes the components of $d_B {\bf h}$ as
\begin{eqnarray}
\gamma(d_B {\bf h}) &=&  \frac{2}{3} \, \Im[\bar\Omega^a Z_a]\,,
\CR
\Lambda_a(d_B {\bf h})
&=&  Z_a( d R) + \mbox{Re}[ \N[\bar \Omega] \Omega_b e^b_i dt^i ] Z_a(R) - \Scal{ Z_a + \N[\Omega]
 c_{abc}\bar \Omega^b \bar Z^c
- \Omega_a\,\Omega_b \bar Z^b}
\CR
G^a{}_b(d_B{\bf h})  &=& c^{ace} c_{bde} \scal{ \bar \Omega^d Z_c - \Omega_c \bar Z^d } + \Omega_b \bar Z^a - \bar \Omega^a Z_b - \frac{2i}{3} \delta^a_b \mbox{Im}[ \bar \Omega^c Z_c ] \label{dh-compon}
\, ,
\end{eqnarray}
where we explicitly separated the terms depending on the derivative of the vector $R$. It is now straightforward (though cumbersome) to evaluate \eqref{GradeDeux}, imposing the above relations and
\begin{eqnarray} 
 c^{ace} c_{bde} \bar \Omega^d \scal{ \N[\Omega] c_{cfg} \bar \Omega^f \bar Z^g - \Omega_c \Omega_f \bar Z^f } -  \bar \Omega^a \scal{ \N[\Omega] c_{bcd} \bar \Omega^c \bar Z^d - \Omega_b \Omega_c \bar Z^c } 
\qquad\qquad
\CR
=- c^{ace} c_{bde} \Omega_c \bar Z^d + \Omega_b \bar Z^a \,, \end{eqnarray}
which can be shown using \eqref{symmetric}. The result one finds is that \eqref{GradeDeux} is
identically satisfied for \eqref{dh-compon}, if and only if the combination of the
first two terms in $\Lambda_a(d_B {\bf h})$ vanishes. This condition implies
\cite{Bossard:2012xsa} that there exists a constant symplectic vector $\hat{R}$, since
\be
d R =-  \mbox{Re}[ \N[\bar \Omega] \Omega_b e^b_i dt^i ] R \ ,
\quad\Rightarrow\quad
R = \frac{\hat{R}}{|Z(\hat{R})|} \label{eq:R-rescale} \ ,
\ee
where the second equation follows from the first by use of standard special
geometry identities. It follows that the generator ${\bf h}_{\text{\tiny C}}$ is in this case
determined by a constant very small projective vector $\hat{R}$ and the scalar fields
are such that
\be \Lambda_a({\bf h_c}) = \frac{ Z_a(\hat{R})}{|Z(\hat{R})|} \, . \label{OmegaRhat}  \ee
One can now return to \eqref{NonBPSfirstOrder}, which becomes a first order flow equation
for the scalars $dU$, $\star d\omega$ and $dt^i$ in terms of the gauge fields and the
constant vector $\hat{R}$. Further details and the characterisation of solutions
to these equations are given in section \ref{sec:composite}.

A special case of this system, namely the single centre subclass, was discussed
in detail in \cite{Bossard:2012xsa}. Indeed, one expects a system describing
multi-centre black hole solutions to contain a consistent subsystem describing single
centre black holes. In the case at hand, one can restrict the above equations
\eqref{NonBPSfirstOrder} to this subsystem by imposing that the scalar momentum
is also of positive grade with respect to the generator
$\frac{1}{2} ( {\bf h}_{\text{\tiny C}} + {\bf h}_*)$ where ${\bf h}_{\text{\tiny C}}$ and ${\bf h}_*$ are
the generators that define the composite system through \eqref{h-com-gen} and the BPS
system through \eqref{h-star-gen} respectively. The result one obtains is that the single
centre non-BPS momenta satisfy \eqref{NonBPSfirstOrder}, for $Z$ and $Z_a$
constrained to satisfy exactly the same phase dependent projection
\eqref{ZAconstraint} above,
\begin{equation} \label{constr-dum}
\bar Z^a - \N[\Omega] \bar \Omega^a \bar Z = e^{-i \alpha} \scal{ c^{abc}
\Omega_b Z_c + \bar \Omega^a ( Z - \N[\Omega]  \bar \Omega^b Z_b )} \,,
\end{equation}
which represents a constraint on the physical degrees of freedom that is necessary to reduce
to single centre solutions. One can interpret this constraint in terms of the decomposition
in \eqref{eq:vec-decomp}, upon identifying the very small vector $\hat R$ in \eqref{OmegaRhat}
with the vector $R$ in section \ref{sec:t-dual-real}. It follows that \eqref{constr-dum}
implies that only the grade $(-1)$ and $(+3)$ components of the gauge fields
(and thus charges) are allowed. We refrain from discussing the single centre system, as we will deal
with the multi-centre case in what follows. However, we will sometimes make use of structures
present already in the single centre case, referring to \cite{Bossard:2012xsa} for an in
depth discussion.

\subsection{Almost-BPS flows}
\label{sec:alm-flow}
We now turn to the derivation of the first order flow equations for the almost-BPS
system \cite{Goldstein:2008fq} of multi-centre black hole solutions. Our treatment is
purely in terms of four/three dimensional quantities, but we still refer to this
system as almost-BPS, as we will find that it contains the original almost-BPS
equations, derived from the BPS conditions of five dimensional supergravity. However, the five dimensional
system is only defined in a fixed frame and does not allow for generic charges at the
centres, while we will present a manifestly duality covariant form of these equations,
as we did for the composite non-BPS system in the previous section. In this sense,
the almost-BPS system below is larger, as it contains the one in five dimensions,
as well as all the charge configurations that can be obtained from it by dualities.
Our covariant system also represents the general form of the almost-BPS system
of \cite{Bossard:2011kz}, where a frame was chosen to match with the five dimensional
results. Closely related first order flow equations have recently been derived using
similar methods in \cite{Ferrara:2012qm} for the $STU$ model.\footnote{\ie for $\mathcal{M}_4\cong SU(1,1) / U(1) \times SU(1,1) / U(1)\times SU(1,1) / U(1)$.}  In this case $\Omega_a$ reduces to three phases, which altogether with $\alpha$ define equivalently the first order equations. Nevertheless, the authors in \cite{Ferrara:2012qm} use the second order equations of motion to determine the evolution of these phases, whereas we consider the first order equations \eqref{AuxilH}, which imply the second order ones by construction.

For the almost-BPS system, the eigenvalue equation \eqref{nil-mom} is defined
by the generator ${\bf h}_{\text{\tiny A}} = {\bf h}_{\text{\tiny C}}  + 2 {\bf h}_*$,
where we use the same generators as defined in \eqref{h-com-gen} and \eqref{h-star-gen}
above. The corresponding graded decomposition of $\mathfrak{sl}_2\oplus \mathfrak{g}_4$
is then given by 
\begin{equation}  \label{decomp-alm-g4}
\mathfrak{sl}_2\oplus \mathfrak{g}_4 \cong  \mathbf{1}^\ord{-4}\oplus
 (\mathds{R}^{n_v} )^\ord{-2} \oplus \scal{ \mathfrak{gl}_1 \oplus \mathfrak{gl}_1 \oplus \mathfrak{g}_5}^\ord{0}
\oplus \overline{(\mathds{R}^{n_v} )}^\ord{2} \oplus \mathbf{1}^\ord{4}\,, 
\end{equation}
which is exactly the same as the one in \eqref{decomp-comp-g4} with respect to the $\mathfrak{g}_4$ components.
The additional notation $\mathds{R}^{n_v}$ and $\overline{\mathds{R}^{n_v} }$ refers to two conjugate
representations of the real group $\g_5$. Similarly, the coset component is decomposed as
\begin{equation}  \label{decomp-alm-cos}
{\bf 2} \otimes \mathds{R}^{2n_v+2} \cong \mathbf{1}^\ord{-5}\oplus 
(\mathds{R}^{\bar n_v} )^\ord{-3} \oplus ({\bf 1}\oplus\mathds{R}^{n_v} )^\ord{-1}
\oplus ({\bf 1}\oplus\mathds{R}^{\bar n_v} )^\ord{1}\oplus 
(\mathds{R}^{n_v} )^\ord{3}\oplus \mathbf{1}^\ord{5}\,, 
\end{equation}
so that $P\in ({\bf 1}\oplus\mathds{R}^{\bar n_v} )^\ord{1}\oplus 
(\mathds{R}^{n_v} )^\ord{3}\oplus \mathbf{1}^\ord{5}$, \ie it lies in the positive grade component.

Due to the presence of the generator ${\bf h}_*$, which by itself defines the BPS
system,\footnote{Note that while the same generator appears in the reduction of the
composite non-BPS system to the single centre class, we treat ${\bf h}_{\text{\tiny A}}$
as the defining object here, instead of imposing a constraint similar to
\eqref{ZAconstraint} on an existing system.} the first order system reduces to the
BPS system \eqref{bps-eigen} for most of the components of the field strength $F$,
except for the ones that define the solution
\begin{equation}
 Z = i\, e^{\frac{i}{2}\alpha} \N[\Omega]^\frac{1}{2}\, {\rm u} \ , \qquad
 Z_a =  i\, e^{\frac{i}{2}\alpha} \N[\overline{\Omega}]^\frac{1}{2}\, \Omega_a\, {\rm u}\,,
\end{equation}
for which the expressions of ${\rm w}$ and $\Sigma^a$ take the opposite sign, where
${\rm u}$ is a real function. The projection to this component is obtained as
\begin{eqnarray}
\mathcal{P} Z &=&  \frac{i}{4} \, e^{\frac{i\alpha}{2}} \N[\Omega]^\frac{1}{2}\,
\mathrm{Im}[ e^{-\frac{i\alpha}{2}} \N[\overline{\Omega}]^\frac{1}{2} Z   + e^{-\frac{i\alpha}{2}} \N[{\Omega}]^\frac{1}{2} \overline{\Omega}^a Z_a ]\,,  \CR
\mathcal{P} Z_a  &=&  \frac{i}{4} \,
e^{\frac{i\alpha}{2}} \N[\overline{\Omega}]^\frac{1}{2}\, \Omega_a \,
\mathrm{Im}[ e^{-\frac{i\alpha}{2}} \N[\overline{\Omega}]^\frac{1}{2} Z
 + e^{-\frac{i\alpha}{2}} \N[{\Omega}]^\frac{1}{2} \overline{\Omega}^b Z_b ]\,,
\end{eqnarray}
for which one easily computes that $\mathcal{P}^2 = \mathcal{P}$.
One can then verify that the first order system obtained by changing the sign of
the components above, given by
\begin{eqnarray}
{\rm w}  & =& e^{-i \alpha}  ( Z - 2   \mathcal{P} Z ) \CR
  & =& \frac{3}{4} e^{- i \alpha} Z + \frac{1}{4} \N[\Omega] \bar Z + \frac{1}{4} \Omega_a \bar Z^a - \frac{1}{4} e^{-i\alpha } \N[\Omega] \overline{\Omega}^a Z_a\,, \CR
 \Sigma^a &=& e^{i\alpha} ( \bar Z^a  - 2  \mathcal{P} \bar Z^a ) \CR
 &=& e^{i\alpha}  \bar Z^a  + \frac{1}{4} \overline{\Omega}^a \scal{ Z + \N[\Omega] \overline{\Omega}^b Z_b - e^{i\alpha} \N[\Omega] \bar Z - e^{i\alpha} \Omega_b \bar Z^b } \,,
\label{flow-almost}
\end{eqnarray}
solves \eqref{nil-mom}, as expected. Using the definitions in \eqref{P-component},
these equations become a first order flow for the four dimensional scalars in terms
of the gauge field strengths.

However, the auxiliary fields $e^{-i \alpha}$ and $\Omega_a$ parametrising
${\bf h}_{\text{\tiny A}}$ are still arbitrary, we therefore need to impose
\eqref{AuxilH} to obtain their dependence on the physical fields. First, the
condition on $d_B {\bf h}_*$ to be of positive grade is the same as the one for
the BPS system in \eqref{bps-phase} and leads to the condition,
\cite{Bossard:2010mv} 
\begin{eqnarray}
d \alpha + Q + \frac{1}{2} e^{2U} \star d \omega + 2\, \Im[ e^{-i\alpha} Z] &=& 0\,,
\label{eq:d-alpha}
\end{eqnarray}
where the last term is now given in terms of scalars by \eqref{flow-almost} and
is therefore more complicated than in \eqref{bps-phase}. Analysing the components
of \eqref{AuxilH} along $\g_4$ to obtain the remaining auxiliary components,
$\Omega_a$, one finds that it reduces again to \eqref{GradeDeux} in
exactly the same way as for the corresponding quantities in the composite non-BPS
system. In fact, one can easily check that $\gamma(d_B {\bf h}_{\text{\tiny A}})$ and
$G^a{}_b(d_B{\bf h}_{\text{\tiny A}})$ are exactly the same as in \eqref{dh-compon}
since they only depend on the algebra \eqref{g4-algebra}. In contrast, manipulating
the expression for $\Lambda_a(d_B {\bf h}_{\text{\tiny A}})$ is sensitive to the
particular form for the scalar momenta in terms of the gauge fields. In order to
write the resulting equation in a suggestive way, it is useful to draw intuition
from the analogous vector in the single centre class\footnote{Note that the non-BPS
single centre system is also contained in the almost-BPS system, but we refrain
from discussing the details of this reduction.}, as well as from known multi-centre
solutions, which indicate that the magnetic dual of the vector described by
$\Omega_a$ is a more convenient variable. We therefore consider the dual very small
vector of mass one, $R^*$, defined as in \eqref{R-star-bas}, \ie we impose
 \be \label{Rst-def-7}
\Omega_a = e^{i \alpha }  \bar Z(R^*) Z_a(R^*)\,.
\ee
With this definition, one can use \eqref{flow-almost} to show that
 \bea
\Lambda_a(d_B {\bf h}_{\text{\tiny A}}) 
&=&  e^{i \alpha } \left( \bar Z(dR^*) Z_a(R^*) +  \bar Z(R^*) Z_a(dR^*)
 +2\, \mbox{Re}[ e^{i \alpha} \Omega_b e^b_i dt^i ]\, \bar Z(R^*) Z_a(R^*) \right)
\CR
&&- \Scal{ Z_a + \N[\Omega] c_{abc}\bar \Omega^b \bar Z^c
  - \Omega_a\,\Omega_b \bar Z^b}
\label{Lambda-almost}
\, .
\eea
As in \eqref{dh-compon}, the second bracket in the last expression satisfies
\eqref{GradeDeux} identically, whereas the first term vanishes upon imposing
that $R^*$ is related to a constant vector $\hat{R}^*$ by
\begin{equation}\label{Rstar-almost}
d R^* =-  \mbox{Re}[ \bar Z(R^*) Z_b(R^*) e^b_i dt^i ] R^* \ ,
\quad\Rightarrow\quad
 R^*=\frac{\hat{R}^*}{|Z(\hat{R}^*)|}\,,
\end{equation}
similar to \eqref{eq:R-rescale} for the composite system. In contrast to
\eqref{dh-compon}, there are two terms containing the derivative of $R^*$ in
\eqref{Lambda-almost}, so that one could expect more general solutions than
the one shown above. However, one can verify that any solution for $R^*$ other
than \eqref{Rstar-almost} is such that it is mutually nonlocal with its
derivative $dR^*$, which is not allowed for an everywhere very small vector.

This concludes our presentation of the derivation of the first order flow
equations for the almost-BPS system. One may now consider solutions to
\eqref{flow-almost}, which is a first order flow equation for the scalars
$dU$, $\star d\omega$ and $dt^i$ in terms of the gauge fields and the
constant vector $\hat{R}^*$, upon using \eqref{Rst-def-7} and \eqref{Rstar-almost}.
In section \ref{sec:almost} we discuss the real form of these equations, show that
they correspond to a linear system and give the characterisation of their solutions
in terms of local functions.

\section{Composite non-BPS system}
\label{sec:composite}

In this section, we present in detail the steps required to characterise
solutions of the flow equations for the composite non-BPS system in terms
of local functions. The starting point is the solution of the nilpotency
condition \eqref{NonBPSfirstOrder}, written explicitly as a first order
flow system for the four dimensional scalars and the metric degrees of
freedom
\bea dU + \frac{i}{2} e^{2U} \star d \omega &=& - \frac{1}{2} e^U \scal{
\Omega_a
\bar Z^a(\star F) - \N[\Omega] \bar Z(\star F)} \label{Grav}
\CR
- e^a_i d t^i &=& c^{abc} \Omega_b  e^U Z_c(\star F) + \tfrac12 \bar \Omega^a
e^U
\scal{ Z(\star F) - \N[\Omega] \bar \Omega^b  Z_b(\star F) }\,, \label{ScalLin}
\eea
where $F$ is the spatial component of the field strengths defined in
\eqref{gauge-decop}. The vector $\Omega_a$ is related to the constant very
small vector, $\hat{R}$, through \eqref{R-def} and \eqref{eq:R-rescale} above.
For later reference, we give the inverse relations for the field strengths
\bea e^U Z(\star F) &=& \frac{1}{2} \N[\Omega] \Scal{ dU - \frac{i}{2} e^{2U}
\star d \omega } - \frac{1}{2} \Omega_b e^b_i d t^i \label{ZstarF}\CR
e^U Z_a(\star F) &=& -c_{abc} \bar \Omega^b e^c_i d t^i
+\frac{1}{2} \N[\bar \Omega] \Omega_a \Omega_b e^b_i dt^i  - \frac{1}{2}
\Omega_a \Scal{  dU - \frac{i}{2} e^{2U} \star d \omega }\,. \eea

In order to solve this system, we first construct the electromagnetic potentials
and use the resulting structure to simplify the equations in section \ref{sec:zeta-5}.
In section \ref{sec:Tdual-5} we exhibit the relevance of the T-dualities
introduced in section \ref{sec:T-duality}. We then proceed to rewrite these
equations as a linear system of differential equations and discuss its integration
in terms of local functions in sections \ref{sec:linsys-5} and \ref{sec:integr-5}
respectively. The reader interested in applications can find a summary of the
final form of the system in section \ref{sec:summ-5}.

\subsection{The electromagnetic potentials}
\label{sec:zeta-5}

In order to solve the system \eqref{ScalLin}, we construct the gauge field
momenta \eqref{cmplx-self} using the derivative of $R$ given by
\eqref{eq:R-rescale},
to obtain
\bea
d\zeta&\equiv& 2\,e^{2U}\Re[\bar Z(\star F) \cV + \bar Z^i(\star F) D_{i}
\cV] \CR
&=& d\,\scal{ e^{U} \Re [ \N[\bar \Omega]  \cV
 -\bar \Omega^i  D_i \cV]}
+\tfrac{1}{2}e^{U}\,\Im[ \N[\bar \Omega] \Omega_i d t^i] \, R
+\tfrac{1}{4} e^{3U} \star d \omega\,R
\label{eq:dzeta}\,,
\eea
where we made extensive use of the special geometry identities in section
\ref{sec:sugra}. The first term is manifestly a total derivative, whereas
the others are along the very small vector $R$ and must therefore combine
into the derivative of a single function. This requirement, along with
\eqref{eq:R-rescale}, leads to the condition
\be
e^{U}\,\Im[\N[\bar\Omega]\, \Omega_i d t^i]
+\tfrac{1}{2} e^{3U} \star d \omega
= M\,e^{3U}\,\Re[ \N[\bar\Omega] \Omega_i d t^i]
- d (M\,e^{3U}) \,,
\label{ImOdt}
\ee
where $M$ is an arbitrary function, so that the gauge field momenta take the form
\be \zeta =  e^U \Re [ \N[\bar \Omega]  \cV
 -\bar \Omega^i  D_i \cV]
- \frac{1}{2} e^{3U} M R\,, \label{Zetas} \ee
with the corresponding central charges given by
\be Z( \zeta) =
\frac{i}{2}\,e^U\, ( 1 + i\,e^{2U} M) \N[\Omega]\,,
 \qquad Z_a( \zeta) = \frac{i}{2}\,e^U\, ( 1 + i\,e^{2U} M) \Omega_a\,,
\label{Zzeta}\ee
for later reference.

The structure of \eqref{Zetas} can be used to show that one vector is always
trivial, simplifying the system. To see that, we compute
\bea \Iprod{\hat{R}}{\zeta}
&=& -2 e^U |Z(\hat{R})|\,,\eea
whereas taking the imaginary part of (\ref{Grav}) one gets
\bea e^{2U} \star d\omega
&=& -\frac{e^U}{2 |Z(\hat{R})|}\, \Iprod{\hat{R}}{\star F}\,. \eea
Finally, using \eqref{gauge-decop}, one finds that
\be \Iprod{\hat{R}}{dw}= 0\,,
\label{Rdw}
\ee
which implies that one vector field is always absent. In terms of the graded
decomposition in section \ref{sec:T-duality}, the vanishing component
is along the very small vector dual to $R$, given in \eqref{R-star-bas},
as will be shown shortly. 

One can now combine \eqref{Zzeta} and \eqref{Rdw} to disentangle the term
proportional to $\star d\omega$ in the definition of the scalar flow equation,
so that \eqref{ScalLin} becomes
\bea\label{eq:nbps-scals}
-e^a_i d t^i &=& c^{abc} \Omega_b  e^U Z_c(\star dw) + \tfrac12 \bar \Omega^a e^U
\scal{ Z(\star dw) - \N[\Omega] \bar \Omega^b  Z_b(\star dw) }  \CR
&&  + \tfrac12  e^{4U} ( -M + i e^{-2U} ) \N[\Omega] \bar \Omega^a
 \star d \omega\,.
\eea
Applying the same procedure on \eqref{ZstarF}, one obtains the inverse relations
of \eqref{eq:nbps-scals} for the central charges $Z(dw)$ and $Z_a(dw)$. The
charge vectors $dw$ can then be straightforwardly constructed with the result
\bea
\star dw&\equiv& 2\,\Im[-\bar Z(dw)\cV + \bar Z^a(dw)D_a\cV] \CR
&=&e^{-U} \Im[- (dU +\tfrac{i}2e^{2U}\star d\omega)\cR_+
 +( i\,\bar Y \,e^{2U}\star d\omega -\N[\Omega] \bar\Omega^i d\bar t_i)\cR_-
\CR
&&\qquad\qquad
-2\, c^{abc} \Omega_b e_c^i d \bar t^i D_a\cV
+ 2\,\N[\Omega] \bar\Omega^i d\bar t_i\bar\Omega^jD_j\cV ]
\,,
\label{dw-mid-5}
\eea
where we used the shorthands
\begin{align}
 \cR_\pm =& \pm\N[\bar\Omega] \cV +\bar\Omega^{i}D_i\cV\,,\\
Y=&\, ( 1 + i\,e^{2U} M) \label{Y-def}\,.
\end{align}
Note that the above results are in direct correspondence with the ones in \cite{Bossard:2012xsa},
where the single centre system was treated.

\subsection{Connection to T-dualities}
\label{sec:Tdual-5}

Given the structure above, it is useful to define a second distinguished very small vector
which is mutually nonlocal with $\hat{R}$, as in \eqref{R-star-bas}-\eqref{Y-def-gen} through
\begin{align}
 \hat{R}^* = |Z(\hat{R})|^{-1}&\,
\mbox{Re}\Bigl[
\bar Y^3\,\N[\bar\Omega]\,{\mathcal V}
+|Y|^2\bar Y\bar \Omega^{i} D_i \cV \Bigr]\,.
\label{Rstar}
\end{align}
Here, we identified the function $Y$ in \eqref{Y-def} with the one in \eqref{Y-def-gen}
and we included an overall rescaling. One can check that $\Iprod{\hat{R}}{\hat{R}^*}=4$ and
that the central charges of $\hat{R}^*$ above satisfy \eqref{small-def}. This new vector is
defined in exactly the same way as the second constant vector used in \cite{Bossard:2012xsa}
in the single centre case, but is not constant in the full composite non-BPS system. 

As explained in \eqref{gen-small}=\eqref{der-small}, a non-constant very small vector with
a constant non-zero symplectic product with $\hat{R}$  must be of the form
\begin{equation}\label{Rstar0-def}
 \hat{R}^*= \exp[\Tgp]R_0^*\,.
\end{equation}
where $R_0^*$ is a constant very small vector satisfying $\Iprod{\hat{R}}{{R}^*_0}=4$ and $\exp[\Tgp]$ is a
T-duality matrix leaving $\hat{R}$ invariant. It then follows that
the derivative of $\hat{R}^*$ takes the form
\begin{equation}\label{dR-generator}
 d\hat{R}^*-d\Tgp \hat{R}^*=0\,,
\end{equation}
which is identically closed by the fact that T-dualities are abelian.
The above are consistent with known solutions \cite{Bossard:2011kz, Bossard:2012ge},
in which $\hat{R}^*$ takes the form of a T-duality whose parameters are
harmonic functions, acting on a constant vector along $p^0$.

Within the composite non-BPS system, one can explicitly compute the
components of the derivative of $\hat{R}^*$ using \eqref{ImOdt} and
\eqref{eq:nbps-scals}, as
\begin{align}
Z(d \hat{R}^*) 
=&\, \Tgp_{\gamma, Q}Z(\hat{R}^*) = Z(\dTgp \hat{R}^*)\,,
\CR
Z_a(d \hat{R}^*) 
=&\,\Tgp_{\gamma, Q}Z_a(\hat{R}^*) =Z_a(\dTgp \hat{R}^*)
\,, \label{DRstar}
\end{align}
where we indicated that the components in this basis are given by the variation of $\hat{R}^*$ itself
under the T-duality transformations \eqref{T-duality}, as shown in \eqref{rstar-flow}. In order to
obtain this result, one has to identify $\hat{R}$ as the grade $(+3)$ very small vector, that is
invariant under all T-dualities $\Tgp$, which we assume henceforth. The explicit values for the
parameters are given by
\begin{align}\label{Tdualstar}
 \gamma_\scp =&\, \tfrac1{6}\,e^{2U} (d M - \star d \omega)\,,
\CR
 \bar Q_{\scp\,a} =&\, -\tfrac{1}{2} \bigl[  Y \N[\Omega] N_a
  + \bar Y c_{abc} \bar \Omega^b \bar N^{c}
  - \tfrac13\,\Omega_a \left( 2\,\bar Y\,\N[\bar\Omega]\Omega_b \bar N^{b}
                       +Y\,\N[\Omega]\bar \Omega^b N_{b}  \right)
\bigr]
\,,
\end{align}
where we used the combination
\begin{equation}
\bar N^a = -e^a_i d t^i +\N[\Omega]\,\bar\Omega^a \left(dU - \frac{i}{2} e^{2U} \star d \omega\right)
\,,\label{scal-comb}
\end{equation}
for brevity. Note that in the single center class of \cite{Bossard:2012xsa} these expressions
were shown to vanish, consistent with the fact that $\hat R^*$ is constant if the T-dualities
are rigid.

Given the above, we can directly apply all considerations of section \ref{sec:T-duality}, since the
presence of the two constant very small vectors $\hat{R}$ and $R_0^*$ implies that the grading shown
in \eqref{eq:vec-decomp} is relevant for the integration of the system. It follows that $R^*_0$ is
identified as the grade $(-3)$ very small vector. As discussed in
\eqref{rstar-flow}-\eqref{ZAconstraint}, a generic T-duality is parametrised
by a rank three grade $(+1)$ vector, which we denote by $\cK$, so that
\eqref{Rstar0-def}-\eqref{dR-generator} become
\begin{equation}\label{Rstar0-real}
 \hat{R}^* = \exp[\Tgp_\csK] R^*_0\,, \qquad \dTgp_\csK R_0^*=d\cK \,.
\end{equation}
The rank of $\cK$ can be verified by checking that the quartic invariant of the vector
$d\hat{R}^*$ in \eqref{DRstar} is vanishing, \ie
\begin{gather}
 I_4(d\hat{R}^*)
= I_4(\dTgp_\csK R_0^*)= I_4(d \cK)=0\,,
\\
\partial_\mu \Tgp_\csK  \partial_\nu \Tgp_\csK{} R_0^*=
 -\tfrac{1}{16}\,I^\prime_4(\partial_\mu \Tgp_\csK R_0^*, \partial_\nu \Tgp_\csK R_0^*, \hat{R})=
 -\tfrac{1}{16}\,I^\prime_4(\partial_{\mu}\cK,\partial_{\nu}\cK, \hat{R})\,.
\label{2TonRs}
\end{gather}
In what follows, we will generally drop the subscript $\cK$ on $\Tgp_\csK$ for simplicity,
since this is the only T-duality appearing throughout.

Finally, it is worth commenting on the difference between the dual very small vector in
\eqref{R-star-bas}, which was used to define $\hat{R}^*$ in \eqref{Rstar}, and the constant
vector $R^*_0$ of grade $(-3)$, that might seem confusing. It is important to realise that
\eqref{R-star-bas} simply defines a possible dual vector, which is not unique. Since two very
small vectors not commuting with $\hat{R}$ are related by exactly a finite T-duality leaving
$\hat{R}$ invariant, one may choose any other vector in that orbit. We will fix this ambiguity by defining the function $\mathcal{K}$ to vanish in the asymptotic region, or equivalently $R^*_0 = \hat{R}^*|_{r\rightarrow \infty}$.
\subsection{The linear system}
\label{sec:linsys-5}

One can now use the vector $\hat{R}^*$ of \eqref{Rstar} in a way similar to the vector
$R$ was used in section \ref{sec:zeta-5}, to project the flow equations and simplify the system.
Indeed, taking the inner product with $dw$ one finds
\bea
\Iprod{\hat{R}^*}{dw}
&=& -\star d(e^{-U}|Z(\hat{R})|^{-1}\,|Y|^2) \equiv -\star dV\,,
\label{V-def}
\eea
where we defined the function $V$. Note that $V$ is not a harmonic function, since $\hat{R}^*$ is not a constant vector.
Combining this with \eqref{eq:R-rescale}, \eqref{ImOdt} and \eqref{eq:nbps-scals}, we
can determine the combination $\Omega_i d t^i$ in terms of $V$, $M$ and the metric components as
\bea
\Re[ \N[\bar\Omega]\, \Omega_i \,d t^i]
&=&|Z(\hat{R})|^{-1} d |Z(\hat{R})|
=-V^{-1} dV -\, dU +|Y|^{-2}\,d |Y|^{2}
 \,,
\CR
\N[\bar\Omega] \Omega_i \,d t^i
&=& (1-2Y)\,dU -\tfrac12 i\, e^{2U}\star d\omega
  +\frac{Y^2}{|Y|^{2}}\,d \bar Y -V^{-1} Y\,dV
 \,.\label{eq:Omega-dt}
\eea

We now use all the above information to write the expression \eqref{dw-mid-5} for $dw$ in a suggestive form.
To this end, we use the expression for the derivative of $\hat{R}^*$ in \eqref{DRstar}, combined with inspiration
drawn from the analogous computation performed in \cite{Bossard:2012xsa} for the single centre class. After
a long but straightforward computation we obtain
\bea
\star dw &-& d\left(\tfrac{M}{V}\, \hat{R}^*\right)
 - \tfrac12\, d V\,\hat{R}
-2\,e^{2U}V\, d(\tfrac{M}{V})\, \Re\!\left[e^{-U}e^{-i\alpha}\cV \right]
\CR
&=& 4\,e^{-U} \mbox{Im}
\bigl[3i\bar Y \N[\bar\Omega] \cV
          -i(2\,\bar Y - Y ) \bar \Omega^i D_i \cV \bigr]
 \,\gamma_\scp
\CR
&&+e^{-U} \Im[ (2\,dU - i\,e^{2U}\star d\omega ) e^{-i\alpha}\cV
-2\,e^{-i\alpha}d t^iD_i\cV
-4\,Q_\scp^a\, D_a\cV
 ]\,.
\label{dw-mid}
\eea
Here, we used \eqref{Y-def-gen} to define a new phase as
\begin{equation}
 e^{-i\alpha}\equiv -\frac{\bar Y^2}{|Y|^2} \N[\bar\Omega]\,,
\end{equation}
which will be useful in what follows. One can compute the derivative of $e^{-i\alpha}$
using \eqref{OmegaRhat} to first obtain
\begin{eqnarray}
\N[\Omega]\, (d - i\,Q) \,\N[\bar \Omega]
&=&  \tfrac{i}2 e^{2U}\star d\omega
  +\frac{2}{|Y|^{2}}\,dY -i\,e^{2U}\,V\,d\left(\tfrac{M}{V}\right)\,,
\end{eqnarray}
from which follows the relation
\begin{equation} \label{alpha-fix2}
d\alpha +Q + \tfrac{1}2e^{2U}\star d\omega
          -\,e^{2U}\, d(\tfrac{M}{V})\, V =0\, .
\end{equation}
We emphasise that these equations are completely analogous to the ones relevant for the
single centre system, but now involve generically non-harmonic functions $M$ and $V$.

Additionally, \eqref{dw-mid} contains the parameters for the particular T-duality
appearing in the derivative of $\hat{R}^*$ in \eqref{Tdualstar}. These can be rewritten
by observing that the action of a T-duality on the symplectic section takes the form
\begin{align}\label{TonSection}
 \dTgp &[2\,e^{-U}\mbox{Im}(e^{-i\alpha}\cV) -\frac{M}{V}\,\hat{R}^*]
\CR
=&\,
2\,e^{-U}\,\mbox{Im}\left[ -3\,i\,\gamma_*\, Y\,e^{-i\alpha}\cV
  +i\,\gamma_*\,(Y - 2\, \bar Y)\bar \Omega^i D_i \cV
  - Q_*^i D_i \cV \right]\,.
\end{align}
Using this relation and \eqref{alpha-fix2} in \eqref{dw-mid} leads to
\begin{align}\label{dw-real}
\star dw =- \left[d -2\,\dTgp \right]
[2\,\mbox{Im}(e^{-U-i\alpha}\cV) - \tfrac12\, V\,\hat{R} -\tfrac{M}{V}\,\hat{R}^*]
\,,
\end{align}
where we also used the fact that $\hat{R}$ is by definition inert under the T-dualities in question.
This is the final form of the flow equations in the real basis, where the local T-dualities have
parameters given by \eqref{Tdualstar} above.

\subsection{Integration and local structure}
\label{sec:integr-5}

Due to the presence of the flat connection for the T-dualities, it is not possible to
solve the system of equations \eqref{dw-real} in terms of harmonic functions only.
However, the scalar and vector fields can be written as
\begin{gather}
 2\,e^{-U}\mbox{Im}(e^{-i\alpha}\cV)  - \tfrac12\, V\,\hat{R} - \tfrac{M}{V}\,\hat{R}^*
=-\cH\,,
\label{scal-vec-H}
\\
 dw = \left(\star d -2\,\star \dTgp \right)\cH \,,
\label{dw-fin-1}
\end{gather}
where the vector of functions $\cH$ is the solution to the non-harmonic equation
\begin{equation}\label{Lapl-H}
 d\star d \cH  - 2\,d\star\dTgp \cH -2\, \dTgp\! \wedge \star d \cH =0 \,.
\end{equation}
In order to disentangle the derivatives on $\cH$ and $\Tgp$ and cast this as a
Poisson equation, we introduce the rescaled vector of functions
\begin{equation}\label{def-H-resc}
\cH_0=\exp[-\Tgp]\cH \,,
\end{equation}
in terms of which we find
\begin{gather}
 2\,e^{-U}\mbox{Im}(e^{-i\alpha}\cV) - \tfrac12\, V\,\hat{R} - \tfrac{M}{V}\,\hat{R}^*
=-\exp[\Tgp]\cH_0 \,,
\label{cH-final}
\\
 dw = \exp[\Tgp]\star d\cH_0 - \star d( \exp[\Tgp] )\,\cH_0\,,
\label{dw-final}
\end{gather}
while \eqref{Lapl-H} takes the form
\begin{equation}\label{LaplOnH}
 d\star d \cH_0  - d\star\dTgp \cH_0 - \dTgp\! \wedge \star \dTgp \cH_0=0
\,.
\end{equation}

It is possible to give a systematic characterisation of the solution for $\cH_0$ and $\exp[\Tgp]$
using the following crucial observation. From the expression \eqref{dw-final} for the vector
fields, we compute for the derivative of $\cH_0$ that
\begin{align} \label{ZdH0}
Z_a(\exp[\Tgp]\star d\cH_0)
&\,-\N[\bar \Omega]\,\Omega_a\, Z(\exp[\Tgp]\star d\cH_0)
\CR=&\,
Z_a(\star dw +\exp[\Tgp]\dTgp \cH_0)
-\N[\bar \Omega]\,\Omega_a\, Z(\star dw +\exp[\Tgp]\dTgp \cH_0)
\CR =&\,
 - \frac{Y^2}{|Y|^2}\left(\bar Q_{\scp \,a} + 2i\,\Omega_a\, \gamma_\scp\right) 
 + e^{i\alpha} N_a
\,,
\end{align}
where $N_a$ is the combination of scalar momenta defined in \eqref{scal-comb}.
One can easily show that the central charges in \eqref{ZdH0} satisfy the reality
constraint \eqref{ZAconstraint}, \ie we have
\begin{align}
e^{i \alpha}&\,\left[ \bar Z^a(\exp[\Tgp] \,d\cH_0) 
- \N[\Omega]\, \bar \Omega^a\, \bar Z(\exp[\Tgp]\, d\cH_0) \right]=
\CR
&\, \quad  c^{abc} \Omega_b Z_c(\exp[\Tgp] \,d\cH_0) 
 + \bar \Omega^a \,\left( Z(\exp[\Tgp]\, d\cH_0) - \N[\Omega]\,\bar \Omega^b Z_b(\exp[\Tgp]\, d\cH_0) \right) \,,
\label{ZAconstraint-5}
\end{align}
This constraint was found to be crucial to describe single centre solutions in
\cite{Bossard:2012xsa}, where it was analysed in some detail. Using \eqref{RProjector}, one computes that it is equivalent to the constraints 
\begin{gather}
 \frac{1}{2} I_4^\prime(\hat{R},\hat{R}^*, \exp[ \Tgp] d \cH_0 ) = 
-4  \exp[ \Tgp] d \cH_0 +3 \Iprod{ \exp[ \Tgp] d \cH_0}{\hat{R}^*} \hat{R} \ , 
\CR 
\Iprod{\hat{R}}{ \exp[ \Tgp] d \cH_0} = 0 \ .
\end{gather}
These equations are manifestly duality covariant, and in particular T-duality covariant.
Therefore, using \eqref{Rstar0-def} one finds that the vector $d \cH_0$ satisfies
the constraints
\be \frac{1}{2} I_4^\prime(\hat{R},{R}^*_0,d \cH_0 ) = -4  d \cH_0 +3 \Iprod{ d \cH_0}{{R}^*_0} \hat{R} \ , \qquad \Iprod{\hat{R}}{ d \cH_0} = 0 \ . \label{sympl-constr-pr} \ee
In a similar fashion, one can show that the vector of functions $\cH_0$ itself satisfies
the same constraint, using \eqref{scal-vec-H}, and therefore only half of its components are
allowed.

Indeed, a vector satisfying this constraint does not contain components of grade $(+1)$ and
$(-3)$, so that in terms of \eqref{eq:vec-decomp}, one finds
\begin{equation}\label{H-grad-5}
\cH_0 \in ({\mathds{R}}^{n_v})^\ord{-1} \oplus \mathds{R}^\ord{3} \ ,
\end{equation}
and therefore describes only $n_v+1$ functions instead of the $2\,(n_v+1)$ one would have
a priori, \ie it lies on a Lagrangian subspace. One of these functions is clearly the function
$V$ in \eqref{V-def}, describing the grade $(+3)$ component of $\cH_0$ along the
direction of $R$. The remaining $n_v$ functions span the grade $(-1)$ vector space, according
to the decomposition imposed by the T-duality $\Tgp$, and are undetermined for the moment.

Using this decomposition in \eqref{LaplOnH}, we find the following grade assignments for each term
\begin{align}\label{Lapl-dec}
&\,\, d\star d \cH_0  - d\star\dTgp \cH_0 - \dTgp\! \wedge \star \dTgp \cH_0=0 \,,
\\
&\, \scriptstyle{(-1)\oplus (+3)} \quad\qquad \scriptstyle{(+1)} \qquad\qquad\quad \scriptstyle{(+3)}
\nonumber
\end{align}
where we used the fact that $\Tgp$ is of grade $(+2)$. Now, since the second term in
\eqref{Lapl-dec} lies in a subspace orthogonal to both the other terms, it is clear that
this equation decomposes in two independent equations, as
\begin{gather}
 d\star d \cH_0  - \dTgp\! \wedge \star \dTgp \cH_0=0\,, \label{Lapl-fin}
\\
d\star\dTgp =0\,. \label{Lapl-K}
\end{gather}
In deriving the second equation we used the fact that applying a T-duality on a generic
vector as in \eqref{H-grad-5} results in a vector of grade $(+1)$, which cannot vanish
for a physical solution\footnote{The non-generic solutions to this equation require constraints
on the vector of T-duality parameters and only exist if $I_4(\cH_0,\cH_0,\cH_0,\hat{R})=0$, which
would lead to solutions with irregular horizons. \label{foot-Hphys}}
unless the T-duality matrix is trivial. We now analyse each of the two equations in turn.

First, it is easy to express \eqref{Lapl-K} in terms of the vector of T-duality
parameters in \eqref{Rstar0-real}, as
\begin{equation}\label{K-def}
 d\star d \cK=0\,,
\end{equation}
so that the parameters $\cK$ are identified as a rank three vector of harmonic functions.
Note that $\cK$ is a priori a generic vector of grade $(-1)$, \ie it lies in the same
Lagrangian subspace as the vector $\cH_0$ above, with the additional restriction of a
vanishing component along $R$, since $\Iprod{\hat{R}^*}{d \hat{R}^*}=0$ by definition.
However, we should note that the constant part of $\cK$ is not physical and can be absorbed
into $R^*_0$ by imposing the boundary condition that $\cK$ vanishes asymptotically. This
choice is useful in the discussion of explicit solutions.

We now turn to the Poisson equation \eqref{Lapl-fin} for the vector $\cH_0$. As shown in
\eqref{Lapl-dec}, the source term of is along the unique grade $(+3)$ component, that is
along the vector $R$. We have therefore identified $\cH_0$ as a vector lying in a Lagrangian
submanifold containing $R$, whose components along the grade $(-1)$ directions are $n_v$
harmonic functions and the component along the direction of $R$ is a single non-harmonic
function, $V$. One can directly compute the source term by varying the combination in
\eqref{TonSection} and using \eqref{cH-final}, to find
\begin{align}\label{R-comp}
\dTgp \wedge \star \dTgp \cH_0=&\,
-\dTgp \wedge \star \dTgp[2\,\mbox{Im}(e^{-i\alpha}\cV) - \frac{M}{V}\,\hat{R}^*]
\CR
=&\, \left( 6\,\gamma_\scp\wedge \star\gamma_\scp - Q_\scp^a \wedge \star \bar Q_{\scp\,a}\right)
        \,R\,,
\end{align}
which is explicitly proportional to the vector $R$, up to a real function.
We can write this result in a simpler form by taking the inner product of
\eqref{dw-final} with $\hat{R}^*$ and comparing with \eqref{V-def}, to obtain
\begin{align}\label{V-comp}
V =-\Iprod{\hat{R}^*}{\cH}=-\Iprod{R^*_0}{\cH_0}\,,
\end{align}
where we used \eqref{Rstar0-def} and \eqref{def-H-resc}.
The Poisson equation for this function can now be found by projecting
\eqref{Lapl-fin} along $R^*_0$, as
\begin{equation}\label{V-Poiss}
 d\star d V = -\Iprod{R_0^*}{\dTgp \wedge \star \dTgp\cH_0}
= -\tfrac1{16}\,I_4(d \cK, \star d \cK, \cH_0, \hat{R}) \,.
\end{equation}
Note that, by \eqref{R-comp}, the right hand side of the first equality is
nonvanishing, since it is the inner product of $R$ with its magnetic dual.
The second equality is a direct consequence of \eqref{Tp2}. We moreover
record the following relations for the expressions involving T-duality
matrices in \eqref{TonSection} and \eqref{R-comp},
\begin{align}
\dTgp[2\,e^{-U}\mbox{Im}(e^{-i\alpha}\cV) -\frac{M}{V}\,\hat{R}^*]
=&\,
\tfrac1{16}\, \exp(\Tgp) I_4^{\prime}(d\cK,\cH_0,\hat{R})\,,
\nonumber\\
\dTgp \wedge \star \dTgp[2\,\mbox{Im}(e^{-i\alpha}\cV) - \frac{M}{V}\,\hat{R}^*]
=&\,
\tfrac1{64}\, I_4(d\cK,\star d\cK,\cH_0,\hat{R})\,\hat{R}\, ,
\label{T-TT-H-5}
\end{align}
which are direct consequences of \eqref{Tp1} and \eqref{Tp2}, in combination
with \eqref{cH-final}.
These equations, along with \eqref{2TonRs}, allow us to evaluate the action of
T-duality generators on the various objects relevant to the system. As the action of
a finite T-duality is expressed as a finite sum due to the nilpotency of $\Tgp$,
it follows that one may also compute the action of finite T-dualities, as alluded
to above. This is especially important when trying to find explicit solutions to
this system \cite{Bossard:tocome}.

The final quantities to be fixed are the angular momentum vector, $\omega$, and
the function, $M$, appearing in the expression for the scalars. In the single
centre class, these are both harmonic and are dual to each other. Indeed, in that
case the T-duality parameters in \eqref{Tdualstar} must vanish and the first of
these equations imposes exactly that $M$ and $\omega$ are harmonic. In the more
general multi-centre case, one has to compute the nontrivial $\gamma_\scp$ to
obtain the analogous equation. This can be done straightforwardly using
\eqref{Tdualstar}, \eqref{def-H-resc}-\eqref{dw-final} and \eqref{T-TT-H-5} to
show that
\begin{align}\label{eq:dom-5}
\star d \omega - d M = \Iprod{\cH}{\star dw}=
\Iprod{\cH_0}{d\cH_0 - \dTgp\,\cH_0}= \tfrac1{16}\,I_4(d\cK ,\cH_0, \cH_0, \hat{R})
\,.
\end{align}
One can alternatively obtain the same expression by manipulating the definition of the
non-harmonic function $M$ in \eqref{ImOdt} using the above results.
The integrability condition on \eqref{eq:dom-5} leads to the following Poisson
equation for $M$:
\begin{align} \label{M-Poisson-5}
d \star d M
=&\,-2\, \Iprod{\cH_0}{ \dTgp \star d\cH_0}
= \tfrac18\,I_4(d\cK, \star d \cH_0, \cH_0, \hat{R}) \,,
\end{align}
which can be solved once the grade $-1$ component of $\cH_0$ and $\cK$ are chosen.

This concludes our duality covariant presentation of the composite non-BPS system
in terms of the real basis. In the next section, we summarise the final
form of the equations to be solved and we comment on some properties of the 
solutions.

\subsection{Summary of results}
\label{sec:summ-5}

In this short section, we summarise all relevant formulae for the composite
non-BPS system in the real basis. All relations presented here were shown
explicitly in the previous sections and we refer to the discussion there
for the details. We find it however useful for future applications to give
a self-contained account of the final form of the system.

The ansatze for the metric and gauge fields are given in \eqref{metricMltc}
and \eqref{gauge-decop} in terms of the function $e^U$, the one-form $\omega$
and the spatial vector fields $dw$, while the electromagnetic potentials are
fixed by \eqref{Zetas}.
The first order flow equation for the composite non-BPS system is given by
\eqref{dw-real}, as
\begin{align} \label{dw-summ}
\star dw =- \left[d -2\,\dTgp_{\scriptscriptstyle \cK} \right]
[2\,\mbox{Im}(e^{-U-i\alpha}\cV) - \tfrac12\, V\,\hat{R} -\tfrac{M}{V}\,\hat{R}^*]
\,.
\end{align}
Here, $M$, $V$, are functions to be specified below, while $\hat{R}$ and
$\hat{R}^*$ are a constant and a non-constant very small vector respectively,
where $\Iprod{\hat R}{\hat{R}^*}= 4$. The non-constant $\hat{R}^*$ is related to a
constant very small vector, $R_0^*$, by \eqref{Rstar0-def} 
\begin{equation} 
 \hat{R}^*= \exp[\Tgp_\csK]R_0^*\,,
\end{equation}
which also satisfies $\Iprod{\hat R}{R_0^*}= 4$. In this and all equations in this section,
$\Tgp_{\csK}$ is a generator of the
T-dualities leaving $\hat{R}$ invariant, parametrised by a vector of
harmonic functions, $\cK$. As discussed in section \ref{sec:T-duality}, the
vector of parameters $\cK$ lies in the grade $(-1)$ component of the vector space
according to the decomposition implied by the T-duality. It is therefore
a three-charge vector satisfying
\begin{equation}\label{grade-1}
\frac{1}{2} I_4^\prime(\hat{R},{R}^*_0, \cK ) = -\Iprod{\hat R}{R_0^*}\, \cK \,,
\end{equation}
which indeed specifies a vector of $n_v$ degrees of freedom.

The solutions to the flow equation \eqref{dw-summ} are simplified by
introducing a vector, $\cH_0$, of grade $(-1)\oplus(+3)$, \ie satisfying
\begin{equation}
 \frac{1}{2} I_4^\prime(\hat{R},{R}^*_0, \cH_0 ) = 
-\Iprod{\hat R}{R_0^*}\,\cH_0 + 3\, \Iprod{  \cH_0}{{R}^*_0} \hat{R} \ . 
\label{sympl-constr-summ} 
\end{equation}
Note that \eqref{grade-1} is trivially a solution of the first equation,
found by setting the grade $(+3)$ component, $\Iprod{R_0^*}{\cH_0}$,
to vanish. The equations resulting from \eqref{dw-summ} take the form
\eqref{dw-final}, 
\begin{gather}
 2\,e^{-U}\mbox{Im}(e^{-i\alpha}\cV) 
=-\exp[\Tgp_{\scriptscriptstyle \cK}]
\left(\cH_0 - \tfrac12\, V\,\hat{R} - \tfrac{M}{V}\,R_0^* \right) \,,
\label{scals-summ-5}
\\
 \star dw = \exp[\Tgp_{\scriptscriptstyle \cK}]
 \left( d\cH_0 - \dTgp_{\scriptscriptstyle \cK}\cH_0\right)\,,
\label{dw-fin-summ}
\end{gather}
where $V$ is now identified with the grade $(+3)$ component of $\cH_0$,
as $V=\Iprod{\cH_0}{R_0^*}$. The compatibility relation for the last
relations leads to the field equation for $\cH_0$, as in
\eqref{Lapl-fin} and \eqref{T-TT-H-5}
\begin{gather}
 d\star d \cH_0 = \dTgp_{\scriptscriptstyle \cK}\! \wedge \star \dTgp_{\scriptscriptstyle \cK} \cH_0
=-\tfrac1{64}\, I_4(d\cK,\star d\cK,\cH_0,\hat{R})\,\hat{R}
\,. \label{Lapl-fin-summ}
\end{gather}
As the right hand side of this relation is only along $\hat{R}$, it follows that
all grade $(-1)$ components of $\cH_0$ are harmonic, whereas $V$ is not, leading
to \eqref{V-Poiss}, as
\begin{equation}\label{V-Poiss-summ}
 d\star d V = -\tfrac1{16}\,I_4(d\cK, \star d\cK, \cH_0, \hat{R}) \,,
\end{equation}
by taking the inner product of \eqref{Lapl-fin-summ} with $R^*_0$.
The final dynamical equation required is the one for the function $M$
in \eqref{dw-fin-summ} and the angular momentum vector $\omega$, both
of which are conveniently given by \eqref{eq:dom-5}, as
\begin{align}\label{dom-summ}
\star d \omega - d M = 
\Iprod{\cH_0}{d\cH_0 - \dTgp_{\scriptscriptstyle \cK}\,\cH_0}= 
\tfrac1{16}\,I_4(d\cK ,\cH_0, \cH_0, \hat{R})\,.
\end{align}
Taking the divergence of this equation, one obtains
a Poisson equation for $M$.

These equations can be seen to be equivalent to the known formulation of
the composite non-BPS system, as given in a fixed duality frame in
\cite{Bossard:2011kz, Bossard:2012ge}, by making a choice for the constant vectors
$\hat R$ and $R^*_0$. In fact, these two papers use different frames for describing
the system, both of which can be reached from our general formulation. The frame of
\cite{Bossard:2012ge} is found by choosing
\begin{equation}\label{sp-frame-5}
 \hat R \propto (0, \delta_I^0)\,, \qquad  R^*_0 \propto (\delta^I_0, 0)\,,
\end{equation}
where we disregard the (arbitrary) normalisation. Similarly the frame in \cite{Bossard:2011kz}
is found by interchanging the expressions for the two vectors in \eqref{sp-frame-5},
corresponding to an electric/magnetic duality.

We close with some comments on the structure of the solutions. First,
the physical scalars and the metric scale factor can be obtained by solving \eqref{scals-summ-5}
in the standard way \cite{Bates:2003vx}. Since all quantities above are appropriate combinations
of the single centre solution in \cite{Bossard:2012xsa}, up to overall T-dualities, it is possible
to use many of the results given there. For instance, the metric scale factor is given by
\begin{equation}
e^{-4U} = I_4(\cH_0 - \tfrac{1}{2} V\,\hat{R} - \tfrac{M}{V}\, R_0^*) \ ,
\end{equation}
where we used the fact that the quartic invariant is by definition invariant under all T-dualities.
This expression is identical to the corresponding one for the single centre class, despite the fact
that the functions $M$ and $V$ are not harmonic in the present context. As in
\cite{Bossard:2012xsa}, one can simplify the expression for $e^{U}$ as follows.
The decomposition of the vector $\cH_0  - \tfrac{1}{2} V\,\hat{R}$ in grades
$(-1)\oplus (+3)$ implies that $ I_4(\cH_0  - \tfrac{1}{2} V\,\hat{R}) $
is linear in $V$, as any other power of the grade $(+3)$ would vanish identically. In
particular we find that
\begin{equation} 
\cH_0 - \tfrac{1}{4} V \hat{R} \in ({\mathds{R}}^{n_v})^\ord{-1} \ ,
\quad\Rightarrow\quad
I_4\scal{ \cH_0 - \tfrac{1}{4} V \hat{R} } = 0 \ , \end{equation}
so that we can show the equality $I_4(\cH_0  - \tfrac{1}{2} V\,\hat{R})=-I_4(\cH_0)$.
It then follows that
\begin{align}
e^{-4U} =&\, I_4(\cH_0  - \tfrac{1}{2} V\,\hat{R})
              - \tfrac{M^2}{V^2}\,\Iprod{R_0^*}{\cH_0  - \tfrac{1}{2} V\,\hat{R}}^2
\CR
=&\,- I_4(\cH_0) - M^2 \ . \label{ScallingFactorInvariant}
\end{align}
This equation implies that the $n_v$ grade $(-1)$ harmonic components in $\cH_0$
must correspond to a rank three charge (\ie a large electric charge in five dimensions),
so that \eqref{ScallingFactorInvariant} leads to a non-degenerate metric.

Based on \eqref{ScallingFactorInvariant}, we conclude that the system above indeed describes
the interactions of black holes that are non-supersymmetric in isolation, since $I_4(\cH_0)$
must be negative globally and in particular at each centre, for a regular geometry. We refrain
from giving an explicit expression for the physical scalars, as these would involve the
action of an arbitrary abelian isometry with parameters $\cK$ on the physical scalars for
the single centre class as given explicitly in \cite{Bossard:2012xsa}. Of course, the
scalars can be computed from the standard formulae in \cite{Bates:2003vx} for any desired
solution.

The characteristic features of all solutions in this class is that each one of the centres
must be of the non-BPS type in isolation, as explained above, and that the charges at all
centres must commute with the vector $\hat{R}$ and must not commute with the vector $R^*_0$,
by regularity. It is then clear that such solutions do not exist for generic non-BPS charges
at the centres. In addition, once an allowed charge configuration is fixed, one cannot
have arbitrary values for the moduli at infinity. This is due to the constraint
on $\cH_0$, which contains only $n_v+1$ asymptotic constants and the fact that some of
these constants, together with the $2n_v$ parameters in $\hat R$ and $R^*_0$, turn out to
parametrise charges that are not described by the poles of $\cH_0$, through
\eqref{dw-summ} and \eqref{dw-fin-summ}. A simple example of this situation is given by
the class of two centre solutions for models with $n_v\ge 3$. In this case, the two non-BPS
charges are in fact arbitrary, since one can always find a choice of $\hat R$ and $R^*_0$
for any pair of non-BPS charges.
However, it is possible to show \cite{Bossard:tocome} that the asymptotic moduli are
constrained to lie on a $(n_v+2)$-dimensional hypersurface of the $2 n_v$-dimensional moduli
space. Note that this is very different from the multi-centre BPS solutions, where solutions
exist a priori everywhere in moduli space and walls of marginal stability arise only when
the constrains implied by global regularity are imposed. In the present case however,
local solutions seem to exist only in certain hypersurfaces of moduli space, while
walls of marginal stability might still arise on these constrained surfaces.

Finally, it is worthwhile commenting on the evaluation of T-dualities appearing as matrices
in the equations above. As we have shown in several examples, one can avoid introducing
explicit matrices, instead computing the action of T-dualities on the relevant vectors
by use of the definition \eqref{K-def} and equations \eqref{T-TT-H-5}. Indeed, we find that
it is possible to reduce all required computations to a recursive application of these
three relations. As the grading involves only four subspaces, this procedure terminates
after at most three steps.

\section{Almost-BPS system}
\label{sec:almost}

In this section we present in detail the characterisation of solutions to the
almost-BPS system of equations, in analogy with the steps taken in the previous
section for the composite non-BPS system. While the discussion here is self-contained,
we will occasionally refer to section \ref{sec:composite}, in order to highlight
similarities and recycle some results. The starting point is the solution to
the nilpotency condition given in \eqref{flow-almost}, which we repeat explicitly
here
\begin{eqnarray}
 dU + \frac{i}{2} e^{2U} \star d \omega
  & =& -\frac{3}{4} e^{- i \alpha} Z -\frac{1}{4} \N[\Omega] \bar Z
       -\frac{1}{4} \Omega_a \bar Z^a + \frac{1}{4} e^{-i\alpha } \N[\Omega] \overline{\Omega}^a Z_a
 \CR
 - e^a_i d t^i
 &=& e^{i\alpha}  \bar Z^a  + \frac{1}{4} \overline{\Omega}^a \scal{ Z + \N[\Omega] \overline{\Omega}^b Z_b - e^{i\alpha} \N[\Omega] \bar Z - e^{i\alpha} \Omega_b \bar Z^b } \,,
\label{flow-almost2}
\end{eqnarray}
where $Z=Z(\star F)$, $Z_a=Z_a(\star F)$ are the central charges of the spatial field
strengths $F$ in \eqref{gauge-decop}. The vector $\Omega_a$ is connected to a constant
very small vector, $\hat{R}^*$, through \eqref{Rst-def-7}-\eqref{Rstar-almost}. For
later reference, we also give the inverse relations, which read
\begin{eqnarray}
 Z  & =& -e^{i \alpha} \left( dU + \frac{i}{2} e^{2U} \star d \omega \right)
 + \frac{i}{2}\, e^{\frac{i\alpha}{2}} \N[\Omega]^\frac{1}{2}\, \mu
\CR
 Z_a &=& -e^{i\alpha} e_{a\,i} d \bar t^i
 + \frac{i}{2}\,e^{\frac{i\alpha}{2}} \N[\overline{\Omega}]^\frac{1}{2}\, \Omega_a \,\mu\,,
\label{Zs-almost}
\end{eqnarray}
where we used as shorthand the one-form
\begin{align}\label{eq:mu-def}
 \mu = \mathrm{Im}\left[
  e^{\frac{i\alpha}{2}} \N[\overline{\Omega}]^\frac{1}{2}
   \left( dU + \frac{i}{2} e^{2U} \star d \omega \right)
 + e^{\frac{i\alpha}{2}} \N[{\Omega}]^\frac{1}{2}\, \overline{\Omega}^a e_{a\,i} d \bar t^i\right]\,.
\end{align}
The final required equation is the compatibility equation \eqref{eq:d-alpha}, which
can be rearranged using \eqref{Zs-almost} and \eqref{eq:mu-def} to obtain
\begin{eqnarray}
\label{d-alph-alm}
d(\alpha-\arg[\N[\Omega]])= -\tfrac2{|Y|^2}\,
\mbox{Im}\left[ Y \left( dU + \frac{i}{2} e^{2U} \star d \omega  -e^{-i\alpha}\Omega^a e^a_i d t^i \right) \right]\,.
\end{eqnarray}
Here, we used the function $Y$ defined in the first equality of \eqref{Y-def-gen}
and the fact that it has unit real part. Similarly, one can show
using the flow equations above, that $\mu$ is also given by the expressions
\begin{align}
 \mbox{Re}\left(e^{\frac{-i\alpha}{2}} \N[{\Omega}]^\frac{1}{2}\right)\, \mu
=&\,  \left( e^{2U} \star d \omega +  2\,\Im[ e^{-i\alpha} Z] \right)\,,
\CR
 =&\,
\tfrac12\,e^{2U}\,d \omega -\mbox{Im}(e^{-i\alpha} \Omega_i\,d t^i) +i\,|Y|^{-2}d \bar Y\,,
\label{eq:mu-3}
\end{align}
that will be used in due time.

In order to solve these equations, we follow a path similar to the last section,
by considering the electromagnetic potentials in section \ref{sec:zeta-7} and using
them to simplify the equations. The connection with T-dualities is shown in section
\ref{sec:Tdual-7}, while sections \ref{sec:linsys-7} and \ref{sec:integr-7} are
devoted to the linear system of equations governing this system and its integration
in terms of local functions respectively. The reader interested in applications can
find a summary of the final form of the system in section \ref{sec:summ-5}.

\subsection{The electromagnetic potentials}
\label{sec:zeta-7}

As a first step towards the solution of the system, we decompose the field strength
$F$ in the electromagnetic potentials and the vector fields that define the conserved
charges. The electromagnetic potentials for this system are computed by their
definition \eqref{cmplx-self} as
\begin{eqnarray}
d\zeta&\equiv& 2\,e^{2U}\Re[\bar Z(\star F) \cV + \bar Z^i(\star F) D_{i}
\cV] \CR
&=& -2\,d\,\scal{ e^{U} \Re [ e^{-i \alpha } \cV ]}
+\tfrac{1}{2}e^{U}\,\mu \, R^*
\label{eq:dzeta-almo}\,,
\end{eqnarray}
where we used \eqref{d-alph-alm} and the definition \eqref{Rst-def-7} to rearrange terms.
The first term in \eqref{eq:dzeta-almo} is already a total derivative, so that the
last term must combine into the derivative a vector, which is necessarily
proportional to $R^*$. Using the relation of $R^*$ to a constant vector in
\eqref{Rstar-almost}, this requirement leads to the condition
\begin{equation}\label{eq:W-def}
 e^{U}\,\mu= - 2\,W\, \mbox{Re}[ \bar Z(R^*) Z_b(R^*) e^b_i dt^i ] + 2\,d W\,,
\end{equation}
where we introduced the a priori arbitrary real function $W$. The result for the
electromagnetic potentials takes the form
\begin{eqnarray}
\zeta &=& -2\,e^{U} \Re [ e^{-i \alpha } \cV ] + W \, R^*
\label{eq:zeta-almo}\,,
\end{eqnarray}
while the corresponding central charges are
\begin{eqnarray}
Z(\zeta) &=& -i\,e^{U} e^{i \alpha } + W \, Z(R^*)
        =  -i\,e^{i \alpha }(e^{U} + i\,W \, e^{\frac{i\alpha}{2}} \N[\bar\Omega]^\frac{1}{2})
\CR
Z_a(\zeta) &=&  W \, Z_a(R^*)
            =   W \, e^{\frac{i\alpha}{2}} \N[\bar\Omega]^\frac{1}{2}\Omega_a\,.
\end{eqnarray}
We can now construct the vector potentials by the definition
\begin{eqnarray}
 Z(dw) & =& Z(F) - Z(\zeta)\,d \omega\,,
\qquad
 Z_a(dw) = Z_a(F) - Z_a(\zeta)\,d \omega\,,
\end{eqnarray}
which leads to the expression
\begin{eqnarray}\label{dw-start-7}
 \star dw &=&
2\,\mbox{Im}\left[ -d(e^{-U} e^{-i \alpha}\,\cV )
+ \left(\tfrac{i}{2}\, e^{-U}\mu +W \,d \omega \right)\, \cR^*
+ i\, e^{-U}\mu\,e^{-\frac{i\alpha}{2}} \N[\bar\Omega]^\frac{1}{2}\,\cV
\right]\,,
\end{eqnarray}
where we use the shorthand in \eqref{eq:mu-def} and
\begin{equation}
 \cR^*= e^{-\frac{3i\alpha}{2}} \N[\Omega]^\frac{1}{2}\,\cV
-\, e^{-\frac{i\alpha}{2}} \N[\Omega]^\frac{1}{2}\,\bar\Omega^a\,D_a\cV\,.
\end{equation}
Here, it is worth pointing out that, unlike in the composite non-BPS system
(cf. \eqref{Rdw}), there is no component of the vector fields that is
vanishing a priori. Nevertheless, the projection of the vector fields
in \eqref{dw-start-7} along the available constant vector $\hat R^*$ is
still relevant and can be computed as
\begin{align}\label{Rstdw-7}
 \Iprod{\hat R^*}{dw}
=&\,2\,\left(1+4\,e^{-U}\frac{W}{|Y|}\right)\star d\left( e^{-U} \,\frac{|Z(\hat R^*)|}{|Y|}\right)
\nonumber\\
&\,
-8\, e^{-U}\frac{|Z(\hat R^*)|}{|Y|} \,\star d\left( e^{-U}\frac{W}{|Y|} \right)\,,
\end{align}
where $Y$ is again as in \eqref{Y-def-gen}.
This can be simplified by imposing consistency of the two expressions for
$\mu$ in \eqref{eq:mu-3} with \eqref{eq:W-def}, combined with \eqref{Rstar-almost},
leading to
\begin{equation}\label{eq:dW-dV}
 \left(1+2\,e^{-U}\frac{W}{|Y|}\right)\,d\left( e^{-U} \frac{|Z(\hat R^*)|}{|Y|} \right)
=2\,e^{-U} \,\frac{|Z(\hat R^*)|}{|Y|} \, d \left(e^{-U}\frac{W}{|Y|}\right)\,.
\end{equation}
Now, it is simple to show that \eqref{Rstdw-7} and \eqref{eq:dW-dV} imply
that the projection $\Iprod{\hat R^*}{dw}$ is given by a harmonic function, $V$,
defined as
\begin{align}\label{V-def-alm}
 \Iprod{\hat R^*}{dw}=&\,-2\,\star d\left( e^{-U} \,\frac{|Z(\hat R^*)|}{|Y|} \right)
 \equiv -\star d V\,,
\end{align}
while the function $W$ is fixed as
\begin{equation}\label{beta-def}
 1+2\,e^{-U}\frac{W}{|Y|}= \beta\,V\,,
\end{equation}
where $\beta$ is an arbitrary constant. Using these results, \eqref{eq:W-def}
simplifies to
\begin{equation}\label{mu-simple}
 \mu= |Y|\,\frac{d V}{V}\,,
\end{equation}
which will be used in due time.

\subsection{Connection to T-dualities}
\label{sec:Tdual-7}

We now discuss the relevance of T-dualities for the almost-BPS system,
which will be important for the integration of the flow equations, as for
the composite non-BPS system in the previous section. In order to exhibit
this, we consider the very small vector
\begin{equation}
 R= \mbox{Im}\left[ -\N[\bar\Omega]\,\cV  + \bar\Omega^a\,D_a\cV \right]\,,
\end{equation}
which is always mutually nonlocal with the constant vector $\hat{R}^*$ and is
not constant in general. As is clear by their definitions these two vectors
are related in exactly the same way as the pair of very small vectors, $R$ and
$R^*$ in \eqref{N-Om-def} and \eqref{R-star-bas} respectively, up to rescalings.
Note that the situation is opposite to the one for the composite non-BPS system,
where $R$ is a constant vector up to rescaling, while $\hat{R}^*$ is not constant.

Based on the discussion in section \ref{sec:T-duality}, we associate $R$ to
the grade $(+3)$ component of the decomposition \eqref{eq:vec-decomp}, while
$R^*$ is identified as the corresponding grade $(-3)$ component.
Moreover, one can check that the normalised very small vector
\begin{equation}\label{R-def-7}
 \hat{R}=2\,|Z(\hat R^*)|^{-1}\,|Y|^{3}\,
\mbox{Im}\left[ -\N[\bar\Omega]\,\cV  + \bar\Omega^a\,D_a\cV \right]\,,
\end{equation}
has a constant inner product with $\hat R^*$, namely $\Iprod{\hat{R}}{\hat R^*}=4$.
As explained in \eqref{gen-small}-\eqref{der-small}, the condition that $\hat{R}$ is a
very small vector can be generally written as
\begin{equation}\label{R0-def}
 \hat{R}= \exp[\Tgm] R_0\,,
\end{equation}
where  $R_0$ is a constant very small vector and $\exp[\Tgm]$ is a
T-duality matrix leaving $\hat{R}^*$ invariant, parametrised by a
grade $(+1)$ vector of functions. It follows that the derivative of
$\hat{R}$ can be expressed as
\begin{equation}\label{dR-7}
 d\hat{R} - \dTgm \hat{R}=0\,,
\end{equation}
which is closed by the property that T-dualities are abelian, exactly
as in the composite non-BPS case. This is consistent with the known formulation of the
almost-BPS system in five dimensions \cite{Goldstein:2008fq, Bena:2009ev, Bena:2009en},
which can be written in terms of a T-duality parametrised by harmonic functions,
acting on the scalar and vector fields.

We can find the relevant T-duality parameters in \eqref{R0-def} by explicitly computing the derivative
of $\hat{R}$, using the flow equations \eqref{flow-almost2} for the
almost-BPS system. After a lengthy but straightforward computation,
we obtain that the derivative of \eqref{R-def-7} is indeed given by
\begin{align}
Z(d \hat{R}) =&\, \Tgm_{\gamma, Q}Z(\hat{R}) = Z(\dTgm \hat{R})\,,
\CR
Z_a(d \hat{R}) =&\,\Tgm_{\gamma, Q}Z_a(\hat{R}) =Z_a(\dTgm \hat{R})
\,, \label{DR}
\end{align}
where we indicated that the result is given by the variation of $\hat{R}$ under the
T-duality transformation in \eqref{T-duality-Rstar}, as shown in \eqref{r-flow}.
The values for the one-form generators are given by
\begin{align}\label{Tdual}
 \gamma_\scm =&\, \tfrac{1}{2}\,
 \,\left[ dU - \Im Y\, e^{2U}\star d\omega -\tfrac13\,\Im( Y e^{-i\alpha} \Omega\cdot dt) \right]\,,
\CR
 \bar Q_{\scm\,a} =&\, -\tfrac{1}{2}\,\bar Y\, \bigl[
 c_{abc} \bar\Omega^b e^{c}_{j} d t^{j} - e^{i\alpha} e_{a\,\bar \jmath} d \bar t^{\, \bar \jmath}
             + \tfrac13\,\Omega_a \left( e^{i\alpha} \bar \Omega^b e_{b\bar \jmath} d \bar t^{\, \bar \jmath}
                  +2\,\tfrac{Y^2}{|Y|^2}\,e^{-i\alpha} \Omega_j d t^{ j}  \right)
\bigr]
\,.
\end{align}
The expression for $\gamma_\scm$ can be rewritten using \eqref{eq:mu-3} in a form
similar to the corresponding T-duality parameter for the composite non-BPS system
in \eqref{Tdualstar}, as
\begin{equation}\label{M-def-7}
  \gamma_\scm = \tfrac16\,e^{2U}\,\left( \star d \omega - d M \right) \,,
\end{equation}
where we used the definition \eqref{Y-def-gen} for the function $M$ in terms of the phases
$e^{i\alpha}$ and $\N[\Omega]$.
In this form, it is manifest that setting the T-duality parameters $\gamma_\scm$,
$Q^a_\scm$ to zero, one finds that the angular momentum is given in terms of a harmonic
function, $M$, while the second of \eqref{Tdual} becomes a reality constraint on
the scalar flow, similar to \eqref{ZAconstraint} in the composite non-BPS system.
This restriction therefore leads to the single centre subclass, which is common
to both the composite non-BPS and almost-BPS systems.

Applying the considerations of section \ref{sec:T-duality} on the system at hand,
we recall that a generic T-duality leaving $\hat{R}^*$ invariant is parametrised
by a rank three grade $(-1)$ vector of parameters, which we denote by $\cK$, so that
\eqref{R0-def}-\eqref{dR-7} become
\begin{equation}\label{R0-real}
 \hat{R}= \exp[\Tgm_\csK] R_0\,, \qquad
 \dTgm_\csK R_0 = d\cK\,.
\end{equation}
In what follows, we will generally suppress the explicit subscript $\cK$ from the
T-duality generators for simplicity. One checks that $\cK$ is indeed a rank three vector, \ie 
\begin{gather}
 I_4(d \hat{R})
= I_4(\dTgm\hat R)=I_4(d\cK)=0\,,
\\
\partial_\mu \Tgm \partial_\nu \Tgm \hat R =
 -\tfrac1{16}\,I_4(\partial_\mu \Tgm\hat R ,\partial_\nu \Tgm \hat R, \hat R^*)=
 -\tfrac1{16}\,I_4(\partial_{\mu}\cK ,\partial_{\nu}\cK, \hat R^*)\, .
\label{pre-source-7}
\end{gather}
These equations are clearly dual to the corresponding equations \eqref{2TonRs} for
the composite non-BPS system.
Alternatively, the same property follows from the fact that $\dTgm R_0$ is
of grade $(+1)$ and the rank of such a vector is at most three.
Indeed, one can directly verify that the vector $d\cK$, as defined in
\eqref{R0-real}, satisfies the constraint \eqref{ZAconstraintStar} and \eqref{ZAConstraintK}, which are equivalent to the real constraint 
\be \frac{1}{2} I_4^\prime(R_0,\hat{R}^*,\cK) = \Iprod{R_0}{\hat{R}^*}  \cK \ .\label{constr-K-7} \ee
 These equations are explicit realisations
of the general situation discussed in section \ref{sec:T-duality}.

Finally, we stress the difference between the constant very small vector $R_0$ of grade
$(+3)$ and the original vector in \eqref{R-def}, which was also used to define $\hat{R}$
in \eqref{R-def-7}. This can seem confusing, especially in view of the fact that we used
the constant vector $R_0$ to define the constraint in \eqref{constr-K-7} and ultimately
the grading of the vector space. However, as already explained in the analogous situation
for the composite non-BPS system, below \eqref{2TonRs}, the grading associated to the T-dualities leaving invariant $\hat{R}^*$ is only defined up to the action of the T-dualities themselves. In this respect, one can chose any constant vector $R_0$ in this orbit to define the grading. Given this redundancy in the definition, it will be convenient to fix $T^-$ to vanish in the asymptotic region, such that $R_0 = \hat{R}|_{r\rightarrow \infty}$.

\subsection{The linear system}
\label{sec:linsys-7}

We are now in a position to use the above results to write the system of flow equations
in the real basis, in terms of the symplectic section, $\cV$, the two very small constant
vectors $\hat{R}^*$, $ R_0$, and the relevant T-duality generators.

To show this, we insert \eqref{beta-def} and \eqref{mu-simple} in
\eqref{dw-start-7} to eliminate the spurious quantities $W$ and $\mu$
in favor of $V$ and $dV$ respectively. Moreover, it is useful to note the
relation
\begin{align}\label{TonSection-7}
 \dTgm\!\left[2\,e^{-U}\,\mbox{Im}(e^{-i\alpha}\cV)\right]
=&\,
2\,e^{-U}\,\mbox{Im}\left[ -3\,i\,\gamma_*\, e^{-i\alpha}\cV
  +i\,\gamma_*\, \bar \Omega^i D_i \cV + Q_*^i D_i \cV \right]\,,
\end{align}
as well as the identity
\begin{equation}
 \left(d -2\,\dTgm\right) V\,\hat{R}= d V\, \hat{R} - V\,d \hat{R} \,,
\end{equation}
which follows from \eqref{dR-7}, and we remind the reader that the
explicit expression for $d\hat{R}$ is given by \eqref{DR} with parameters
as in \eqref{Tdual}.
One can then verify that the expression
\begin{align}\label{dw-real-7}
\star dw =- \left(d -2\,\dTgm\right)
\left[2\,\mbox{Im}(e^{-U-i\alpha}\cV) - \tfrac12\, V\, \hat{R} -\tfrac{M}{V}\,\hat{R}^*\right] 
- \beta\,\star d \omega \, \hat{R}^* \,,
\end{align}
is equivalent to \eqref{dw-start-7} above. This result is manifestly duality
covariant in the real basis, as it is written in terms of real symplectic vectors
only. In particular, note that \eqref{dw-real-7} is completely analogous to the
corresponding result \eqref{dw-real} for the composite non-BPS system, up to the
term explicitly proportional to the angular momentum. However, it is simple to
shown that this term is unphysical, after considering the electromagnetic
potentials as well. These are computed by combining the result \eqref{eq:zeta-almo}
with \eqref{beta-def} for the function $W$, to find
\begin{equation}
 \zeta = -2\,e^{U} \Re [ e^{-i \alpha } \cV ] +(\beta - \tfrac1{V}) \, \hat R^*\,, \label{BPSpbeta}
\end{equation}
which leads to the following expression for the total spatial field strengths
\begin{align}
F= \zeta d\omega + dw = &\,
-\left(2\,e^{U} \Re [ e^{-i \alpha } \cV ] + \tfrac1{V} \, \hat R^* \right)  d \omega
\CR
&\,
- \star (d -2\,\dTgm)
\left[2\,\mbox{Im}(e^{-U-i\alpha}\cV) - \tfrac12\, V\,\hat{R} - \tfrac{M}{V}\,\hat{R}^*\right] \,.
\end{align}
Given that the constant $\beta$ does not appear in the gauge invariant
total field strengths, we conclude it corresponds to a residual gauge transformation
of the type $A \rightarrow A + \beta\, \hat{R}^*\,dt$ and therefore is unphysical.
Henceforth we set $\beta=0$ in all relations, for simplicity. With this choice,
\eqref{dw-real-7} is formally exactly the same as its counterpart in the
composite non-BPS system in \eqref{dw-real}, up to changing the relevant
T-dualities from those leaving $R$ invariant to those leaving $R^*$ invariant.

Note however that this choice of $\beta$ is not the most convenient one for all purposes,
as for example in showing that the above equations describe multi-centre BPS solutions as
a particular case. It can be shown that this is the case when $V$ is a constant, but one only
recovers the standard form of BPS solutions after imposing $\beta =  \frac{1}{V}$, as is
clear from equation \eqref{BPSpbeta}.

\subsection{Integration and local structure}
\label{sec:integr-7}

The presence of the T-duality connection $\dTgm$ in \eqref{dw-real-7} does not
allow for a straightforward solution in terms of harmonic functions, but one can
follow steps similar to the composite non-BPS system in order to solve the system
in terms of local functions. We can write the scalar and vector fields as
\begin{align}\label{scal-vec-H-7}
 2\,e^{-U}\mbox{Im}(e^{-i\alpha}\cV) - \tfrac12\, V\,\hat{R} -\tfrac{M}{V}\,\hat{R}^*=- \cH \,,
\qquad
dw = \star \left(d -2\,\dTgm\right) \cH \,,
\end{align}
where the vector of functions $\cH$ is the solution to the non-harmonic equation
\begin{equation}
 d\star d\cH -2\,d\star \dTgm \cH -2\,\dTgm\wedge \star d\cH=0 \,.
\end{equation}
This can be simplified and cast as a Poisson equation after introducing a
rescaled vector, as
\begin{equation}\label{def-H-resc-7}
 \cH_0 = \exp[-\Tgm]\cH\,,
\end{equation}
which in turn is the solution to the equation
\begin{equation}\label{LaplOnH-7}
 d\star d \cH_0 - d\star \dTgm \cH_0 -\dTgm\wedge\star \dTgm \cH_0 = 0 \,.
\end{equation}
This can be formally obtained from \eqref{LaplOnH} upon exchange of T-duality transformations.
In terms of the new vector, $\cH_0$, the scalar and vector fields are given by
\begin{gather}
 2\,e^{-U}\mbox{Im}(e^{-i\alpha}\cV) 
=- \exp[\Tgm] \left( \cH_0 - \tfrac12\, V\,R_0 -\tfrac{M}{V}\,\hat{R}^* \right) \,,
\CR
 dw = \exp[\Tgm]\star d\cH_0 - \star d( \exp[\Tgm] )\,\cH_0\,,
\label{dw-final-7}
\end{gather}
which is the final form of the system in the real basis.

The solutions to the above system can be characterised using the fact that
the components of the vector $\cH_0$ are restricted, in the following way.
The form \eqref{dw-final-7} of the vector fields allows us to compute
\begin{align} \label{ZdH0-7}
Z_a(\exp[\Tgm]\star d\cH_0)
&\,-\N[\bar \Omega]\Omega_a\, Z(\exp[\Tgm]\star d\cH_0)
\CR=&\,
Z_a(\star dw +\exp[\Tgm]\, \dTgm \cH_0 )
-\N[\bar \Omega]\Omega_a\, Z(\star dw + \exp[\Tgm]\, \dTgm \cH_0 )
\CR =&\,
 2\,\bar Q_{\scm\,a} - e^{i\alpha} N_a
  + 2i\,e^{-U} \Omega_a\, \gamma_\scm\, \Scal{3\,\frac{Y}{|Y|^2}-2}
\,.
\end{align}
The crucial observation is that the central charges in \eqref{ZdH0-7}
satisfy the reality constraint \eqref{ZAconstraint}, as
\begin{align}
e^{i \alpha}&\,\left[ \bar Z^a(\exp[\Tgm]\, d\cH_0) 
- \N[\Omega] \bar \Omega^a \bar Z(\exp[\Tgm]\, d\cH_0) \right]=
\CR
&\, \quad  c^{abc} \Omega_b Z_c(\exp[\Tgm]\, d\cH_0) 
 + \bar \Omega^a \left( Z(\exp[\Tgm]\, d\cH_0) - \N[\Omega]  \bar \Omega^b Z_b(\exp[\Tgm]\, d\cH_0) \right)
\label{dH-cons-7} \,,
\end{align}
which restricts the components of the corresponding vector to lie on a particular
Lagrangian subspace \cite{Bossard:2012xsa}. The same constraint holds for
the integrated vector $\cH_0$, for which
\begin{align} \label{ZH0-7}
Z_a(\exp[\Tgm]\cH_0)
-\N[\bar \Omega]\Omega_a\, Z(\exp[\Tgm]\cH_0)
=&\, - \Omega_a\, e^{-U} Y
\,.
\end{align}
The real form of this relation is the same as for the composite non-BPS system \eqref{sympl-constr-summ}, which we recall in this section for completeness 
\be \frac{1}{2} I_4^\prime(R_0,\hat{R}^*, \cH_0 ) = -\Iprod{R_0}{\hat{R}^*}    \cH_0 + 3 \Iprod{  \cH_0}{\hat{R}^*} {R}_0 \ , \qquad \Iprod{{R}_0}{  \cH_0} = 0 \ . \label{sympl-constr-pr-7} \ee
where we used \eqref{R0-real} to undo an overall T-duality on all terms in this equation.
The constraints  \eqref{sympl-constr-pr-7} are exactly dual to
\eqref{constr-K-7}, as they are related by replacing $\hat{R}^*$ and $R_0$.
We therefore conclude that $\cH_0$ and $\cK$ lie in opposite Lagrangian subspaces, \ie
the two vectors have no common directions and span $2 n_v +1$ independent components in the
$2(n_v+1)$-dimensional vector space.

To be more precise, the vector $\cK$ is of grade $(+1)$, as explained in
section \ref{sec:Tdual-7}, while the vector $\cH_0$ and its derivative lie in the
Lagrangian subspace composed by grade $(-1)$ and $(+3)$ components
in the decomposition \eqref{eq:vec-decomp}, as
\begin{equation}\label{H-grad-7}
\cH_0 \in ({\mathds{R}}^{n_v})^\ord{-1} \oplus \mathds{R}^\ord{3} \ ,
\end{equation}
exactly as in \eqref{H-grad-5}. Applying this to \eqref{LaplOnH-7}, the
following pattern arises for the various terms
\begin{align}\label{Lapl-dec-7}
&\,\, d\star d \cH_0  - d\star\dTgm \cH_0 - \dTgm\! \wedge \star \dTgm \cH_0=0 \,,
\\
&\, \scriptstyle{(-1)\oplus (+3)} \quad\,\,\,  \scriptstyle{(-3)\oplus (+1)}
 \qquad\qquad \scriptstyle{(-1)}
\nonumber
\end{align}
where we used the fact that $\Tgm$ lowers the grade of a vector by $(-2)$.
In direct correspondence with \eqref{Lapl-dec}-\eqref{Lapl-K} for the composite
non-BPS system, we find that \eqref{Lapl-dec-7} decomposes into two equations
according to its graded decomposition, as
\begin{gather}
d\star d \cH_0  - \dTgm\! \wedge \star \dTgm \cH_0=0 \,,
\label{Lapl-fin-7}
\\
d\star\dTgm =0\,, \label{Lapl-K-7}
\end{gather}
where in the second equation we used the property that no T-duality $\Tgm$ leaves
the vector \eqref{H-grad-7} invariant. This is true because the grade $3$ component
of $\cH_0$ can be identified as the nowhere vanishing harmonic function $V$ defined in
\eqref{V-def-alm}, as
\begin{align}\label{grade-3-alm}
 \Iprod{\hat R^*}{\cH_0} = - V\,,
\end{align}
as can be seen by contracting \eqref{dw-final-7} by $\hat{R}^*$.
We now analyse each of the two equations \eqref{Lapl-fin-7}-\eqref{Lapl-K-7} in turn.

The solution to \eqref{Lapl-K-7} is equivalent to the condition
\begin{equation}\label{K-def-7}
 d\star d \cK=0\,,
\end{equation}
where we used \eqref{R0-real}. It follows that the vector of parameters $\cK$ is a
generic grade $(+1)$ vector of harmonic functions, $\cK$, which is of rank three.
Note that, in this system, the poles of $\cK$ represent new independent physical
charges, since this vector is by definition linearly independent from $\cH_0$.
In fact, the poles of this function are the only relevant information, since one
may always absorb the constant part of $\cK$ into $R_0$ in \eqref{R0-real}.

We now turn to the Poisson equation \eqref{Lapl-fin-7} and observe that the source term
is of grade $(-1)$, according to \eqref{Lapl-dec-7}. It follows that only $n_v-1$ out
of the $n_v$ components of $\cH_0$ are sourced, leading to an equal number of non-harmonic
functions. The remaining  component is the harmonic function $V$, already identified in
\eqref{grade-3-alm} above. The source term in \eqref{Lapl-fin-7} can be computed explicitly
using \eqref{T-duality-Rstar}, with the result
\begin{align}\label{Lapl-source-7}
Z\left(-\exp[\Tgm]\,\dTgm \wedge \star \dTgm \cH_0]\right)
 =& -\frac{i}{64}\,V\,\frac{\partial I_4}{\partial \bar{Z}}(\hat{R}^*, d\hat{R}, \star d\hat{R})\,,
\nonumber\\
Z_a\left(-\exp[\Tgm]\,\dTgm \wedge \star \dTgm \cH_0 \right)
 =& \frac{i}{64}\,V\,\frac{\partial I_4}{\partial \bar{Z}^a}(\hat{R}^*, d\hat{R}, \star d\hat{R})
\,.
\end{align}
\ie it is proportional to the vector defined in \eqref{pre-source-7}. This vector is of
grade $(-1)$ by construction and it can be verified to satisfy the constraint in \eqref{dH-cons-7}.
We can now rewrite \eqref{Lapl-fin-7} as
\begin{equation}\label{Lapl-H-7}
 d\star d \cH_0 = \tfrac1{64}\,V\, I^{\prime}_4(\hat{R}^*, d\cK, \star d\cK)\,,
\end{equation}
where we used the definition of $\cK$ in \eqref{R0-real} and \eqref{pre-source-7}.

Finally, we present the covariant form for the equation determining the angular
momentum and the function $M$ in \eqref{Y-def-gen} and \eqref{M-def-7}. The starting
point is the first of \eqref{Tdual}, which upon use of \eqref{eq:mu-3} can be written
as
\begin{align}
\star d \omega - d M = \Iprod{\exp[\Tgm]\cH_0}{\star dw}=
\Iprod{\cH_0}{d\cH_0 - \dTgm\,\cH_0}\,,
\end{align}
where we also used \eqref{dw-final-7} in the second equality. Explicit computation of
the last expression using the definition \eqref{Tm1} leads to the alternative form
\begin{equation}\label{do-dm-7}
 \star d \omega - d M = -\tfrac12\,V\,\Iprod{\cH_0}{d \hat R}
   = -\tfrac12\,V\,\Iprod{\cH_0}{d \cK}\,.
\end{equation}
Taking the divergence and the curl of this equation one obtains the relevant equations
for the function $M$ and the angular momentum respectively. The resulting Poisson
equation for $M$ reads
\begin{align}\label{Poiss-M-7}
d \star d M
=&\,
\tfrac12\,d\scal{V\,\Iprod{\cH_0}{ \star d \cK}}
\,,
\end{align}
and can be solved once $\cH_0$ and $\cK$ are specified. Note that upon setting the
parameters, $\cK$, of the T-dualities to vanish, these equations imply that $M$
is a harmonic function, while $\omega$ is the corresponding dual one-form, consistent
with the single centre class.

This concludes our duality covariant presentation of the almost-BPS system
in terms of the real basis. In the next section, we summarise the final
form of the equations to be solved and we comment on some of the properties of
solutions.

\subsection{Summary of results}
\label{sec:summ-7}

In this short section, we summarise the relevant formulae for the 
almost-BPS system in the real basis. All relations presented here were
shown explicitly in the previous sections and we refer to the discussion
there for further details. We find it however useful, both for clarity and
for future applications, to give a as self-contained as possible account of
the final form of the system.

The ansatze for the metric and gauge fields are given in \eqref{metricMltc}
and \eqref{gauge-decop} in terms of the function $e^U$, the one-form $\omega$
and the spatial vector fields $dw$, while the electromagnetic potentials are
fixed by \eqref{BPSpbeta}.
The first order equation for the almost-BPS system is given by
\eqref{dw-real-7}, as
\begin{align} \label{dw-summ-7}
\star dw =- \left[d -2\,\dTgm_{\scriptscriptstyle \cK} \right]
[2\,\mbox{Im}(e^{-U-i\alpha}\cV) - \tfrac12\, V\,\hat{R} -\tfrac{M}{V}\,\hat{R}^*]
\,.
\end{align}
Here, $M$, $V$, are functions to be specified below, while $\hat{R}^*$ and
$\hat{R}$ are a constant and a non-constant very small vector respectively,
where $\Iprod{\hat{R}}{\hat{R}^*}= 4$. Here, the non-constant $\hat{R}$ is
related to a constant very small vector, $R_0^*$, by \eqref{R0-def} 
\begin{equation} 
 \hat{R}= \exp[\Tgm_\csK]\,R_0\,,
\end{equation}
which again satisfies $\Iprod{R_0}{\hat{R}^*}= 4$.
In all equations, $\Tgm_{\csK}$ is a generator of the
T-dualities leaving $\hat{R}^*$ invariant, parametrised by a vector of
harmonic functions, $\cK$. As discussed in section \ref{sec:T-duality}, the
vector parameter $\cK$ lies in the grade $(+1)$ component of the vector space
according to the decomposition implied by the T-duality. It is therefore
a three-charge vector satisfying
\begin{equation} 
\frac{1}{2} I^\prime(R_0,\hat{R}^*,\cK) = \Iprod{R_0}{\hat{R}^*}\, \cK \ ,
\label{grade-1-7} 
\end{equation}
which indeed specifies a vector of $n_v$ degrees of freedom.

The solutions to the flow equation \eqref{dw-summ-7} are simplified by
introducing a vector, $\cH_0$, of grade $(-1)\oplus(+3)$, \ie satisfying
\be 
\frac{1}{2} I_4^\prime(R_0,\hat{R}^*, \cH_0 ) = 
-\Iprod{R_0}{\hat{R}^*}\, \cH_0 + 3 \Iprod{  \cH_0}{\hat{R}^*} {R}_0 \ . 
\label{sympl-constr-summ-7} \ee
Note that \eqref{grade-1-7} follows from a similar constraint, obtained
by interchanging $\hat{R}^*$ with $R_0$, that projects to the
$(+1)\oplus(-3)$ component of the vector space. The equations resulting
from \eqref{dw-summ-7} upon use of $\cH_0$ in \eqref{sympl-constr-summ-7}, take the
form \eqref{dw-final-7}, 
\begin{gather}
 2\,e^{-U}\mbox{Im}(e^{-i\alpha}\cV) 
=-\exp[\Tgm_\csK] \left(\cH_0 - \tfrac12\, V\,R_0 - \tfrac{M}{V}\,\hat{R}^* \right) \,,
\label{scals-summ-7}
\\
 \star dw = \exp[\Tgm_\csK]
 \left( d\cH_0 - \dTgm_\csK\cH_0\right)\,,
\label{dw-fin-summ-7}
\end{gather}
where $V$ is now identified with the grade $(+3)$ component of $\cH_0$,
as $V=\Iprod{\cH_0}{\hat{R}^*}$. The compatibility relation for these
relations leads to the field equation for $\cH_0$, as in
\eqref{Lapl-H-7}
\begin{gather}
 d\star d \cH_0 
= \dTgm_\csK\! \wedge \star \dTgm_\csK
=\tfrac1{64}\,V\, I_4(d\cK,\star d\cK, \hat{R}^*)
\,. \label{Lapl-fin-summ-7}
\end{gather}
As the right hand side of this relation is a vector of grade $(-1)$, the
corresponding components of $\cH_0$ are not harmonic, whereas $V$ is, as
can be seen by taking the inner product of \eqref{Lapl-fin-summ-7} with
$\hat{R}^*$
\begin{equation}\label{V-Poiss-summ-7}
 \Iprod{\hat{R}^*}{d\star d \cH_0} = d\star d V = 0 \,,
\end{equation}
where we used \eqref{grade-1-7}.
The final dynamical equation required is the one for the function $M$
in \eqref{scals-summ-7} and the angular momentum vector $\omega$, both
of which are conveniently given by \eqref{do-dm-7}, as
\begin{align}\label{dom-summ-7}
\star d \omega - d M = 
\Iprod{\cH_0}{d\cH_0 - \dTgm_\csK\,\cH_0}= 
\tfrac12\,V\,\Iprod{\cK}{\cH_0} \,.
\end{align}
Taking the divergence of this equation, one obtains
a Poisson equations for $M$.

The equations above can be seen to be equivalent to the known formulation of
the almost-BPS system, as given in five dimensional supergravity
\cite{Goldstein:2008fq, Bena:2009ev, Bena:2009en}, by making a choice for the
constant vectors $R_0$ and $\hat R^*$. Indeed, upon choosing
\begin{equation}\label{sp-frame-7}
 R_0 \propto (0, \delta_I^0)\,, \qquad  \hat R^* \propto (\delta^I_0, 0)\,,
\end{equation}
where we disregard the (arbitrary) normalisation, one can show a complete
equivalence of the above to the original system in \cite{Goldstein:2008fq}.
This particular frame is convenient in that it allows to lift to five dimensional
solutions that are locally but not globally supersymmetric. However, our formulation
of the almost-BPS system is closed under four dimensional dualities and includes
all duals of the system in \cite{Goldstein:2008fq}.
More recently, it was shown in \cite{Hristov:2012nu} that some of the BPS structure
is preserved in four dimensions as well, upon reinterpreting the constant vector
$R^*_0$ as Fayet-Iliopoulos terms in a gauged theory.

We close with some comments on the structure of the solutions. First, the physical scalars
and the metric scale factor can be obtained by solving \eqref{scals-summ-7} in the standard
way \cite{Bates:2003vx}, once $\cH_0$ and $M$ are solved for. Since all quantities above are
appropriate combinations of the single centre solution in \cite{Bossard:2012xsa}, up to
overall T-dualities, it is possible to use many of the results given there. For instance,
the metric scale factor is given by \eqref{ScallingFactorInvariant}, as
\begin{align}
e^{-4U} =\,- I_4(\cH_0) - M^2 \ ,
\end{align}
in exactly the same way as for the composite non-BPS system. However, in this case
the situation is richer and more complicated, in view of the fact that the
grade $(-1)$ components of $\cH_0$ are not harmonic and its grade $3$ component $V$ does not necessarily carry a pole at all centres. Indeed, it turns out that not
all black holes described by the almost-BPS system are non-supersymmetric in isolation.
On the contrary, the presence of both BPS and non-BPS types of centres, is the
distinguishing property of this system, as shown in
\cite{Goldstein:2008fq, Bena:2009ev, Bena:2009en}. Clearly, the fact that the harmonic
functions $\cK$ lie in a subspace independent of the one where $\cH_0$ lives is the
crucial ingredient that allows for both BPS and non-BPS types of charges to exist
simultaneously. 

As seen in the case of the composite non-BPS system, solutions do not exist for all
charge configurations and this holds also in the almost-BPS system. Moreover, it is not
possible to obtain arbitrary asymptotic moduli for a given allowed charge configuration,
for exactly the same reasons explained in section \ref{sec:summ-5}. Indeed,
\eqref{dw-summ} and \eqref{dw-fin-summ} have exactly the same
structure in both cases, so that some of the $n_v+1$ asymptotic constants in $\cH_0$
and the parameters of $\hat R$ and $R^*_0$ will correspond to charges rather than moduli.
We once again refer to \cite{Bossard:tocome} for more details on the structure of
almost-BPS solutions in four dimensions and for explicit examples.

\section{Conclusion}
\label{sec:concl}

In this paper, we gave a comprehensive treatment of the flow equations describing
multi-centre under-rotating black holes in $\N=2$, $D=4$ supergravity coupled to
vector multiplets with a symmetric scalar manifold. In particular, we considered
the non-linear sigma model obtained after timelike dimensional reduction to three dimensions
and derived the general, frame independent, flow equations for two systems
of multi-centre non-BPS black holes, namely the composite non-BPS and almost-BPS systems.

This represents a generalisation of the systems given in specific frames in
\cite{Bossard:2011kz, Goldstein:2008fq}, to systems that are closed under
electric/magnetic duality.
The resulting structure for the vector fields and scalars in terms of real symplectic
vectors turns out to be very similar for both systems. In particular, both systems
are described in terms of space-dependent transformations along abelian subgroups of
isometries on the scalar target space. In terms of the natural embedding to string
theories, these subgroups of the full duality group are conjugate to the so-called
spectral flow transformations, that are combinations of T-dualities with gauge
transformations on the $p$-form gauge fields. In this paper, we refer to them simply
as  T-dualities for brevity.

The main distinction between these solutions and the BPS multi-centre solutions, is that the electromagnetic vector fields are not harmonic anymore, but satisfy instead 
\be  d \star d w - 2 d\Tgpm_{\csK} \wedge \star d w = 0 \ , \ee
where the functions $\cK$ are themselves harmonic. Note that the consistency of this equation requires that the generators $\Tgpm$ are indeed abelian, as for T-dualities. It follows from this equation that the poles of the harmonic functions $\cK$ contribute to the electromagnetic charges in a non-linear way. Despite the interpretation of these functions as parameters of abelian isometries of the scalar manifold, they are not associated to a gauging of the theory. 

The crucial property that makes a general discussion in terms of covariant objects possible
is that the action of general T-dualities can be given explicitly using the quartic invariant,
$I_4$, of symmetric special K\"ahler geometry.
Indeed, as summarised in sections \ref{sec:summ-5} and \ref{sec:summ-7},
all relevant equations are written in terms of this invariant only, evaluated for the real
vectors parametrising the solutions. In this form, these duality covariant systems are not significantly 
more complicated than the equations given for the composite non-BPS system in
\cite{Bossard:2011kz} and for the almost-BPS system in \cite{Goldstein:2008fq} and can
be solved in exactly the same way.

The main advantage of the formulation displayed in this paper is that one need not define
solutions in a fixed duality frame in terms of generic parameters, and only compute the
electromagnetic charges and asymptotic moduli a posteriori, as in the constructions of \cite{Bena:2009ev,Bena:2009en,Bossard:2011kz,Bossard:2012ge}. In contrast, one can start
from any configuration of physical charges satisfying the required criteria associated to each system
and construct the corresponding solution, using the results summarised in sections \ref{sec:summ-5} and \ref{sec:summ-5}. This is in particular very useful for studying the domain of stability of these solutions in moduli space. Using this formulation, one can start from a given set of electromagnetic charges consistent with the system (\eg they have to all mutually commute with a common very small vector $R$ in the composite non-BPS system), and parametrize the most general very small vectors $R$ and $R^*$ satisfying the corresponding constraints. Using the formulation of this paper, one can then determine the most general solution associated to a given 
charge configuration and define the domain of existence of such solutions in moduli space. As opposed to BPS solutions, the domain of existence of such solutions in moduli space will be restricted to a hypersurface of non-zero co-dimension. The normal directions to the hypersurface are probably not forbidden physically, but rather push us out of the domain where we know how to describe the solution. For example, the BPS solutions within the almost-BPS system only exist on a co-dimension one hypersurface, but only due to the fact that the charges at all centers
must be compatible with a single constant vector $R^*_0$, leading to a subset of all BPS solutions. Nonetheless, one can still wonder if there are walls of marginal stability within the hypersurfaces defined by each of
the two systems described in this paper, \ie whether there are boundaries of the domain of existence of such solutions at finite values of the moduli. We intend to study two-centre configurations in the aim of exhibiting (or not) walls of marginality for non-BPS solutions in a forthcoming publication.

From a more general point of view, the unified description of the two known
non-BPS systems and its relative simplicity are encouraging for further uncovering
the structure of non-BPS solutions in supergravity. In particular, the isometries of
the scalar manifold seem to play a crucial role not only in the effective
three-dimensional theory, but also in the real formulation in four dimensions. It
would be interesting to understand the role of these isometries in the reduction of
the equations of motion to first order systems, which is not clear from our treatment
in terms of nilpotent orbits. In fact, it is known that higher orbits, describing
more complicated systems of non-BPS solutions, exist and one might hope that similar
structures as the ones described in this paper appear in those cases as well.

\section*{Acknowledgement}
We thank Hermann Nicolai for pointing out to us reference \cite{Faulkner:1971}. This work was supported by
the French ANR contract 05-BLAN-NT09-573739, the ERC Advanced Grant no. 226371
and the ITN programme PITN-GA-2009-237920. The work of SK was supported in part
by the ANR grant 08-JCJC-0001-0, and by the ERC Starting Independent Researcher
Grant 240210-String-QCD-BH.

\begin{appendix}
 
\section{$\N=2$ supergravity and symmetric special K\"{a}hler geometry}\label{sec:sugra}
The bosonic Lagrangian of $\N=2$ supergravity coupled to $n_v$ vector multiplets
reads \cite{deWit:1984pk, deWit:1984px}
\begin{eqnarray}\label{eq:Poincare-4d}
8\pi\,e^{-1}\, {\cal L} &=&
  - \tfrac12  R -i \, \Iprod{D^{\mu} \bar \cV}{D_{\mu} \cV}
-\tfrac{\mathrm{1}}4\, F^I_{\mu\nu}\, G_I^{\mu\nu} \,.
\end{eqnarray}
Here, the $F^{I}_{\mu\nu} = \partial_\mu A^I_\nu - \partial_\nu A_\mu^I$ for
$I=0,\dots n_v$ encompass the graviphoton and the
gauge fields of the vector multiplets and $G_I^{\mu\nu}$ are the dual field strengths,
defined in terms of the $F^I_{\mu\nu}$ though the scalar dependent couplings,
whose explicit form will not be relevant in what follows. The gauge field
equations of motion and Bianchi identities can then be cast as a Bianchi
identity on the symplectic vector
\begin{equation}\label{eq:dual-gauge}
 \cF_{\mu\nu}=\begin{pmatrix} F_{\mu\nu}^I\\ G_{I\, \mu\nu}\end{pmatrix}\,,
\end{equation}
whose integral over any two-cycle defines the associated electromagnetic charges
through
\begin{equation}
\Gamma=\begin{pmatrix} p^I\\ q_{I}\end{pmatrix}
 =\frac{1}{2\pi} \,\int_{S^2} \cF\,.
\end{equation}

The physical scalar fields $t^i$, which parametrize a special K\"ahler space $\mathcal{M}_4$ of complex dimension $n_v$, only appear in \eqref{eq:Poincare-4d} through the section, $\cV$, of a holomorphic $U(1) \times Sp(2n_v+2,\mathbb{R})$ bundle over $\mathcal{M}_4$.
Choosing a basis, this section can be written in components in terms of scalars
$X^I$ as
\begin{equation}\label{eq:sym-sec}
\cV=\begin{pmatrix} X^I\\ F_I\end{pmatrix}\,, \qquad
F_I= \frac{\partial F}{\partial X^I}\,,
\end{equation}
where $F$ is a holomorphic function of degree two, called the prepotential,
which we will always consider to be cubic
\begin{equation}\label{prep-def}
F=-\frac{1}{6}c_{ijk}\frac{X^i X^j X^k}{X^0} \equiv  -\frac{\N[X]}{X^0} \,,
\end{equation}
for completely symmetric $c_{ijk}$, $i=1,\dots n_v$, and we introduced the cubic
norm $\N[X]$. The section $\mathcal{V}$ is subject to the constraint
\begin{equation}
  \label{eq:D-gauge}
  \Iprod{\bar{\mathcal{V}}}{\mathcal{V}} =  i  \,,
\end{equation}
and is uniquely determined by the physical scalar fields $t^i=\frac{X^i}{X^0}$ up to a local $U(1)$ transformation. The $U(1)$ gauge invariance of \eqref{eq:Poincare-4d} is ensured by the appearance of the K\"{a}hler connection $Q_\mu$ in the covariant derivative. The K\"{a}hler potential on $\mathcal{M}_4$ is defined up to an arbitrary holomorphic function $f(t)$ as
\begin{equation} \label{Kah-pot}
\cK = - \mbox{ln}\scal{   i \,  \N[t-\bar t]}  + f(t) + f( \bar t)  
\end{equation}
and we fixed the $U(1)$ gauge invariance in terms of K\"ahler transformations by requiring that the K\"{a}hler connection is determined by the K\"{a}hler potential as
\begin{equation} Q = \Im[ \partial_ i \cK dt^i] \label{Kah-conn}\ ,\end{equation}
such that
\begin{equation}
 \label{Kah-metr-der}
g_{i\bar \jmath}= \partial_i\partial_{\bar \jmath} \cK\,, \qquad
D_\mu\cV=(\partial_\mu +i\,Q_\mu)\cV = D_i \cV \, \partial_\mu t^i
  = (\partial_i \cV + \tfrac{1}{2}\partial_i \cK\,\cV) \, \partial_\mu t^i\,,
\end{equation}
where $D_i\cV$ is the corresponding K\"ahler covariant derivative on the components of the
section. With the prepotential \eqref{prep-def}, the special geometry identities \cite{Ceresole:1995ca} reduce to 
\begin{equation}
 \bar D_{\bar \jmath} D_i \cV=\,g_{i\bar \jmath} \cV\,, \qquad
  D_i D_{j} \cV= i  e^{\cK} \,c_{i j k} g^{k\bar k} \bar D_{\bar k} \bar \cV\,, \label{SpecialGeoId} 
\end{equation}
which are used throughout the main text.

We introduce the following notation for any symplectic vector $J$
\begin{align}\label{central-ch-def}
Z(J) = \Iprod{J}{\cV} \,,\qquad
Z_i(J) = \Iprod{J}{D_i \cV} \,,
\end{align}
with the understanding that when the argument is form valued, the operation is
applied component wise. For instance, the central charge of the gauge field is\begin{equation} Z(\cF)
 = e^{\frac{\cK}{2}} \Scal{ G_0 + t^i G_i +
\frac{1}{2} c_{ijk} t^i t^j F^k - \N[t] F^0 }\,, \end{equation}
for the prepotential \eqref{prep-def}. With these definitions it is possible to introduce a scalar dependent complex
basis for symplectic vectors, given by $(\cV,\, D_i\cV)$, so that any vector
$J$ can be expanded as
\begin{equation}\label{Z-expand}
J = 2 \Im[- \bar{Z}(J)\,\cV + g^{\bar \imath j} \bar{D}_{\bar \imath}  \bar{Z}(J)\, D_j \cV]\,,
\end{equation}
whereas the symplectic inner product can be expressed as
\begin{equation}\label{inter-prod-Z}
\Iprod{J_1}{J_2} = 2 \Im[- Z(J_1)\,\bar{Z}(J_2)
   + Z_a(J_1) \, \bar{Z}^a (J_2)]\,.
\end{equation}
Finally, we introduce the notion of complex selfduality of the gauge fields
\eqref{eq:dual-gauge}, which satisfy the identity
\begin{equation}
 \mathrm{J}\,\cF=-*\cF\,,\label{cmplx-sdual}
\end{equation}
where $\mathrm{J}$ is a scalar dependent complex structure defined as
\begin{equation}\label{CY-hodge}
\mathrm{J}\cV=-i \cV\,,\quad
\mathrm{J} D_i\cV=i  D_i\cV\,.
\end{equation}

\end{appendix}

\bibliography{PaperG} \bibliographystyle{JHEP}

\end{document}